\documentstyle[12pt,urthesis,epsfig]{report}

\newcommand{\bec}[1] {\begin{equation}\label{#1} }
\newcommand{\eec} {\end{equation} }

\newcommand{\beq}{\begin{equation} }
\newcommand{\eeq}{\end{equation}}

\newcommand{\bea}{\begin{eqnarray}}
\newcommand{\eea}{\end{eqnarray}}

\newcommand{\at}{\bar{t}}
\newcommand{\ab}{\bar{b}}

\newcommand{\la}{\mbox{$\lambda$}}

\def\as{\alpha_S}

\def\GeV{{\rm GeV}}
\def\lapprox{\lower .7ex\hbox{$\;\stackrel{\textstyle <}{\sim}\;$}}
\def\gapprox{\lower .7ex\hbox{$\;\stackrel{\textstyle >}{\sim}\;$}}

\begin{document}
\pagenumbering{roman}

\title{
The Top Quark at Linear Colliders: \\
Quantum Chromodynamics Corrections
}

\author{ Cosmin Macesanu }
\department{Physics and Astronomy}
\degree{Doctor of Philosophy}
\supervisor{Professor Lynne H. Orr}
\degreeyear{2001}
\maketitle

\pagestyle{headings}

%------------- dedication

\vspace*{4.cm}
\hspace*{8.cm} {\it to my family}

\newpage

% ------------ CV
\begin{center}
{ \large \bf Curriculum Vitae }
\end{center}

\vspace{1.cm}

The author was born in Bucharest, Romania, on May 13 1973. He attended
the Faculty of Physics at the University of Bucharest from 1991 to 1996,
and graduated with a Bachelor of Sciences degree in 1996. He came to 
the University of Rochester in the fall of 1996 and began graduate 
studies in the Department of Physics and Astronomy. He received a 
Robert Marshak Fellowship in 1996 and 1997. He pursued his research in the 
field of elementary particle physics under the direction of 
Professor Lynne H. Orr and received the Master of Science degree in 1997.
\newpage

{ \large \bf Acknowledgments}

\vspace{1.cm}

First, I want to thank my adviser, Professor Lynne H. Orr, for her support
of my work on this thesis and for suggesting the topic in the first place.
I would also like to express my gratitude for the cooperation and 
help of my colleagues and of many postdoctoral fellows who performed
research at the University of Rochester these past five years. 
Among these, I would
like to give special thanks to Dr. Doreen Wackeroth, who spend many hours
discussing with me the details of the evaluation of the electroweak
corrections to the $W$ pair production and decay process, and to Dr.
Rob Szalapsky, who helped me with the computation of Passarino-Veltman
functions and with the computational details of my work in general.
Moreover, I would like to thank Dr. Oleg Yakovlev for a discussion concerning
the diagrams contributing to the top production and decay process, 
and Dr. Alexander Chapovsky for discussions on the DPA approximation
and for providing me with routines for the evaluation of nonfactorizable
corrections in the soft gluon approximation.
\newpage

\abstractpage

{\def\baselinestretch{2.}\normalsize
We present a computation of QCD next-to-leading order corrections
to the top production and decay process at linear colliders. The 
top quarks are allowed to be off-shell and
the production and decay subprocesses are treated together, thus
allowing for interference effects. We consider
the case of real gluon radiation, as well as virtual corrections to
the tree level amplitude. The framework employed for our computation
is the double pole approximation (DPA). We describe the implementation of
this approximation for the top production and decay process
and compare it with the implementation of DPA for the evaluation
of QED corrections to the $W$ pair production at LEP II. Similarities
and differences between the two cases are pointed out. \par}

% apparently you hahe to do it this way.
{\def\baselinestretch{2.}\normalsize
The theoretical approach we present in this thesis is implemented in a 
Monte Carlo generator. The total amplitude is separated into 
gauge invariant parts which can be associated with radiative corrections 
to production and decay subprocesses, and interference between these.
The contributing amplitudes are computed using spinor
techniques, and include all top width effects, spin correlations and
$b$ quark mass effects. The results discussed for the real gluon 
radiation case include studies of the gluon radiation properties and
the effects of this radiation on top mass reconstruction. We also
examine the effects of interference between production- and decay- stage
radiation, whose magnitude is sensitive to the value of the top quark
width. After the computation of virtual corrections, we present results for 
the total top production cross sections, and we analyze the magnitude 
of nonfactorizable (interference) corrections.
We study the impact these corrections can have on the 
top invariant mass distributions. \par}

\endabstractpage

\addcontentsline{toc}{chapter}{Table of Contents}
\tableofcontents

\newpage
\listoftables
\addcontentsline{toc}{chapter}{List of Tables}
\newpage

\addcontentsline{toc}{chapter}{List of Figures}
\listoffigures
%\newpage

%\newpage

% ----- for 1 and a half spacing
%\linespread{1.3}

% ----- for double spacing
%\linespread{1.6}
% apparently, these commands do not work

%\pagestyle{headings}

\chapter{Introduction}
\pagenumbering{arabic}

Since ancient times, people have been looking for explanations for 
the natural phenomena surrounding us. With the birth of the 
modern scientific method in the 15th - 16th century, great steps 
forward have been made in our understanding of the natural laws.
The body of accumulated knowledge has naturally coalesced into
theories which deal with different aspects of reality. Examples of such
theories are Newtonian mechanics, which deals with the interaction of
normal bodies, optics, which deals with properties of light, and
electricity and magnetism, which deals with the electric and magnetic properties
of materials. As time has passed, there has been a tendency to 
look for ways to combine
these disparate theories into more fundamental ones, which encompass
and explain all the phenomena previously dealt with separately 
(Maxwell's combination of electricity, magnetism and optics into
electromagnetism is such an example).
In the past century, these searches 
have coalesced into a search for a unified theory which underlies all
the physical reality around us.

The result is the Standard Model of particle physics. This 
theory deals with the fundamental constituents of matter 
and their interactions (except gravity). According to our present knowledge, 
all ordinary matter is built from elementary particles: six
leptons and six quarks (spin 1/2 particles, or fermions), 
which are arranged into three families:
\bec{fermions}
\begin{array}{lccc}
\hbox{Leptons}: & \left( \begin{array}{c} \nu_e \\ e \end{array} \right)
	& \left( \begin{array}{c} \nu_{\mu} \\ \mu \end{array} \right)
	& \left( \begin{array}{c} \nu_{\tau} \\ \tau \end{array} \right) \\
\hbox{Quarks}: & \left( \begin{array}{c} u \\ d \end{array} \right)
	& \left( \begin{array}{c} c \\ s \end{array} \right)
	& \left( \begin{array}{c} t \\ b \end{array} \right)
\end{array} 
\eec
The interactions of these particles are mediated by the gauge bosons:
\bec{bosons}
 \hbox{Gauge bosons}: \ \ \gamma \ ,\ Z_0\ , \  W^{\pm}\ , \ g
\eec
which are spin 1 particles.
The mathematical framework  which describes these interactions 
is gauge field theory; according to this theory, each generator of the 
gauge symmetry group of the Lagrangian corresponds to one gauge force 
carrier. Thus, the photon ($\gamma$), $ Z_0$ and  $ W^{\pm}$ are carriers
of the electroweak gauge force (symmetry group $U(1) \times SU(2)$),
which is felt by all particles,
%mediates the interaction of leptons among themselves and
%the interactions of the leptons with the quarks, 
while the gluons ($g$) are the carriers of the strong gauge force 
(symmetry group $SU(3)$) which mediates the interactions of the 
quarks among themselves solely.

So far, the successes of the Standard Model are impressive;
some of its  predictions have been tested 
with a remarkable degree of accuracy, and, at this time, not a single
piece of experimental evidence contradicts it convincingly.
Still, there are pieces 
missing; for example, 
the exact mechanism of electroweak symmetry breaking
(which is responsible for giving masses to the particles in 
Eqs. \ref{fermions}, \ref{bosons}) is still unknown.
It is surmised that the breaking of electroweak symmetry is
driven by yet undiscovered bosons: either the Higgs particle(s), in the 
most promising models, or maybe top quark condensates 
(in technicolor models \cite{technicolor}). 
Supersymmetric (SUSY) models \cite{susy}, which solve some
theoretical problems in the Standard Model, also predict 
a whole slew of new particles (supersymmetric partners).
 In general, there is a great interest today in  what lies beyond the
Standard Model. (Since it does not include gravity, 
we know that it cannot be the final theory).
 It is expected that,  in coming years, experiments at 
 higher energies will help answer these questions.

%which are 
%expected to be produced at energies at most an order of magnitude
%larger than those available today.   
%There are theoretical models which account for this mechanism;
%The Higgs boson (the particle 

The study of the top quark might shed light on the answer to at least some 
of these questions.
The characteristics of the top quark  make it one of the most interesting 
elementary particles discovered so far. 
Its mass is quite large, about 175 GeV (we use natural units in which
$\hbar = c = 1$);
correspondingly, its Yukawa coupling (the coupling to the Higgs boson)
is of order unity; and this might be an indication
that the top plays a special role in the electroweak symmetry breaking
process. Moreover, the top Standard Model width is about 1.5 GeV, thus being
much larger than $\Lambda_{QCD} \sim 200$ MeV. 
This means that the top quark decays before
having time to hadronize \cite{top_had}, therefore providing us with a unique
opportunity to study the interactions of a bare quark. All these properties,
together with the fact that,
since the energies involved in the top production and decay processes are
large,
 we can use perturbative QCD for reliable theoretical predictions, 
insure that the study of the top quark will be one of the main
goals of particle physics for the next decade.

The top quark was discovered in 1995 at the Fermilab Tevatron
Collider \nolinebreak[3] \cite{top_discovery}.
Only a couple hundreds events have been identified
so far.
Due to limited statistics, the only precise information available on 
the top quark so far is its mass \nolinebreak[3]
\footnote{There are also results for other top quark parameters, like
the production cross section and its
couplings, but the values obtained have large statistical uncertainities.}; the
latest analysis \cite{top_mass} gives the value  $174.3 \pm 5.1$ GeV.
The Tevatron Run II, starting this summer, is expected to provide us 
with a sample of top events about an order of magnitude larger than
the one available so far; this will result in more precise determination 
of the top mass, width and couplings. 
The Large Hadron Collider, once it starts operating
in 2006, will be a top factory, producing more than 8 million top-antitop pairs
anually \cite{cern}.
However, the analysis of the data coming from hadron colliders is
complicated by uncertainties in the initial state, large QCD backgrounds,
etc. Thus, even with the large top sample provided by the LHC, the 
uncertainity
in the top mass measurement, for example, will be of order 1 to 2 GeV (due
mostly to systematic uncertainties \cite{cern}).

An $e^+ e^-$ linear collider with center of mass energy greater than the
$t \bar{t}$ production threshold (350 GeV) would be an ideal machine for 
the precision study of the top quark (among other things). 
At such a collider, the electron and positron in the initial state annihilate
and create a top-antitop pair, which in turn each decay into a $W b$ pair.
So far, 
this machine is in the design stage, with a projected date of completion
 not before
the year 2010. At this time, there are two main designs; 
the European TESLA
and the American-Japanese NLC-JLC (for details about the parameters
of these machines, see for example \cite{orange}).
Both machines are linear accelerators,
about 30 Km in length, operating at energies up to 1 TeV (800 GeV for TESLA)
with luminosities of order $10^{34} cm^{-1}s^{-1}$ (which means 100 inverse
femtobarns per year).

One of the main advantages of an $e^+ e^-$ collider versus a hadronic
one is that the energy of the initial state is well determined. Thus, 
it is possible to perform threshold studies for the production of  a
$t \bar{t}$ pair. The shape of the production cross section near the 
threshold is quite sensitive to a number of top quark parameters.
In this energy range, current analysis indicates 
that it is possible to measure the top mass with an accuracy of about 40 MeV
using only $10\ fb^{-1}$ \cite{orange}. With a larger data sample, 
measurements of the top decay width, Yukawa coupling and 
strong coupling constant at the several percent level can also be achieved.
Going to higher energies, we can 
 study the V-A structure of the top quark 
couplings to the gauge bosons ($ \gamma, Z$ and $W$) \cite{anom_coup}.
The information about couplings can be obtained by using spin correlations:
the top quarks are produced in certain spin states, as dictated by the
top - $ \gamma, Z$ couplings. Since the top decays before hadronization,
the spin states of the top directly influence the angular 
distributions of its decay products. Simulations show that by analyzing
kinematic variables of final state particles we can measure
top couplings at the several percent level \cite{orange}.

Of course, in order to obtain information about
fundamental parameters like the top quark mass
and couplings, experimental data is only a part of the equation. The other
part is a good theoretical understanding of the physical processes
which are studied. For the threshold region, comprehensive theoretical
studies (NNLO computation with resummation of large logarithms,
careful treatment of the renormalon ambiguity) have already been performed
\cite{top_thres}. Above threshold, the theory lags behind.
At 500 GeV center of mass energy, with a 500 $pb^{-1}$ integrated luminosity,
we will get about $3\times 10^5$ top quark pairs created; as a consequence, 
the experimental accuracy is better than one percent. Ideally, we would like 
to have a similar (or better) precision in the theoretical predictions.

As for most physical processes,
precise theoretical predictions for the top production and decay
require the computation of next to leading order corrections 
(either virtual or real).
These can be split into two categories: electroweak corrections, due
to the radiation of a photon or weak interaction gauge boson ($Z$ or $W$),
and QCD corrections, due to virtual or real gluon radiation. Since the
top is a quark, the QCD corrections are the most important ones, and these
corrections will be the subject of the present paper.

First, we start by reviewing previous work done in the area.
In the first approximation, the top production and decay processes
can be considered separately.
%in the approximation in which the decay is independent of production.
%naturally split into separate parts : production and decay.
 The QCD radiative corrections to 
individual subprocesses have been comprehensively studied.
For the top production subprocess:
$$ e^+ e^- \rightarrow t\ \bar{t}\ (g) $$ 
 there are computations for  virtual and soft (low energy) gluon radiation
\cite{jersak},
as well as for hard (higher energy) gluon radiation 
(\cite{jersak}, \cite{korner1}, \cite{parke1}, \cite{arndt1}
are just some examples).
Similarly, the top decay subprocess:
$$ t \rightarrow b\ W \ (g) $$
has been computed taking into consideration virtual and
hard gluons together (\cite{jeza1}, \cite{oakes}, \cite{czarn}).

Using these results, one can try to approximate the top production and
decay process by assuming that the intermediate tops are on the 
mass-shell (narrow
width approximation)  and treating the subprocesses separately
\cite{schmidt}. This assumption is usually reasonable;
the result for the total cross section is valid up to terms of 
order $\Gamma_t/m_t \sim 1 \%$. 
However, finite top width effects, which can be thought of
as interference between production and decay subprocesses, can be
important in some differential cross sections. Therefore, for precision
better than \% level, it is necessary to treat the production 
and decay together, by allowing the top to go off-shell. Thus, in
our computation we take into account the interference terms (also
known as nonfactorizable corrections).

 We shall perform this computation and present the results at the parton 
level only. Since the final state of the top quark pair 
production process is $ b W^+ \ab W^-$ (at lowest order),
the experimental signature  
is either 6 jets (if both $W$'s decay hadronically) or lepton(s) +
jets + missing energy (if one or both $W$'s have semileptonic decays). 
We assume that the issues related to jet reconstruction and 
identification have been solved, and our final state contains two $W$ bosons
\footnote{ At the Monte Carlo level, we actually allow the on-shell $W$'s
to decay, either semileptonically or into a pair of massless quarks, but 
in the latter case we do not take 
into consideration QCD corrections to the W decay.},
two $b$ quarks and possibly a gluon.
Even at this level, the complete computation of all the diagrams contributing
to this final state (Born and next to leading order) is a very difficult
task. Therefore, we shall employ the double pole approximation (DPA) which
means taking into account only the diagrams which contain a top - antitop pair.

It is worth mentioning a resemblance between the 
process of interest to us:
$$ e^+ e^- \rightarrow t\ \bar{t}\ (g) \rightarrow b\ W^+\ \bar{b}\ W^- \ (g)$$
and the $W$ pair production and 
decay process ($e^+ e^- \rightarrow W^+\ W^- \rightarrow 4f) $  
with QED corrections at LEP.
The issues which arise in the two computations are similar,
because in both cases we are dealing with the production and
decay of heavy unstable particles.
Our treatment is largely similar to the one used for the electroweak 
process \cite{racoon}, \cite{ddr}, \cite{BBC}. But there are
some differences, both in the implementation of the DPA approximation
and  in the number and type of terms which contribute to the final result
(the latter being due to the fact that in our case the
intermediate off-shell particles are fermions, and not bosons).  
These differences will be pointed out in the course of our discussion.

The outline of the thesis is as follows. In Chapter {\bf 2} we lay out the 
general framework in which we perform our computation. This includes 
a description of the double pole approximation (DPA) method for the top production
and decay case, with an overview of the Feynman diagrams contributing to 
this process. We also discuss the computation of 
differential cross sections, with 
some details on the treatment of the infrared singularities, and we
review some salient features of the treatment of the widths of unstable 
particles.
 
 Chapter  {\bf 3} is dedicated to an analysis of the top production 
and decay process in conjunction with the radiation of a real (hard) gluon.
We start with a description of the computation of the amplitudes contributing
to  this process. Since we perform a full off-shell computation, a prescription
for separating the total amplitude into parts which can be associated
with radiation in the production or decay stage is also given. Finally,
we study the properties of gluon radiation, and the impact of the gluon on
top mass reconstruction. Special attention is paid to the issue of interference
between production- and decay-stage radiation, and the effect which it 
can have on differential distributions.

The computation of virtual corrections is the subject of Chapter {\bf 4}.
The evaluation of NLO amplitudes in the double pole approximation is presented
and discussed in some detail. The results for the amplitudes
corresponding to interference diagrams are similar to results 
previously obtained
for the $W$ pair production process, while for the vertex corrections and
fermion self-energy diagrams there are differences between the two cases. 
In order to facilitate comparisons with the on-shell approach, we also
formulate our results in terms of correction to the production and 
decay subprocesses and interference contributions. The gauge invariance
of the total and partial amplitudes in DPA is manifest in this formulation.
Results for this section include values for the total cross section and
an analysis of the impact of interference effects on 
the invariant top mass distribution. Comparisons between our results
and results obtained in an alternative approach (in which the real  
gluon interference terms are computed analytically) are also made.

We end in Chapter {\bf 5} with a summary and the conclusions. The Appendix
contains some technical details related to our computation: we present 
a short description of the spinor techniques used to evaluate the 
amplitudes contributing to our process, expressions for these amplitudes,
and details about the evaluation of virtual corrections with the help 
of Passarino-Veltman functions. 

%---------------------------------------------------------------------------
\chapter{Computational Approach}

Our aim in this thesis is the evaluation of next-to-leading order
amplitudes and differential cross sections for the
top production and decay process at linear colliders. In this section,
we start with a description of the general framework in which we perform our
computation. The Feynman diagrams contributing to the process are introduced,
and the double pole approximation is explained. Some general issues 
related to the evaluation of the cross section, the treatment of infrared
singularities and the treatment of the top width are examined.
Details about the evaluation of
amplitudes and a comprehensive analysis of the specific issues
arising in this computation can be found in later sections.

\section{Amplitudes in DPA}

\begin{figure}[ht!] % fig 2
\centerline{\epsfig{file=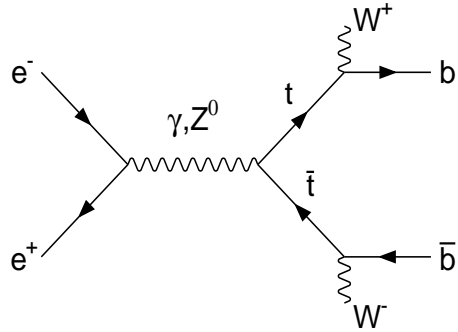,height=2.in,width=2.5in}}
\caption{The top-antitop diagrams contributing to the process (\ref{proc1}).}
\label{tree_level}
\end{figure}

\begin{figure}[ht!] % fig 3
\centerline{\epsfig{file=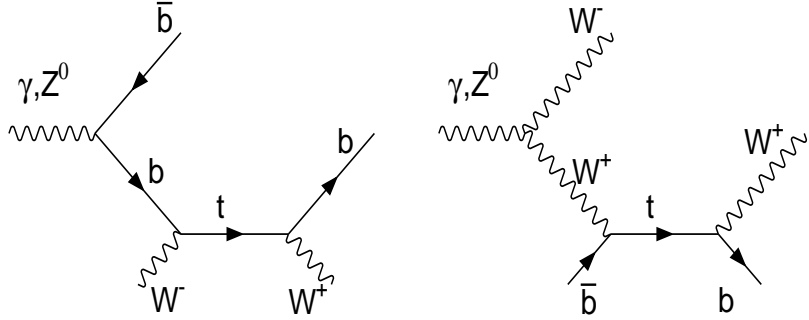,height=2.in,width=4.5in}}
\caption{Single top diagrams contributing to the process (\ref{proc1}).}
\label{sin_res_diag}
\end{figure}

In top pair production at linear colliders,
 what is actually observed experimentally is the process
\bec{proc1}
 e^+ e^- \rightarrow\ b\ W^+\ \bar{b}\ W^- 
\eec
There are many diagrams contributing to this process. At tree level,
they can be split into 3 classes: diagrams which contain a top-antitop
pair (Fig. \ref{tree_level}), diagrams which contain either one top quark
or a top antiquark (Fig. \ref{sin_res_diag} and charge conjugates),
 and diagrams which 
do not contain any top (there are about 50 such diagrams). Although not an easy 
task, it is possible to perform the evaluation of all these amplitudes 
(by using one of the automated tree level amplitude computation programs,
like MADGRAPH \cite{madgraph}). The computation of QCD corrections to all
tree level diagrams increases the degree of complexity by quite a lot, 
and is probably not feasible yet.

Fortunately, we can make use of the fact that, looking
for top pair production, we are interested in a specific region
of the final state phase space of (\ref{proc1}). This region 
is defined by the requirement that the invariant mass of the $W\ b$
pair is close to the top mass: 
$ p_t^2,\ p_{\bar{t}}^2  \approx m_t^2$, where
$ p_t = p_{W^+} + p_b  $ and $ p_{\bar{t}} =  p_{W^-} + p_{\bar{b}} $ 
\nolinebreak[5] \footnote{
 $p_i$ are the four-momenta of the particles involved in the process; 
 $p^{\mu} = (E;p_x,p_y,p_z)$ and 
 $p^2 = p^{\mu} p_{\mu} = E^2 - p_x^2 -p_y^2 -p_z^2$ is a quantity invariant
 under Lorenz transformations.
  We also denote 
the antiquarks by a bar over the corresponding quark symbol. }.
In this region, the amplitudes corresponding to the top-antitop diagrams
are enhanced by the two resonant propagators coming from the two intermediate
top quarks: 
\bec{fdask} {\cal{M}} \sim  \frac{1}{p_t^2 - \bar{m}_t^2}\ 
\frac{1}{p_{\bar{t}}^2 - \bar{m}_t^2}\ \eec
For this reason, the first class of diagrams (Fig. \ref{tree_level}) are called
doubly resonant diagrams. Correspondingly, the second class 
(Fig. \ref{sin_res_diag})  and the third class of diagrams,
which contain only one top quark propagator or none, 
are called singly resonant and non-resonant diagrams respectively.  

Note that the contributions coming from singly resonant diagrams 
(whose corresponding amplitudes contain a single resonant propagator)
are {\it reduced}
by a factor $\Gamma_t/m_t$ with respect to the doubly resonant contributions.
Therefore, in the first approximation we can neglect the singly-resonant
and non-resonant diagrams; we shall use the {\it double pole approximation}
(DPA) which means keeping only the amplitudes which have a doubly resonant 
behavior when the top and the antitop go on-shell. 

\begin{figure}[t]
\begin{center}
%\mbox{\epsfig{figure=diagrams.ps,width=16.0cm}}
\hspace*{-3cm}\mbox{\epsfig{figure=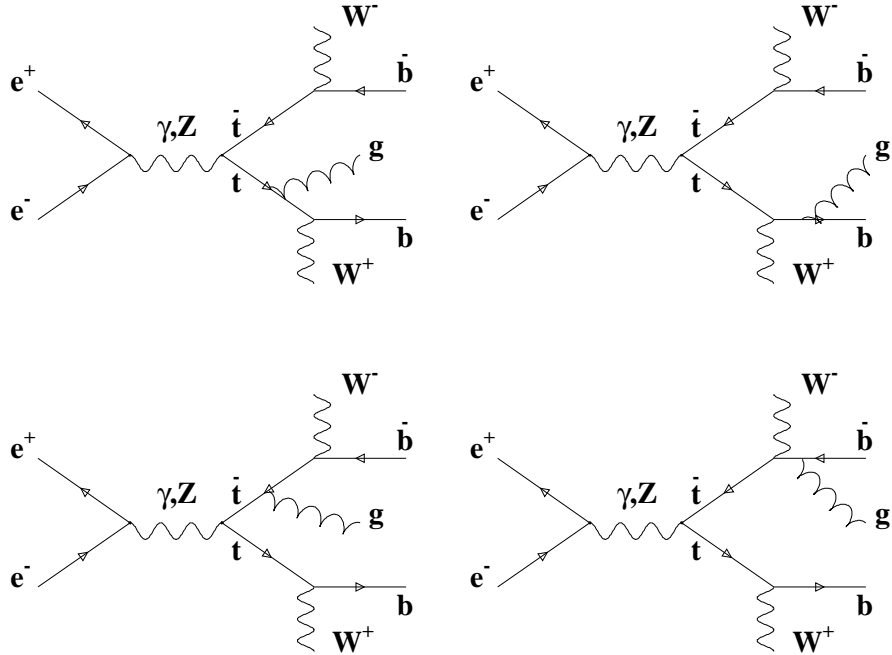,width=11.0cm}}
\caption{Feynman diagrams for gluon radiation in top production and decay.}
\label{diagrams}
\end{center}
\end{figure}

\begin{figure}[t!] % virtual corrections diagrams
\centerline{\epsfig{file=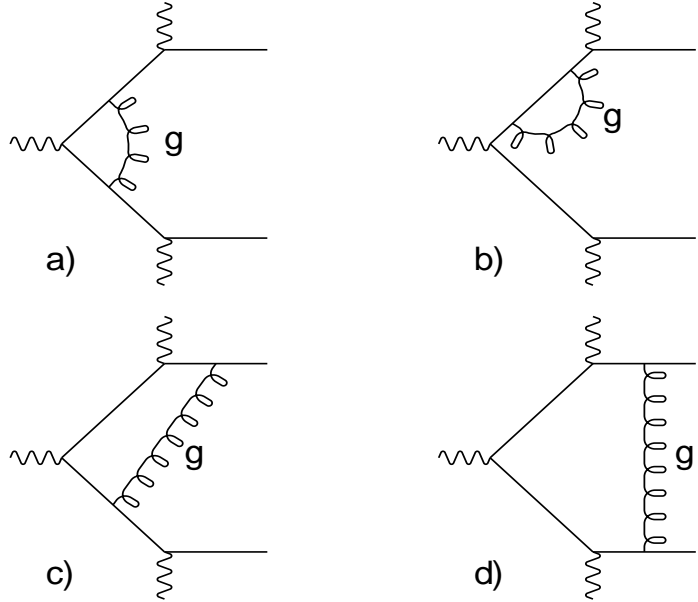,height=9.cm,width=11cm}}
\caption{Feynman diagrams for virtual gluon corrections to top production and decay.}
\label{vir_diag}
\end{figure}

At next to leading order, the diagrams contributing to the top production 
and decay process are presented in Figures \ref{diagrams} and \ref{vir_diag}. 
In Figure \ref{diagrams} we present the doubly-resonant 
diagrams contributing to the process with a real gluon in the final state:
\bec{fksa} e^+ e^- \rightarrow\ b\ W^+\ \bar{b}\ W^-\ g \eec
The DPA amplitude for this process can be written as a sum of four terms :
\bec{mrg_amp} {\cal{M}}^{rg} = {\cal{M}}_t + {\cal{M}}_{\at}
 + {\cal{M}}_b + {\cal{M}}_{\ab}\eec
each term corresponding to one of the diagrams in Figure \ref{diagrams}
(the subscript refers to the quark the gluon couples to).

Some of Feynman diagrams for virtual corrections to the tree level process
are presented in Figure \ref{vir_diag}. These diagrams can be divided into
two classes : corrections to particular subprocesses -- the vertex and 
fermion self-energy diagrams in Fig. \ref{vir_diag} a) and b) 
respectively -- and 
interference type corrections (Fig. \ref{vir_diag} c) and d)). Strictly
speaking, the vertex and self-energy diagrams also contribute to interference
between subprocesses; but, for the sake of brevity,
we shall refer to the diagram in Fig. \ref{vir_diag} a) as the production
vertex correction diagram, and so on. Also, in the following, we will denote
the tree level amplitude (Fig. \ref{tree_level}) by ${\cal{M}}^0$, and
the amplitude for the first order virtual corrections by ${\cal{M}}^{vg}$:
\bec{mvg_amp}
 {\cal{M}}^{vg} = {\cal{M}}_{t\bar{t}} + {\cal{M}}_{t b} +
  {\cal{M}}_{\bar{t}\bar{b}} + 
{\cal{M}}_{b\bar{t}} + {\cal{M}}_{t\bar{b}} + {\cal{M}}_{b\bar{b}} 
\eec 
Here, the  first three terms correspond to the three off-shell vertex
corrections (which include in a suitable way the fermion self-energies,
as described in section {\bf 4.1.3}), 
and the last three terms come from the interference diagrams.

\section{Cross sections and infrared singularities}

The partial amplitudes corresponding to the contributing Feynman diagrams
are evaluated using spinor techniques (see Appendix {\bf A} for details).
The differential cross sections
are obtained by taking the square of the total amplitude (sum of partial 
amplitudes) and summing over the polarizations (spins) of particles in the
final state:
\bec{crs_1}
 d\sigma^{0,vg,rg}\ \propto \ \sum_{polarizations} | {\cal{M}}^{0,vg,rg}|^2
\eec
The proportionality constant contains phase space and kinematic factors
and it is given in Appendix {\bf B}.
Our approach allows the computation of cross
sections for particles of definite polarizations in the final state;
but since these polarizations
cannot be experimentally observed, all our results will be given with the
spin sums in the final states performed.

Our computations are relevant to two different experimental situations:
the case when the final state contains only the $b, \ab, W^+ $ and $ W^- $
particles, and the case when besides these particles there is also a gluon.
In the latter case, the only amplitudes which contribute are the amplitudes
in Figure \ref{diagrams}; the total cross section can be written as: 
\bec{}
\sigma^{rg}\ \propto \ \int_{\epsilon_g > E_{cut}} | {\cal{M}}^{rg}|^2 \ 
d\Omega_{0+g}
\eec
(where $\Omega_0$ is the phase space of the tree level process, and 
$\Omega_{0+g}$ is the phase space of the process with a real gluon radiated).
Note that the integral is not performed over the entire phase space, but
a lower limit $E_{cut}$ on the gluon energy $\epsilon_g$ is imposed.
 The reasons
for this cut are twofold: first, experimental: only gluons with energies 
greater than a certain threshold can be observed; second, theoretical:
when the gluon energy goes to zero, the amplitude ${\cal{M}}^{rg}$ goes to 
infinity, and the cross section itself diverges. This is the well known
issue of infrared singularities in the radiation of a massless boson.
The way to deal with this divergent behavior is to note 
that it is not possible to observe
gluons with arbitrarily low energy, no matter how good our detector
performance.
Therefore, the contribution to the cross section coming from infinitesimally
small gluon energies   
should be added to the cross section for the process without a gluon in the
final state
% ( cite Bloch  Nordsieck?).
\footnote{The solution to the problem of infrared divergences 
is due to Bloch and Nordsieck \cite{bloch}; for a pedagogical introduction
see \cite{peskin_schroeder}; a complete treatment including methods
for evaluation the soft gluon integral in Eq. \ref{fac_form}
can be found in \cite{yfs}.}.
Thus the infrared contributions coming from soft real gluon radiation 
cancel out the equal but opposite-sign infrared quantities which
appear in the evaluation of the virtual corrections.

In consequence, the tree level diagrams (Fig. \ref{tree_level}), the virtual
corrections diagrams (Fig. \ref{vir_diag}) and the real gluon radiation
diagrams (Fig. \ref{diagrams}) all contribute to the NLO cross section
for the process with the $b, \ab, W^+, W^- $ final state :
\bec{a}
\sigma^{1}\ \propto \ 
\int \ \left( | {\cal{M}}^0 |^2 + 2\hbox{Re}[ {\cal{M}}^0 ({\cal{M}}^{vg})^*] \right)\ d\Omega_0\ +\ 
\int_{\epsilon_g < E_{cut}} | {\cal{M}}^{rg}|^2 \ 
d\Omega_{0+g}
\eec
The cancellation of infrared divergences between the real gluon radiation and
the virtual corrections part is performed by using the phase space slicing
method. This method amounts to introducing a technical cut-off
$\epsilon$ for the gluon energy; if the gluon energy is smaller than 
$\epsilon$, the contribution  ${\cal{M}}^{rg}$ is evaluated in the soft 
gluon approximation. In this approximation
the amplitude factorizes to the tree-level amplitude times an eikonal factor:
\bec{mrg_ir}
{\cal{M}}^{rg} = {\cal{M}}^0 \ \left( \frac{p_b^{\mu}}{k p_b} - 
\frac{p_{\ab}^{\mu}}{k p_{\ab}} \right)\ (-i g_s)\ \epsilon_{g\mu}
= {\cal{M}}^0 \ B^{\mu}\ (-i g_s)\ \epsilon_{g\mu}
\eec
(only gluons radiated from the on-shell b quarks contribute to the 
infrared singularity, as opposed to the case of the on-shell approximation,
when gluons radiated from the top quarks also give rise to a singular behavior).
The phase space also factorizes: 
$ d\Omega_{0+g} = d\Omega_0 \times d\Omega_g$; therefore
\bec{fac_form}
\int_{\epsilon_g < \epsilon} | {\cal{M}}^{rg}|^2 \ 
d\Omega_{0+g} = \int | {\cal{M}}^0 |^2 d\Omega_0 \ \times \left(
\frac{\as}{\pi}\int_{\epsilon_g < \epsilon} \frac{d^3k}{4\pi \epsilon_g}
(-1)B^{\mu}B_{\mu} \right)
\eec
The term in parentheses (let's call it $Y_{b\ab}$), which 
describes the effects of soft gluon radiation, can then be 
computed independently of the tree level amplitude.  
Since it is infrared divergent,  $Y_{b\ab}$ is usually evaluated 
with the help of some regularization procedure (we use mass regularization,
which means assuming that the gluon has an infinitesimally small mass $\mu$).
The same regularization procedure is used to compute the virtual corrections
diagrams which are infrared divergent (in our case, the
decay-decay interference diagram \ref{vir_diag} d) and corrections to the
self-energies of the $b$, $\ab$ quarks). The final result
\bec{d}
\sigma^{1}\ \propto \ 
\int \ \left( | {\cal{M}}^0 |^2 (1 + Y_{b\ab}) 
+ 2\hbox{Re}[ {\cal{M}}^0 ({\cal{M}}^{vg})^*] \right)\ d\Omega_0\ +\ 
\int_{\epsilon < \epsilon_g < E_{cut}} | {\cal{M}}^{rg}|^2 \ 
d\Omega_{0+g}
\eec
is finite and independent of the value of the regularization parameter $\mu$.

Finally, we shall make some comments on the choice of the 
$\epsilon$ and $E_{cut}$
parameters. The $E_{cut}$ is meant to separate between the experimentally 
observable gluons and those gluons soft enough to be undetectable in the 
experimental setup under use. For the detectors planned for an $e^+ e^-$
linear collider this would mean a value for $E_{cut}$ of order 5 to 10 GeV.
\footnote{More about our choice (as well as other cuts intended to select observable
gluons) can be found in section {\bf 3.2}}
On the other hand, $\epsilon$ is an unphysical technical 
parameter needed to solve
the infrared singularities problem. The only requirement on its value is
that it is small enough so that the soft gluon approximation works. As
we shall argue in section  {\bf 4.4}, this means that $\epsilon$ should
be much smaller than the energy of the gluons which contribute
to the interference (about 1.5 GeV).  We therefore shall use 
for the technical cut $\epsilon$ values of order 0.1 GeV.

\section{Top width and gauge invariance}

The implementation of finite width for unstable particles is a delicate 
problem. If we use the zero order propagator for the top quark:
$$ {\cal{S}}_0 (p) = \frac{i}{\not{p} - m_0} $$
the amplitude for our process will have an unphysical non-integrable singularity
in the region of the phase space where the top propagator becomes resonant.
Hence the need to regularize this singularity by taking into account the
unstable particle's width. 
However, the introduction of finite width tends to spoil gauge invariance.
The reason for this is that our theory is gauge invariant {\it in fixed order
in perturbation theory}. The introduction of the width, performed
naturally through Dyson resummation
\footnote{ Here $ m_0$ is the bare mass of the particle, while
$\Sigma (\not{p})$ stands for the one-particle irreducible self-energy.
%A more detailed discussion of the evaluation of the NLO top quark propagator
%can be found in section ***
}:
\bec{t_prop}
{\cal{S}} = {\cal{S}}_0 + {\cal{S}}_0 (-i \Sigma (\not{p})) {\cal{S}}_0
+ \ldots = \frac{i}{\not{p} - m_0 - \Sigma (\not{p})}
\eec
amounts to taking into account contributions coming from higher orders in 
perturbation theory. Therefore, gauge invariance problems might arise.

These issues have been first recognized and dealt with
in processes where the role
of unstable particles is played by gauge bosons. There are a variety of methods
to restore gauge invariance in these cases; the fermion-loop scheme
\cite{f_loop} is a favored one, although it is difficult to implement 
for complicated processes. For the $W$ pair production process at LEP, 
a simpler prescription is available \cite{racoon}, \cite{ddr}, \cite{BBC}. 
In the computation of the tree level DPA amplitude:
\bec{WWproc}
{\cal{M}}_{DPA} = \frac{ \tilde{\cal{M}} (\ p_{W^+}^2, p_{W^-}^2)}
{(p_{W^+}^2 -M_W^2 + i M_W\Gamma_W)(p_{W^-}^2 - M_W^2 + i M_W\Gamma_W)}
\eec
one evaluates the normalized amplitude $ \tilde{\cal{M}} $ at the poles 
(that is, for on-shell $W$'s). The residue $ \tilde{\cal{M}} (M_W^2, M_W^2)$
is gauge invariant, and the difference between
this residue and the exact result is obviously non-doubly resonant,
therefore it can be ignored within the DPA. Thus, in this approach the
consequences of having off-shell particles are restricted to the denominators
of the $W$ gauge bosons propagators. Moreover, 
since the DPA radiative corrections 
to this process are proportional to the tree level amplitude,
gauge invariance at NLO is insured also.

However, the issue of gauge invariance in the $W$ pair production process
appears even at tree level because of the fact that $W$ {\it is a gauge boson}.
In other words, the gauge invariance we are concerned with in Eq. \ref{WWproc}
is invariance with respect to the $W$ gauge. For the top production process,
this issue does not arise; therefore we can compute the tree level amplitude
${\cal{M}}_0$ with off-shell momenta, and the result will be gauge independent
because the top quark propagator does not depend on any gauge. 

At next to leading order, though, the top quark can radiate a gluon
(either real or virtual), and
the problem of gauge invariance with respect to the gluon gauge arises.
However, this is a quite different problem from the one discussed above; 
even in the $W$ pair production case, at NLO gauge invariance with 
respect to the gauge of the radiated photon  has to be 
treated separately. For QCD corrections to the top pair production
case, we shall address this problem for both the case of real gluon radiation 
and virtual corrections in the appropriate chapters below. 

There is one more issue related to the treatment of the unstable particle
width. The width extracted from the imaginary part of the particle self-energy
in Eq. \ref{t_prop} is a quantity dependent on the top energy -- called the
running width. In the $W$ pair production process, it has been shown that
the use of the running width generates problems at higher energies
\cite{racoon}. Another
option is to use the fixed width obtained by evaluating the top decay
width from the process $t \rightarrow b W (g)$, and using this constant
value in the top propagator. In the $W$ case, it has been found preferable 
to use the {\it complex-mass scheme}, which replaces $M_W^2$ with the 
complex mass $\bar{M}^2 = M_W^2 - i M_W \Gamma_W$ not only in the denominator
of the propagator, but also in the couplings of the particles to the gauge 
bosons (since these couplings can be treated as dependent on the $M_W$ mass).
In the top production case, the couplings do not depend on the top mass; so
in our implementation, the top mass is replaced by the complex mass
$\bar{m}_t^2 = m_t^2 - i m_t \Gamma_t$ 
(here $\Gamma_t$  is the constant next-to-leading order top width) 
in the top propagator denominators only.

%\begin{subequations}
%\end{subequations}
%----------------------------------------------------------------------

\chapter{Real gluon radiation}

Future high energy lepton colliders --- $e^+e^-$ and $\mu^+\mu^-$ --- can
provide relatively clean 
environments in which to study top quark physics.  Although top production
cross sections are likely to be lower at these machines than at hadron
colliders, the color-singlet initial states  give lepton machines some
advantages.  Furthermore, the fact that the 
laboratory and hard process center-of-mass frames coincide greatly
simplifies the reconstruction of final states.  In addition, many of 
the top quark's  couplings, 
especially those to the photon and $Z^0$ boson, can be easily studied there.

The potential for precision studies of top  physics at such colliders 
requires precision predictions from the theory, beyond leading order
in perturbation theory.  
In particular, QCD corrections must be taken into account.  One effect 
of the QCD interaction is the radiation of a real gluon. This process
can have a sizable impact on the analysis of 
the top production and decay final state variables.
Jets from radiated gluons can be indistinguishable from quark jets, 
complicating identification of top quark events from reconstruction of
top's  decay products.  To make matters worse, 
emission may occur in either the top production or decay
processes, so that radiated gluons may or may not themselves be 
products of the decay.  Subsequent  mass measurements can 
be degraded, not only from misidentification of jets but also
from subtle effects such as jet broadening when gluons are 
emitted near other partons.

In this section we study the effects of real gluons radiated in
top quark production and decay at $e^+e^-$ colliders
\cite{hardglu}.  We consider 
collision energies well above the top pair 
production threshold, so although for definiteness we will refer
to electrons in the initial state, our parton-level results
apply equally well to $\mu^+\mu^-$ collisions at the same energy.
We allow for the top quarks to be off-shell, 
keeping the full width-dependent top propagator and retaining all
spin correlations.
Gluon radiation for off-shell top has been
treated previously in the soft gluon approximation  \cite{jikia,kos}.
Here we give an exact treatment for arbitrary gluon energies.
We study properties of the radiated gluons,  top
mass reconstruction, and effects of interference between 
production- and decay-stage gluons that can be sensitive to
the top quark width.

%\newpage
\section{Amplitude evaluation}

Our aim is to evaluate the differential cross section for  real
gluon emission in top quark production and decay: 
\begin{equation}
e^+e^- \rightarrow \gamma^*, Z^* \rightarrow t\bar{t} (g)
\rightarrow bW^+ \bar{b}W^-g\; .
\end{equation}
In top events at $e^+e^-$ colliders, there are no gluons radiated from the 
color-singlet initial state.  Final-state gluon emission can occur in 
both the production and decay processes, with gluons radiated from the
top or bottom quarks (or antiquarks), as shown in Figure \ref{diagrams}.

For purposes related to top mass reconstruction, it is desirable to
be able to associate the radiated gluon with either the top production
 process or the decay process. While  trivial in 
the on-shell approach, it is not possible to make this differentiation
with 100\% certainty
if the top is allowed to be off-shell.
However, we can use a sensible, {\it ad hoc}, definition: if the top is
closer to its on-shell mass after the gluon has been emitted, call
that production stage radiation; otherwise, call it decay stage radiation.
Thus, emission from the top quark contributes 
to both production- and decay-stage
radiation, depending on when the top quark goes on-shell.  Emission 
from the $b$ quarks contributes to decay-stage radiation only.  The
separation of these contributions at the amplitude level 
will be discussed below.

We compute the exact quantum mechanical amplitudes 
(also called matrix elements) for the diagrams shown in Figure 
\ref{diagrams} with all spin correlations and the bottom mass
included,  using the helicity methods
of Kleiss and Stirling     \cite{kleiss}.  
Working at the matrix element, rather than
 the matrix element squared, level has the usual advantages of 
numerical efficiency, and in our case has the additional 
advantage that we can identify individual contributions as well as 
interference between them.  The explicit 
expressions for the matrix elements are complicated and not particularly 
illuminating, so we do not reproduce them here\footnote{A FORTRAN 
program containing the matrix elements  
can be obtained from the author.}. A general overview
including some specific details concerning this computation
%the particular definitions of massive spinors we use 
can be found in the appendix.

We do not assume the top quark to be 
on-shell; therefore we keep the 
finite top width $\Gamma_t$ in the top quark propagator and include
all interferences between diagrams.  We use exact kinematics in
all parts of the calculation.
We do not include radiation from hadronic $W$ decays; the 
$W$ bosons are assumed to decay leptonically and we integrate over 
the decay products in the results presented here. 
In practice, radiative hadronic $W$ decays should probably be taken care of
at the level of jet fragmentation simulations;
but this requires a separate study.

%--------------------------------------------------------------------
\subsection{Production--decay decomposition}

As mentioned above, calculating at the amplitude level allows us to 
identify contributions from individual processes and 
their interferences.  We are particularly interested in 
distinguishing between contributions from gluons
radiated in the top quark production and decay stages.  
This is directly related to reconstruction of the top quark
momentum  from its decay products,
which in an experiment 
allows us both to identify top events and to measure $m_t$. 
 The presence of gluon radiation complicates 
the reconstruction because 
the emitted gluon may or may not be part of the top decay.
If the gluon is not part of the decay, then it is represents
a correction to top production and  should not be included in
the top momentum reconstruction:
\beq
m_t^2\approx p_t^2=(p_b+p_W)^2\equiv p_{bW}^2 \; .
\label{mtprod}
\eeq
(The use of the $\approx$ sign in the above equation signifies
that the sum squared of the 4-momenta of the top decay products
-- for which we will use the shorthand $p_{Wb}$ -- is close to the 
top mass squared.)
If on the other hand the gluon is part of the decay, then it 
{\it should} be  included in top reconstruction:
\beq
m_t^2\approx p_{tg}^2=(p_b+p_W+p_g)^2\equiv p_{bWg}^2 \; . 
\label{mtdecay}
\eeq
Being able to make this distinction turns out to be useful for purposes
of efficient phase-space integration as well.

Although this production--decay distinction cannot be made absolutely 
in an experiment\footnote{If the interference between processes is 
large this distinction is not even meaningful.}, 
the various contributions can be separated in the calculation.  
For radiation from the $b$ and $\bar{b}$ quarks, the assignment is 
easy:  these contributions, corresponding to the two right-hand 
diagrams in Fig.~\ref{diagrams}, are clearly part of the top
quark decay.  However, as noted
above,  gluon emission from the top quark (or antiquark) contributes
to both the production and decay stages; which is which depends on 
whether the  top was closer to its mass shell  before or after  
emitting the gluon.  This condition corresponds to which of the two 
propagators from the top that emitted the gluon is numerically larger.

We can make the separation in our calculation
as follows \cite{kos}.  For definiteness, we consider gluon emission from
the top quark, shown in the upper left diagram in Fig.~\ref{diagrams}.
The matrix element for this diagram contains propagators for the 
top quark both before 
and after it radiates the gluon.  The matrix element therefore 
contains the factors 
\begin{equation}
{\cal{M}}_t \, \propto \,\left( {{1}\over{p_{Wbg}^2-m_t^2+im_t\Gamma_t}}\right)
\left( {{1}\over{p_{Wb}^2-m_t^2+im_t\Gamma_t}}\right)\; .
\label{twoprop}
\end{equation}
The right-hand side can be rearranged to give 
\begin{equation}
{\cal{M}}_t\, \propto \,{{1}\over{2p_{Wb}*p_g}}
\left( {{1}\over{p_{Wb}^2-m_t^2+im_t\Gamma_t}} -
{{1}\over{p_{Wbg}^2-m_t^2+im_t\Gamma_t}}\right)\; .
\label{decomp}
\end{equation}
This separates the production and decay contributions to the matrix element.
The first term in parentheses contains a propagator that 
peaks when  $p_{Wb}^2=m_t^2$, which corresponds to the condition for
production stage, as in Eq.~\ref{mtprod}.  Similarly, the 
second term peaks for $p_{Wbg}^2=m_t^2$, which 
corresponds to decay emission as in Eq.~\ref{mtdecay}.  

The complete amplitude in Eq. \ref{mrg_amp}
can now be rewritten schematically as
\footnote{Expressions for the partial amplitudes can be found in 
Appendix {\bf B}.} 
\bec{mrg_gi}
{\cal{M}}^{rg} = {\cal{M}}_{prod} + {\cal{M}}_{tdecay}+ {\cal{M}}_{\bar{t}decay}\; .
\eec
The cross section, obtained from taking the absolute square of 
${\cal{M}}^{rg}$,
then contains separate production and decay contributions, from 
$|{{\cal{M}}}_{prod}|^2$ and  $|{\cal{M}}_{tdecay}|^2$,  $|{\cal{M}}_{\bar{t}decay}|^2$,
respectively.  It also contains cross terms representing the interferences,
which in principle confound the separation but in 
practice are quite small.

The interference terms {\it are}  interesting in their own right, although
not for top reconstruction.  In particular, the interference between 
production- and decay-stage radiation can be sensitive to the top
quark width $\Gamma_t$  \cite{jikia,kos}, which is  1.42 GeV in the 
Standard Model at ${\cal{O}}(\alpha_s)$  \cite{cern}.
The interference between for example the two propagators shown 
in Eq.~\ref{decomp}  can be 
thought of as giving rise to two overlapping Breit-Wigner resonances.  The
peaks are separated roughly by the gluon energy, and each curve
has width $\Gamma_t$.  Therefore when the gluon energy becomes comparable 
to the top
width, the two Breit-Wigners overlap and there can be substantial
interference.
In contrast, if the gluon energy is much larger than $\Gamma_t$, the 
overlap --- and hence the interference ---  is negligible. 
Therefore  the amount of interference 
serves as a measure of the top width.  We will explore this more below.

%--------------------------------------------------------------------
\subsection{Gauge invariance}

Since gauge bosons contribute to the processes analyzed in this thesis,
it is  natural to consider the issue of gauge invariance. 
In order to perform our computations, it is necessary to choose some gauge 
for the bosons involved (for example, the propagators for the
$\gamma, Z_0$ bosons are computed  the Feynman gauge, while the 
definition of the gluon polarization vector depends on a gauge parameter
as described in the appendix).
While our choice of gauge may vary, the final results
should be independent of this choice.

Due to the fact that the electrons in the initial state are considered
massless, there is no amplitude dependence on the gauge in which we evaluate 
the $\gamma, Z_0$ propagators. However, this is not true anymore for the
case of the gluon gauge. This has to do with the fact that the top 
quarks can be off-shell; it is an easy exercise showing that amplitudes
in the on-shell approximation are independent of the gluon gauge.
More precisely, the reason that gauge invariance is lost is that the physical 
process we actually compute is:
\bec{eettg}
 e^+ e^- \rightarrow\ b\ W^+\ \bar{b}\ W^- g
\eec
and to obtain a gauge invariant result we should take into account all the 
Feynman diagrams contributing to this final state.
Since our computation includes only diagrams with two intermediate top quarks,
the final result is not strictly gauge invariant.

However, the diagrams we take into consideration are the only ones 
from the set contributing to (\ref{eettg}) which 
have a doubly resonant structure. So, if one considers the gauge invariant
result as being a sum of doubly-resonant, singly resonant and nonresonant
terms, the only diagrams which can contribute doubly resonant terms are the 
 diagrams containing two top quarks in Figure 
\ref{diagrams}. Therefore, the amplitude in (\ref{mrg_gi}) differs from the
gauge invariant result by singly or non-resonant terms only. 
This means that by subtracting non-doubly
resonant terms from ${\cal{M}}_{tot}$ (which is allowed in DPA) 
 we can obtain a gauge invariant answer.

One way to perform these subtractions is as follows.
Consider the diagram where the gluon is radiated by the top quark
(the first diagram in Fig.~\ref{diagrams}). 
The contribution of this diagram to the production amplitude is:
\beq
\label{gi_tprod1}
 {\cal{M}}_{prod}^{(t)} \sim \frac{1}{2k p_t}\ 
 \bar{u}(b)\ \not{\epsilon}_{W}\
 \frac{\not{p}_{Wb} + m_t}{p_{Wb}^2 - \bar{m}_t^2}\
\not{\epsilon}_{g} \ (\not{p}_{Wb}+\not{k} + m_t) \ldots v(\bar{b})\; .
\eeq
By commuting $\not{\epsilon}_{g}$ to the right, this can be written 
\beq
\label{gi_tprod2}
 {\cal{M}}_{prod}^{(t)} \sim \frac{1}{2k p_t}\ 
 \bar{u}(b)\ \not{\epsilon}_{W}\
 \frac{(\not{p}_{Wb} + m_t)
( 2 \epsilon_{g} \cdot p_{Wb} + \not{\epsilon}_{g} \not{k} ) -
(p_{Wb}^2 - m_t^2) \not{\epsilon}_{g}
 }{p_{Wb}^2 - \bar{m}_t^2}\
 \ldots v(\bar{b})\; .
\eeq                     
The term which breaks gauge invariance here is 
the one proportional to $( p_{Wb}^2-m_t^2 )$. 
This is a non-resonant term, regardless of the gluon being
radiated in the production or decay stage (in other words, regardless of
$ p_{Wb}^2 \approx m_t^2$ or $p_{Wbg}^2 \approx m_t^2 $); therefore, in 
keeping
with the approximation used, we can neglect it. 

A similar analysis works for the contribution of this diagram to the top decay 
amplitude. Starting with the expression: 
\bec{gi_tdec1}
 {\cal{M}}_{tdecay}^{(t)} \sim \frac{-1}{2k p_t}\
  \bar{u}(b)\ \not{\epsilon}_{W}\
 \frac{\not{p}_{Wbg} -\not{k} + m_t}{p_{Wbg}^2 - \bar{m}_t^2}\
\not{\epsilon}_{g} \ (\not{p}_{Wbg} + m_t) \ldots v(\bar{b})\; 
\eec 
by commuting $\not{\epsilon}_{g}$ to the left, we obtain:
\beq
\label{gi_tdec2}
{\cal{M}}_{tdecay}^{(t)} \sim \frac{1}{2k p_t} \bar{u}(b) \not{\epsilon}_{W}
 \frac{ \not{\epsilon}_{g} (p_{Wbg}^2 - m_t^2) - 
 (2 \epsilon_{g} \cdot p_{Wbg} - \not{k} \not{\epsilon}_{g}) 
 	(\not{p}_{Wbg} + m_t)}   {p_{Wbg}^2 - \bar{m}_t^2}
 \ldots v(\bar{b})\;
\eeq
%with $p_{Wb}$ replaced by  $p_{Wbg}$
and in this case we drop the term proportional to $( p_{Wbg}^2-m_t^2 )$.
Finally, the amplitudes corresponding  to the diagram in which the gluon
originates from the $\bar{t}$ can be computed in the same manner. The 
final result is gauge invariant, and differs from the exact
result by non-doubly-resonant terms, as we have shown.

We have implemented the above computation in the Monte Carlo program, and 
have checked numerically that the difference between the gauge invariant result
and the exact result is very small (of order 0.01\% of the total cross 
section,  and 
order 1\% with respect to the interference terms). This indicates that the 
other non-doubly resonant contributions (coming from diagrams with a 
single top or none) are also small; a more detailed study is in progress.

Finally, we note  that this method for restoring gauge invariance
is not unique. We could, for example, have instead  replaced the
top mass in the top propagator numerator with the invariant masses:
$\sqrt{p_{Wb}^2}$ in the production amplitude, and  $\sqrt{p_{Wbg}^2}$ in 
the decay amplitudes. The result obtained with this method 
is also gauge invariant, and also differs
from the exact result by non-resonant terms.

\subsection{Monte Carlo and phase space integration}

The integration over the final state phase space to obtain the cross
section involves an integrand that contains multiple Breit-Wigner
peaks from the top quark propagators as well as infrared singularities
when the gluon energy becomes small.  Even with 
cuts on $E_g$, the rapid variation of the integrand can spoil the integration
procedure. To eliminate this problem, we tailor the momentum generator
to the production of a gluon in association with two massive particles
($\gamma^*, Z^* \rightarrow t \bar{t} g$ or $t  \rightarrow b W g $).
The multiple Breit-Wigner peaks are taken into account by using a 
multi-channel approach that integrates separately over the 
individual production and decay contributions; the Breit-Wigner
behavior is smoothed out in the phase space generation.  The 
interference terms, which have products of Breit-Wigners that
peak in different places, much like in Eq.~\ref{twoprop},
 are integrated using
a combination of the three main channels.

%----------------------------------------------------------------------
%%%%%%%%%%%%%%%%%%%%%%%%%%%%%%%%%%%%%%%%%%%%%%%%%%%%%%%%%%%%%%%%%%%%%%%
%----------------------------------------------------------------------
\section{Numerical Results}

In this section we show results of the numerical calculation described
above.  We present the cross section for 
$b\bar{b}W^+W^-g$ production in
$e^+e^-$ collisions at a 500 GeV center-of-mass 
energy, with a few exceptions which are clearly identified.  
The calculation is entirely at the parton level, and we do not 
include initial state radiation, beam energy spread, or beamstrahlung.
We use the 
following numerical values of parameters:  $m_t=175\ \GeV$,
$m_b=5\ \GeV$, $M_W=80\ \GeV$, $\Gamma_t=1.42\ \GeV$, and $\alpha_s=0.1$.
Note that for the results presented in this section 
$\alpha_s$ appears simply as an overall factor, because
all of our events contain a gluon.  

Unless otherwise indicated, we use the following cuts.  We require
$E_g>5\ \GeV$ to eliminate the infrared singularity and because
we intend for the gluon to be detectable.  In addition we wish
the gluon to be separable from the $b$ and $\bar{b}$ quarks;
this is implemented via the requirement $m_{bg}, m_{\bar{b}g}> 10\ \GeV$,
which we shall identify below as ``$m_{bg}$ cuts.''
(Separation could also be achieved with a cut on the gluon's 
transverse energy $E_T$
with respect to the $b$ or $\bar{b}$;  the choice
makes little difference in the resulting distributions.)
In order to
 make sure that we do not get contributions to our results
from regions of the phase space where non-doubly-resonant diagrams 
might be important, we require
\beq
160\ \GeV \,\leq \,m_{bW}\, \leq \,190\ \GeV \;\;\;\;\; {\rm or} \;\;\;\;\;
160\ \GeV \,\leq \,m_{bWg}\, \leq \, 190\ \GeV 
\eeq
and the same thing for the $\bar{b}$.  These conditions will be identified
as ``$m_t$ cuts'' below.

\begin{figure}[ht]	
\vskip -.25 cm
\hspace*{2cm}
\mbox{\epsfig{figure=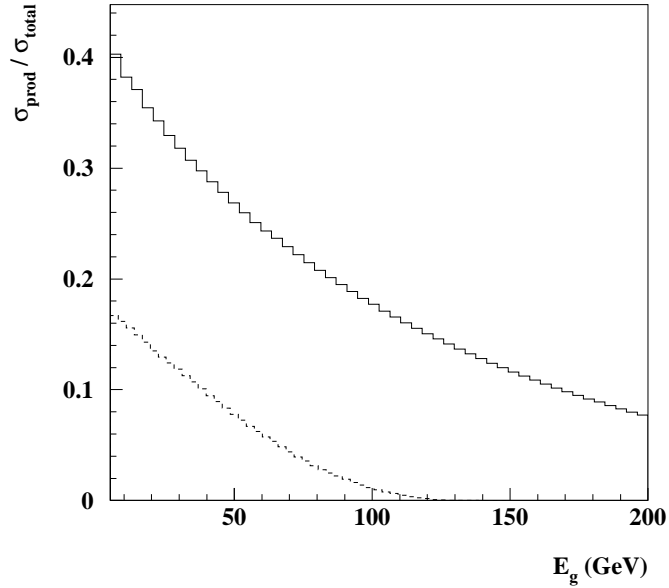,width=11.0cm}}
\caption{
\label{prodfrac}
\small The fraction of gluon emissions radiated in the production stage, as
a function of minimum gluon energy, for center-of-mass energy 1 TeV (solid
line) and 500 GeV (dashed line), with no cuts.}
\end{figure}
 
\subsection{Characteristics of the gluon radiation}

We begin  with the  relative contributions 
of production-
and decay-stage radiation to the total cross section.  Figure \ref{prodfrac}
shows the fraction of the total cross section due to production stage 
emission in events with an extra gluon, as a function of the 
minimum energy of the gluon.   This figure contains no cuts 
besides that for gluon energy and is simply meant to
illustrate how radiation is apportioned in top
production and decay for different center-of-mass energies; 
the solid line corresponds to c.m.\ 
energy 1 TeV, and the dashed line is for 500 GeV.  
Both curves fall off 
as the minimum gluon energy increases; this reflects the decrease in 
phase space for gluons radiated in the production stage.  We see that 
 the production fraction is always higher at 
1 TeV collision energy than at 500 GeV.  This too reflects phase
space --- for a given gluon energy there is more phase space
available to produce gluons in association with top
pairs  at the higher c.m.\ energy.
However both fractions remain below 0.5; decay-stage radiation always 
dominates at these energies.

\begin{figure}[!t]	
%\vskip -.25 cm
\mbox{\epsfig{figure=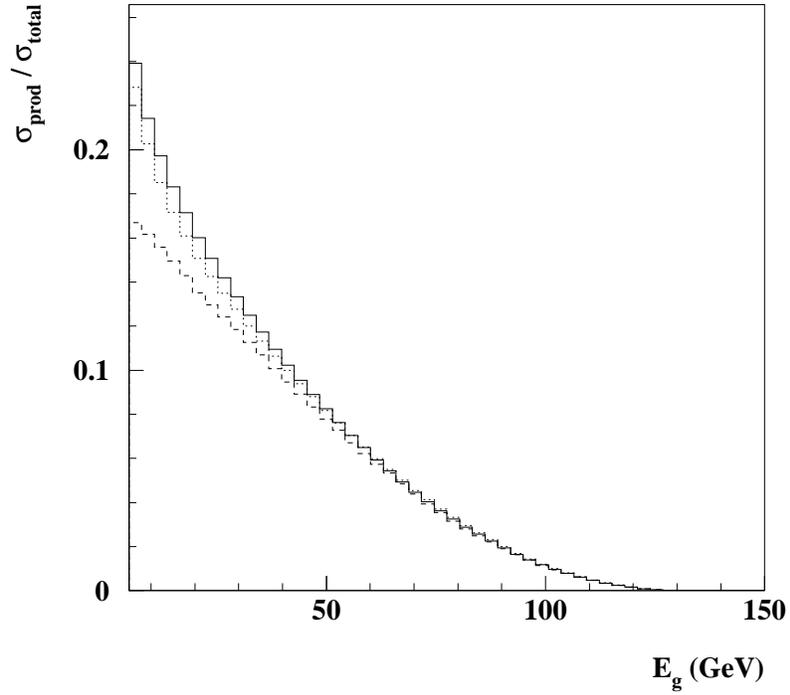,width=13.0cm}}
\vskip -.5 cm
\caption{
\small The fraction of gluon emissions radiated in the production stage, as
a function of minimum gluon energy, for center-of-mass energy  500 GeV,
with no cuts (dashed line), $E_T(g,b)>3\ \GeV$ (dotted line),
and $m_{bg}>10\ \GeV$ (solid line).}
\label{prodfraccuts}\end{figure}

Figure \ref{prodfraccuts} shows for a 500 GeV center-of-mass
energy the effect on the production fraction
of separation cuts between the gluon and $b$ quarks.  The dashed line
shows the fraction with no cuts.  The dotted line corresponds to
requiring that the transverse energy of the gluon with respect to 
the $b$ and $\bar{b}$ --- which we denote $E_T(g,b)$ ---
be greater than 3 GeV.  The solid line corresponds to the 
cut $m_{bg}>10\ \GeV$ where the $b$ can either be a quark or antiquark.
The effect of both of these cuts is to eliminate gluons that are
soft and/or close to one of the bottom quarks; since these
contributions tend to come from decay-stage radiation, their effect
is to increase the fraction of production-stage radiation. If the
$b$ were massless there would be a collinear singularity in the decay
contribution; this does
not happen in our case but the decay distribution still peaks when
the $b$-quark--gluon angle is small.   
The effects of both cuts become smaller
with increasing gluon energy.

\begin{figure}[!t]	% in second brace, h=here, t=top, b=bottom	
%\vskip -.5 cm
\mbox{\epsfig{figure=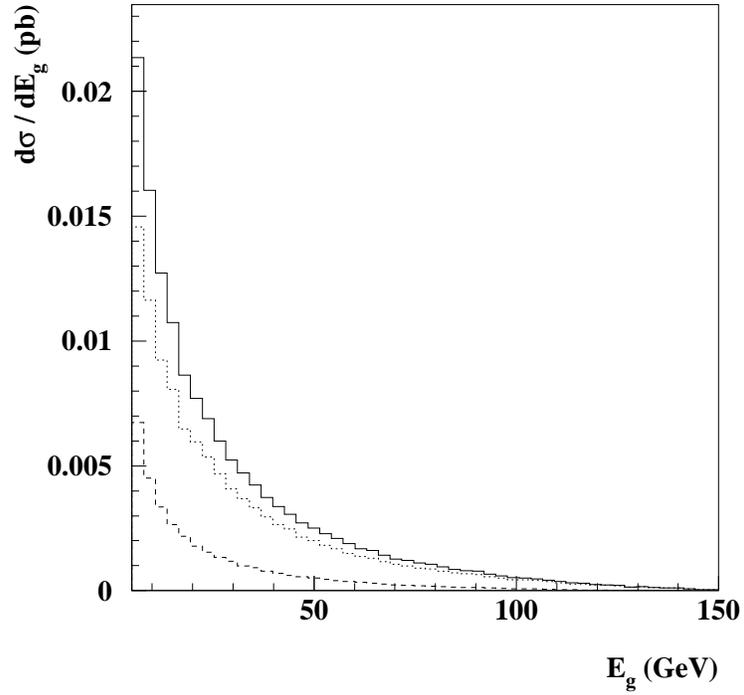,width=13.0cm}}
\vskip -.75 cm
\caption{
\label{energy}
\small The  spectrum of radiated gluons as a function of gluon energy in
GeV for center-of-mass energy 500 GeV, with $m_{bg}$ and $m_t$ cuts
(see text).  Dashed histogram:  production-stage radiation.  Dotted histogram: 
decay-stage radiation.  Solid histogram:  total.}
\end{figure}

Figure \ref{energy} shows the total gluon energy spectrum for a
collision energy of 500 GeV along with its decomposition into production
(dashed histogram) and decay (dotted histogram) contributions. 
The interferences between the two are negligible and are not shown; this 
will be true for all subsequent figures until  we consider the 
interference explicitly.  Included in this figure are the $m_t$ and 
$m_{bg}$ cuts discussed above.  As indicated in the previous figures, 
radiation from the top decays dominates.  
Otherwise the spectra are not vastly 
different; both exhibit the rise at low energies due to the 
infrared singularity characteristic of gluon emission, and both fall
off at high energies as phase space runs out. 

\subsection{Mass Reconstruction}

\begin{figure}[!t]		
\mbox{\epsfig{figure=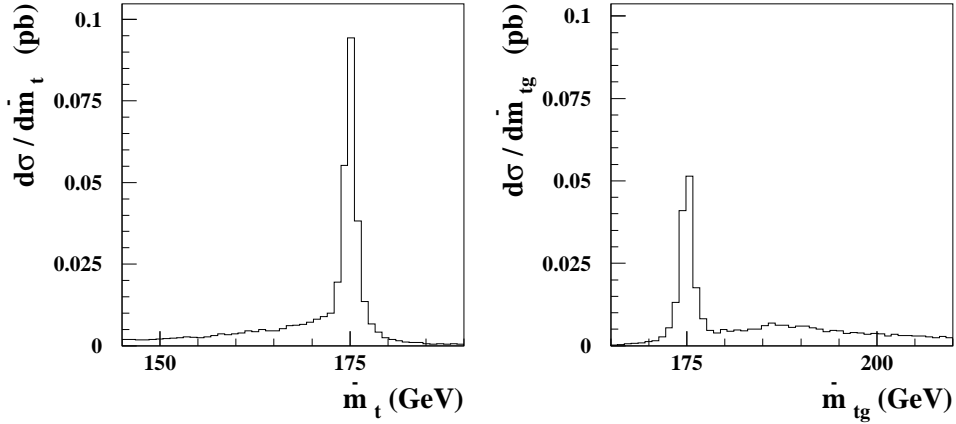,width=13.0cm}}
\vskip -6cm
\caption{
\label{masses}
\small The  top invariant mass spectrum without (left) and with (right) 
the gluon momentum included, for center-of-mass energy 500 GeV, with
$m_{bg}$ cuts and $E_g>5 \GeV$.}
\end{figure}

We now turn to the question of top mass reconstruction in the case
when there is a gluon in the final state (more about measuring the top mass
can be found in the next chapter).  
Figure \ref{masses} shows top invariant mass distributions with and without
the extra gluon included; the first plot shows the distribution in 
$m_{bW}$ and the second shows $m_{bWg}$.  We have imposed $m_{bg}$ cuts
and required $E_g>5\ \GeV$.  
  In both cases there is a clear peak  at the 
correct value of $m_t$.  Note that the peak in the 
first plot contains the production contribution as expected,
but the radiative decay part contributes as well.  
This is because even for decay-stage radiation, only one of 
the produced $t$ quarks decays radiatively; the other still has 
$p_t^2=p_{bW}^2$ and therefore contributes to the $m_{bW}$ peak.
The long tails in the two distributions are from misassignments of
the gluons.  
In the left-hand plot, where the gluon
is not included in the reconstruction, we see a low-side tail due to 
events where the gluon was radiated in the decay but was not included
in the reconstruction.  Similarly, 
in the right-hand plot we see a high-side tail due to events where the 
gluon was radiated in association with production, and was included when it 
should not have been.

\begin{figure}[!t]		
\mbox{\epsfig{figure=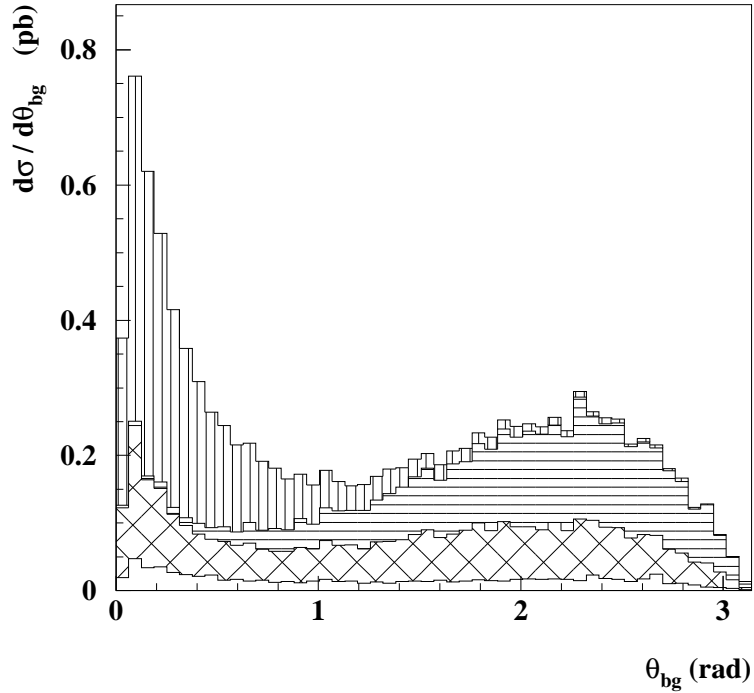,width=13.0cm}}
\vskip -1cm
\caption{
\label{bgangle}
\small The  distribution in the angle between the 
 gluon and the $b$ quark for center-of-mass energy 500 GeV, with
$E_g>5\ \GeV$. The various contributions are as described in the text.}
\end{figure}

We wish to define a single distribution for the top mass that 
combines both types of events yet omits wrong combinations as much 
as possible.  One possibility is to cut on the angle between
the gluon and the $b$ quark, whose distribution we show in
Figure \ref{bgangle}.  This is motivated by the fact that 
gluons radiated from the $b$ quarks are always part of the 
decay, and such gluons tend to be emitted close to the 
$b$ quark direction.  
As we have mentioned, the mass
of the $b$ quark prevents a collinear singularity, but the
gluon distribution still peaks close to the $b$, as can be 
seen in the figure.  Because we wish to define cuts that 
give a narrow top invariant mass distribution, the
distribution in $\theta_{bg}$ is decomposed into various
invariant mass regions.  (Here we refer to the $b$ quark only,
and not the $\bar{b}$.)    
 Using the variables $\tilde{m}_t=m_{bW^+}$, 
 $\tilde{m}_{tg}=m_{bW^+g}$, $\tilde{m}_{\bar t}=m_{\bar{b}W^-}$ and
 $\tilde{m}_{{\bar t}g}=m_{\bar{b}W^-g}$  
  we define four types of events: \\
\hspace*{0.5in} type 1 : $172\ \GeV < \tilde{m}_{tg} , \tilde{m}_{\bar{t}} <178\ \GeV$ 
(vertical hatching)\\
\hspace*{0.5in} type 2 : $172\ \GeV < \tilde{m}_{t} , \tilde{m}_{\bar{t}g} <178\ \GeV$ 
(horizontal hatching)\\
\hspace*{0.5in} type 3 : $172\ \GeV < \tilde{m}_{t} , \tilde{m}_{\bar{t}} <178\ \GeV$ 
(cross hatching)\\
    \hspace*{0.5in} type 4 : any other event  
(no hatching)\\
Type 1 events are dominated by contributions from radiative $t$ decays,
and we can see that they do tend towards the $b$ quark
direction.  
Type 2 events (horizontal hatching) are in turn
dominated by radiative $\bar{t}$ decays; gluons in
this case tend to cluster near the $\bar{b}$ direction, and since
the $b$ and $\bar{b}$ tend to appear in opposite hemispheres,
type 2 gluons are mostly found at large angles to the $b$.  
Events of type 3 (cross hatching) are 
mostly production-stage contributions; their distribution is more or less
uniform, independent of the $b$ quark direction.
Finally, events of type 4 (no hatching)
get contributions from both production and decay, with no compelling 
evidence for one over the other.

\begin{figure}[!t]		
\mbox{\epsfig{figure=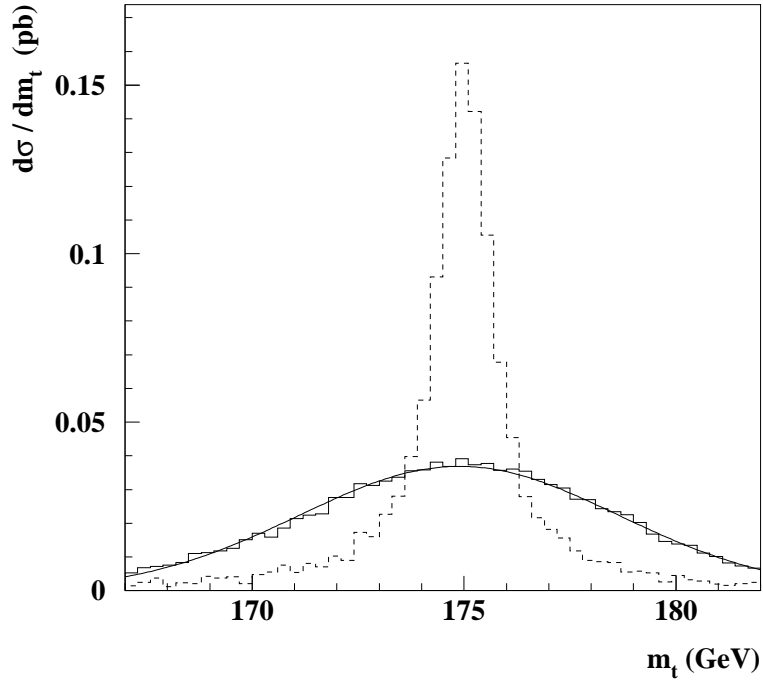,width=13.0cm}}
\vskip -1.25cm
\caption{
\label{masssmear}
\small The  top invariant mass spectrum with $b$-gluon angle selection
criteria (dotted histogram), for center-of-mass energy 500 GeV,
minimum gluon energy 5 GeV, and $m_{bg}$ cuts.  
The solid curve and histogram show the effects of energy smearing.}
\end{figure}

Using this figure we can make the following conventions:\\
\hspace*{0.5in} if $\theta _{bg} < 0.7 \ {\rm rad}, $ assign gluon to $t$ decay \\
 \hspace*{0.5in} if $\theta_{\bar{b}g} < 0.7 \ {\rm rad},$ assign gluon to $\bar{t}$
  decay \\
\hspace*{0.5in} if $\theta_{bg}, \theta_{\bar{b}g} >1\ {\rm rad}, $ 
 assign gluon to $t\bar{t}$ production.\\
 With these cuts on the proximity of the gluon to the 
$b$ quark, we construct  the top mass distribution  presented
in the dotted histogram in Figure \ref{masssmear}.

Of course an important reason the cuts are so effective is that we 
work at the 
parton level.  The experimentalists do not have that luxury, and, as one
would expect, hadronization and detector effects are likely to cloud
the picture.  The solid histogram in Fig.~\ref{masssmear} shows the
mass distribution after including energy smearing; the solid curve is a 
Breit-Wigner fit.   
The spread in the measured momenta of the final state particles is 
parametrized by Gaussians with
widths $\sigma=0.4 \sqrt{E}$ for quarks and gluon, and $ 
\sigma=0.15 \sqrt{E}$ for the $W$'s.  We see that the central value does not 
shift, but the distribution becomes significantly wider.  

These results are meant to give an indication of the effects of hard gluon
radiation 
on mass reconstruction and how they might be dealt with.  Other variables
to consider in choosing the cuts are $m_{bg}$,  the transverse energy
of the gluon with respect to the $b$ or $\bar{b}$, or some combination
of energies and angles as defined in the various algorithms used
in jet definitions for $e^+e^-$ colliders.    At tree
level and with partons only, the exact choice is not very important.
We will revisit the question in more detail when we include virtual
corrections in a full NLO calculation.

\subsection{Interference and Sensitivity to $\Gamma_t$}

Finally, we return to the subject of interference.  As mentioned above, the 
interference between the production- and decay-stage radiation can 
be substantial for gluon energies close to the 
total width of the top quark $\Gamma_t$; the interference 
is therefore  sensitive to the value of $\Gamma_t$.  
However, because this interference is in general small, we need to find
regions of phase space where it is enhanced.  This question was considered
in  \cite{kos} in the soft gluon approximation\footnote{See also 
 \cite{jikia,siopsis}}, where it was found that 
the interference was enhanced when there was a large angular separation
between the $t$ quarks and their daughter $b$'s.  

\begin{figure}[!t]		 
\mbox{\epsfig{figure=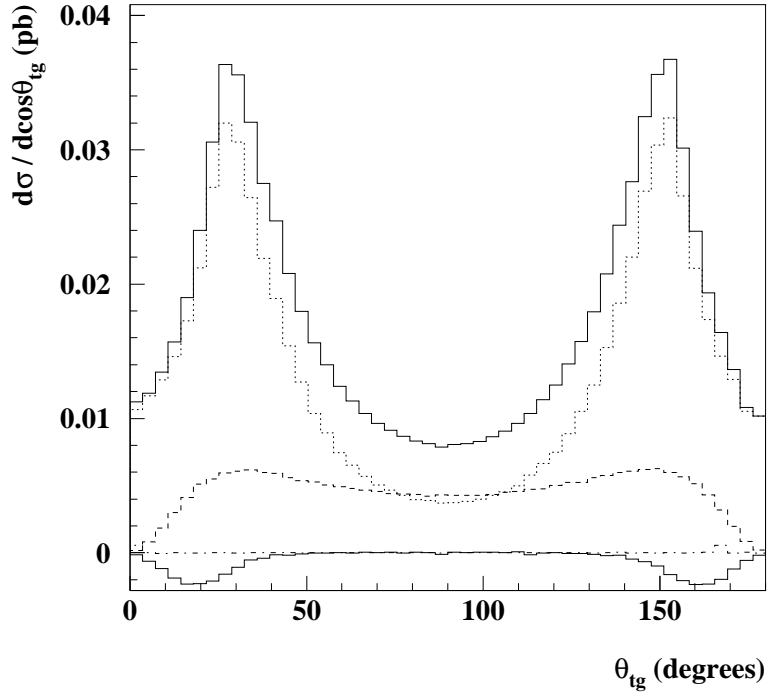,width=13.0cm}}
\vskip -1 cm
\caption{
\label{thetagt}
\small The  distribution in angle between the top quark and the gluon for 
gluon energies from 5 to 10 GeV, 
$\cos\theta_{tb},\cos\theta_{\bar{t}\bar{b}}<0.9$, $m_t$ cuts,
 and 750 GeV collision
energy.  The upper solid histogram is the  total and the other histograms
represent the individual contributions:  dotted:  decay; dashed: production;
dot-dashed: decay-decay interference; solid:  production-decay interference.}
\end{figure}

Here we examine whether
the result of \cite{kos}, which considered a  fixed final-state
configuration, 
survives the exact calculation and phase space integration.  
Figure \ref{thetagt}
shows that it does.  There we plot the distribution in the angle between 
the 
emitted gluon and the top quark for gluon energies between 5 and 10 
GeV and with $\cos\theta_{tb}<0.9$ and $m_t$ cuts.  
The center-of-mass energy is 750 GeV. This c.m.\ energy is chosen because
for there to be significant interference between production
and decay-stage radiation, both contributions must be sizable.  
At 500 GeV, we see from Figs.~\ref{prodfrac} and \ref{prodfraccuts}
that the production contribution is suppressed compared to that from decay;
as a result, the interference is very small.  Increasing the 
energy increases the production-stage contribution.  We note that 
the distributions at 750 GeV and 1 TeV do not differ substantially.

The  histograms in Fig.~\ref{thetagt} 
show the decomposition into the various contributions. The  production-stage
radiation is shown as a dashed histogram; we see that it reaches its
largest values at relatively small and large angles.  Small angles
correspond to the $t$ direction, and large angles more or less to the 
$\bar{t}$ direction, since for the small gluon energies of interest here, the $t$ 
and $\bar{t}$ are nearly back-to-back.  The dotted histogram represents
the decay-stage contribution; it dominates the cross section and 
 peaks in the same region as the production contribution.  This leads
to substantial production-decay interference, shown as the negative solid
histogram.  This interference is destructive, so that it serves to 
{\it suppress} the total cross section, shown as the positive solid
histogram.  This effect would be enhanced if
we lowered the gluon energies to values closer to $\Gamma_t$, but 
jets from very low energy gluons are not likely to be observable, so we 
cut off the gluon energy at 5 GeV.  Finally, interference
between the emissions in the $t$ and $\bar{t}$ decays are shown
as a dot-dashed histogram, but as there is very little overlap between 
the two phase space regions even with these cuts, this contribution is 
negligible.

\begin{figure}[!t]		 
\mbox{\epsfig{figure=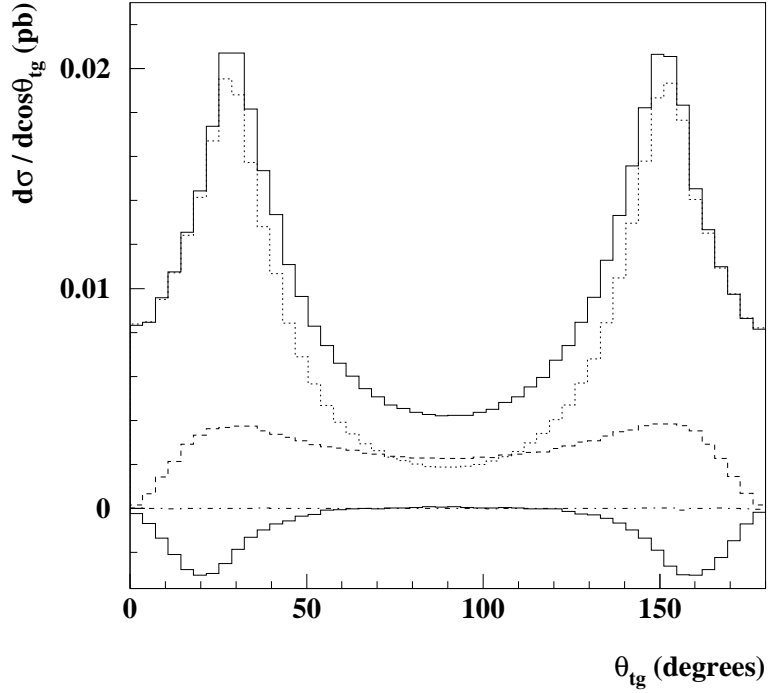,width=13.0cm}}
\vskip -1 cm
\caption{
\label{thetagtcuts}
\small As in Fig.~\ref{thetagt}, with the addition of the cuts given
in Eqs.~\ref{interferencecutsa},\ref{interferencecutsb}.}
\end{figure}

The cuts we have used are fairly generic; we can further enhance
the interference terms with a judicious choice of additional cuts. 
If we 
examine their  behavior in more detail in various
regions of phase space, we find that 
the sign of the interference terms depends on the value of the
invariant mass of the top quark.    Since we  integrate
over this  mass, we get cancellations (a similar effect
ensures cancellations of non-factorizable corrections in inclusive 
quantities).

Consider the interference between radiation in the production stage and 
the top decay
stage. The product of the two Breit-Wigner peaks is proportional to the  factor
\beq 
f_{t} = (p_{Wb}^2 - m_t^2)*(p_{Wbg}^2 - m_t^2) + m_t^2 \Gamma_t^2 
\eeq
This factor will multiply a quantity which, upon integration over angles,
is negative. Therefore, for invariant mass values such that $f_{t}$ is 
positive, the interference terms are negative, while 
for negative $f_{t}$,  the interference terms are positive.
We can impose cuts that take advantage of this:  if we require
the invariant masses to satisfy
\beq
f_{t} > 0\ , \ \mbox{if}\ \theta_{bg} <  \theta_{\bar{b}g} ; 
\label{interferencecutsa}
\eeq
\beq
 f_{\bar{t}} > 0\ , \ \mbox{if}\ \theta_{bg} >  \theta_{\bar{b}g}\; , 
\label{interferencecutsb}
\eeq
we obtain the distribution shown in Figure \ref{thetagtcuts}.  The 
interference effects are enhanced, though at the cost of a substantial decrease 
in cross section.

\begin{figure}[!t]		
\mbox{\epsfig{figure=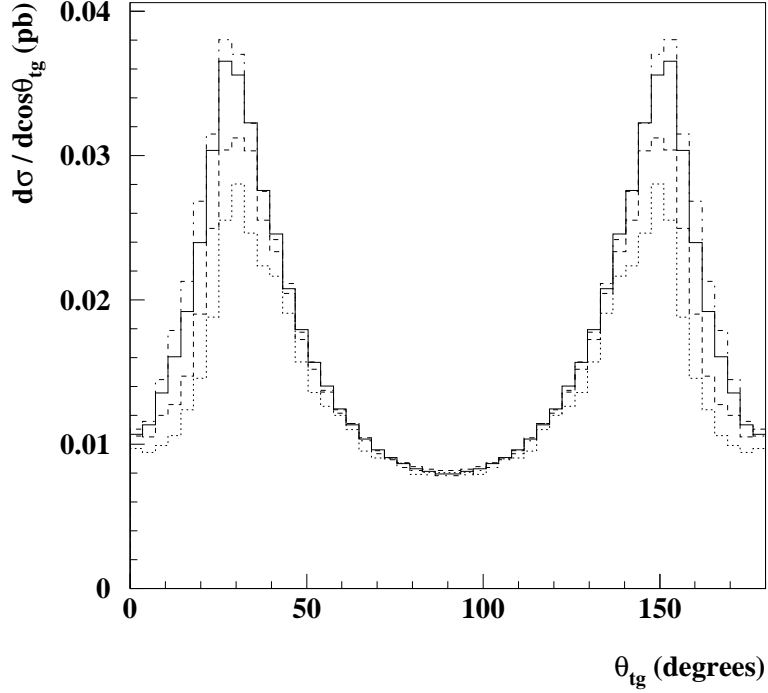,width=13.0cm}}
\vskip -1 cm
\caption{
\label{width}
\small The  distribution in angle between the top quark and the gluon for 
gluon energies from 5 to 10 GeV, 
$\cos\theta_{tb},\cos\theta_{\bar{t}\bar{b}} <0.9$, 
and 750 GeV collision
energy.  The histograms correspond to different values of the top 
width $\Gamma_t$:  dot-dashed: 0.1 GeV; solid: 1.42 GeV (SM); 
dashed: 5. GeV;
dotted: 20 GeV.}
\end{figure}

Because the production-decay interference is destructive, increasing
the top width would further suppress the total distribution.  The height
of the peaks, then, is sensitive to the value of $\Gamma_t$.  
This is illustrated in Figure \ref{width}, which
shows the cross section (without the cuts of 
Eqs.~\ref{interferencecutsa},\ref{interferencecutsb}) 
for different values of the top 
width\footnote{The histograms here are scaled so that they all would have
the same normalization
in the absence of interference effects.  Without this rescaling, changing
the width while keeping the $tbW$ coupling fixed changes the total
cross section, which behaves like $1/\Gamma_t^2$ for small $\Gamma_t$.}, 
ranging from 0.1 GeV to 20 GeV.  The SM case ($\Gamma_t=1.42\ \GeV$) is
shown as a  solid line.  It is interesting to note that 
in the context of perturbative gluon radiation, the SM top width
is actually a small quantity.   There are several other points to note.
In principle, this sensitivity to $\Gamma_t$ gives us
a method to measure the top quark's total  width,
independent of decay mode, above the top production threshold.
Although in practice statistics would surely limit the 
possible precision of such a measurement, the total top width is not
so easy to measure directly by any means.  Furthermore, the effects 
illustrated here arise from simple quantum-mechanical interference,
and finding experimental evidence for interference between the 
radiation at the various stages is an interesting goal by itself.

\chapter{Virtual corrections}

 The subject of this chapter is the computation of virtual corrections
to  the process:
\bec{tree_p}
 e^+ e^- \rightarrow t\ \bar{t}\ \rightarrow b\ W^+\ \bar{b}\ W^- 
\eec
Although real gluon radiation plays an important role
in top production and decay, 
the four particle final state in Eq. \ref{tree_p} will be the predominant signal
observed experimentally. Therefore, it is of utmost interest to have good
theoretical predictions for this process, and this means that 
the virtual corrections to the tree level amplitude should be computed.

While in studies of top quark production near threshold most of the
information is extracted from the shape of the total cross section as a
function of the beam energy, above threshold
the information will be extracted from 
the analysis of kinematical variables of the decay products. Consequently,
an analytic computation is much less useful in this case (and much less
feasible, too), so the approach we will use to perform our computations
will be Monte Carlo simulation.

One of the most important quantities to measure at future
collider experiments is the top quark mass. A precise measurement 
of this parameter will allow stringent tests of the Standard Model
of particle physics (or give information on the supersymmetric
model, if SUSY has been discovered by that time). Indeed, it 
is well known that, at the present time, the measured top quark
mass together with 
values for the $W$ boson mass and other electroweak parameters can be used
to constrain the Higgs mass in the Standard Model (these constraints are not
very strong so far, but they will be greatly enhanced by improvements  
in the top quark mass measurements \cite{top_constr}).

One way to measure the top mass at energies above the production threshold
is by determining the position of the peak in the distribution of the invariant
mass of the top decay products $\sqrt{(p_W + p_b)^2}$ (this is the way the 
top mass has been measured at the Tevatron). The quantity thus obtained is
called the pole mass; the threshold studies will measure a quantity
called the threshold mass (closely related to the $\bar{MS}$ mass)
\footnote{Since a quark is not a stable particle, its mass is an elusive
quantity; there are several theoretical definitions for the mass of the 
top quark; among these, the most common are the pole mass and the 
$\bar{MS}$ mass.}. A theoretical relation exists among these quantities;
hence a measurement of both of them will allow a test of our understanding
of the underlying theory.

The extraction of the top quark mass from the $bW$ invariant mass distribution
is the most natural and straightforward method; therefore, there 
are a lot of studies in this area (and we shall present some results 
of relevance to this case, too). However, it suffers from a fundamental 
limitation on the precision it can achieve: since the physical observable 
is not a color singlet, there will be uncertainties of order $\Lambda_{QCD}$
in the measured quantities (what this actually means is that we cannot measure
the momenta of the $b$ jet - of any jet generated by 
a parton which has color, for that matter- with a precision
greater than $\Lambda_{QCD}$). Therefore, the minimum error achievable
through this method is about 200 MeV. In order to circumvent this limit, 
it might be desirable to try to extract the top mass from 
distributions which involve color singlet physical observables; for example,
end points in distributions of quantities like $p_W(p_b+p_{\ab})$, 
or the line shape of the decay leptons energies
\cite{orange}. However, since these methods
are quite involved, and would also require a good understanding of 
hadronization and detector issues, we will not discuss them further.
 
%Some comments on the magnitude of virtual corrections to production; also 
%about how we take into account interference, too

This kind of precision measurement obviously cannot be performed without
a good theoretical understanding of the underlying process. Hence the need
to compute the QCD corrections to the top production and decay process. 
A quick analysis shows that these corrections are indeed important. It is
well known that the total NLO QCD corrections to the production of a
pair of massless quarks is $\alpha_s/\pi \times$ the Born cross section.
For the top production cross section, this would correspond to the high
energy limit (beam energies of order TeV). On the other hand, near the 
top production threshold, the QCD corrections are quite large; from the
results presented in \cite{jersak}, we have:
$$ \sigma(e^+e^-\to t \at (g)) \ \longrightarrow
\ \sigma^{\hbox{Born}} \left(1 + 
\alpha_s \frac{2\pi}{3 \beta} \right) \ \hbox{, ~~~~ as} \ \beta \to 0
$$
where $\beta$ is the top quark speed
\footnote{Near the production threshold, fixed order NLO computation 
does not work anymore; rather, we need to perform a 
resummation of the large log$\beta$ logarithms. 
Following the suggestion advanced in \cite{jersak}, we can
assume that the NLO computation gives reasonable results if 
$\alpha_s/\beta <1/4$, which means that
the  collision center-of-mass energy should be bigger than about 380 GeV.}.
This indicates that the NLO corrections to top production are rather
large for CM energies smaller that 1 TeV; as we shall see below, they are
about 20\% of the Born cross section at 500 GeV, and 7\% at 1 TeV.

In the following, we shall present a NLO computation for the top 
production and decay process which takes into account not only 
corrections to the production and decay subprocesses, but also interference
contributions. The framework is that of the double pole approximation, 
as described in Chapter {\bf 2}. The contributing 
amplitudes (Eq. \ref{mvg_amp}) are computed using off-shell momenta, and 
numerical integrations for the cross section are performed over the 
full off-shell phase space. This approach allows for further development
of our Monte Carlo (taking into account singly resonant terms, for example)
to be performed in an easy and natural way. 
We shall also discuss
an alternative computational method (on-shell DPA) and compare the 
results with those obtained in our approach.
Some results on the total 
cross section for top production and decay and the impact of 
interference corrections on top mass reconstruction are presented at the end.

%-----------------------------------------------------------------------
\section{Resonant structure of partial amplitudes}

%Some general comments about the resonant structure of diagrams and evaluation
%in DPA.

In this section, we shall discuss the evaluation of 
NLO amplitudes contributing to the process \ref{tree_p}. Some of the 
relevant Feynman diagrams are presented in Figure \ref{vir_diag}. As
mentioned before, these diagrams can be split into two classes: 
(i) vertex and fermion self-energy diagrams, and (ii) interference diagrams.
In the following, we shall look at each of these types separately,
starting with the interference diagrams.

However, let's first consider some general properties of these amplitudes.
In keeping with the approximation used (DPA), all of them have a doubly
resonant behavior in the phase space region where the invariant mass
of the $W b$ pair is close to the top mass. However, the exact type of 
behavior is different from diagram to diagram; while for the zero order 
amplitude the resonant behavior is of type $pole \times pole$:
$$ {\cal{M}}_0 \sim  \frac{1}{p_t^2 - \bar{m}_t^2}\ 
\frac{1}{p_{\bar{t}}^2 - \bar{m}_t^2}\ $$
for the interference amplitudes the resonant behavior is formally
of type $pole \times logarithm$; for example:
$${\cal{M}}_{b\bar{t}} \sim  \hbox{log}(p_t^2 - \bar{m}_t^2) \ 
\frac{1}{p_{\bar{t}}^2 - \bar{m}_t^2} $$
The replacement of a $pole$-type singularity with a $log$-type 
singularity can be traced back to the integration over the momentum 
of the gluon in the virtual loop. Moreover, this kind of term does not appear
only in the purely interference diagrams, but also in the vertex and 
self-energy diagrams. The amplitude 
corresponding to the top decay vertex correction diagram, for example,
can be written as a sum of terms:
$${\cal{M}}_{tb} = \frac{1}{p_t^2 - \bar{m}_t^2}\ 
\frac{1}{p_{\bar{t}}^2 - \bar{m}_t^2}\ \tilde{\cal{M}}_{tdec} \ + \
\hbox{log}(p_t^2 - \bar{m}_t^2) \ 
\frac{1}{p_{\bar{t}}^2 - \bar{m}_t^2}\ {\cal{M}}_{tb}' $$
where the terms with the double pole structure can be thought of as being the correction  to the top decay process, and the term with a $pole \times log$
structure contributes to interference between the production and 
top decay process.

Another property of the amplitudes for the interference diagrams is
that, in the DPA, they are proportional to the tree level amplitude
${\cal{M}}_0$. We say that, in the DPA, the interference amplitudes
factorize. This is also true for the case of the $W$ pair production process 
(\cite{ddr}, \cite{BBC}); moreover, for this process it has been shown
that {\it all} the interference contributions (including 
those coming from vertex correction diagrams, for example) factorize.
Therefore, the total interference can be written as a scalar factor times the 
Born amplitude, which makes the evaluation of these contributions really easy.
However, this is not true in our case; for the top production and decay
process, there are interference contributions coming from the vertex
and top self-energy correction diagrams which are not proportional to the
tree level amplitude, as we shall see below.

\subsection{Interference diagrams}

In this section we will examine the behavior of interference diagrams.
Consider, for example, the $\at - b$ interference diagram. 
(Fig. 2c).
The amplitude associated with this diagram is:
{\small
$$ {\cal{M}}_{b\bar{t}} =
 \bar{u}(b) \left[ (-i g_s^2)
 \int \frac{d^4 k}{2\pi^4}\ \frac{1}{k^2 + i\epsilon}\
\gamma^\mu \ \frac{\not{p}_b-\not{k} + m_b}{(p_b-k)^2 - m_b^2}\
\not{\epsilon}_{W^+} \ \frac{\not{p}_t-\not{k} + m_t}{(p_t-k)^2 - \bar{m}_t^2}\
{\Gamma}_{\gamma,Z_0}\ \right.
$$
\bec{mbbart}
\left.
\frac{-\not{p}_{\bar{t}} - \not{k} + m_t}{(p_{\bar{t}}+k)^2 - \bar{m}_t^2}\
\gamma_\mu \right] \ 
\frac{-\not{p}_{\bar{t}} + m_t}{p_{\bar{t}}^2 - \bar{m}_t^2}\
\not{\epsilon}_{W^-} \ v(\bar{b})
\eec }
where the
Feynman gauge is used for the gluon propagator.

The evaluation of this amplitude is obviously quite a difficult task.
However, the only terms of interest to us in DPA are those which
have resonances at the top and antitop quark propagator poles. This
simplifies our task greatly. 
The doubly resonant terms can be extracted with the help of the following
observation: if the virtual gluon in the loop is hard, then 
the quantity in  brackets does not have any singularity, and the overall
resonant structure for this diagram is given only by the pole due to 
the antitop propagator:
$ {\cal{M}}_{b\bar{t}}(\hbox{hard gluon}) \propto 1/(p_{\bar{t}}^2 - \bar{m}_t^2)$.
This means that any doubly resonant terms contribution to
$ {\cal{M}}_{b\bar{t}}$ are entirely due to soft 
virtual gluons. 
Therefore, we can neglect the $\not{k}$ terms in 
the numerator of (\ref{mbbart}).
Following \cite{ddr}, we shall call this approximation 
the {\it extended soft gluon approximation} (ESGA)
\footnote{In the standard soft gluon approximation, $k^2$
terms in the denominator of top quark propagators would also
be neglected; we do not do this here for computational reasons
 (see also \cite{kos}, \cite{ddr}).}.

 With the help of the transformations:
\bec{trans_1}
\gamma^\mu \ (\not{p}_b + m_b) = ( -\not{p}_b + m_b)\ \gamma^\mu + 2p_b^\mu
\rightarrow 2p_b^\mu
\eec
$$ 
(-\not{p}_{\bar{t}} + m_t)\ \gamma_\mu = \gamma_\mu \ (\not{p}_{\bar{t}} + m_t)
- 2 p_{\bar{t} \mu} \rightarrow - 2 p_{\bar{t} \mu}
$$
(the term  $(\not{p}_{\bar{t}} + m_t)$ 
on the second line is neglected, since 
it would lead to a singly resonant contribution),
the following result is  obtained for the amplitude (\ref{mbbart}):
$$ {\cal{M}}_{b\bar{t}}(DPA+ESGA) = 
\frac{\alpha_s}{4\pi}\
 {\cal{M}}_0 \ * \ (-4 p_b p_{\bar{t}})( p_t^2-\bar{m}_t^2 ) \ *
$$
\bec{mbbtr}
%\left[
\int \frac{d^4 k}{i\pi^2}\ \frac{1}{k^2+ i\epsilon}\ \frac{1}{k^2-2kp_b}\  
\frac{1}{(p_t-k)^2 - \bar{m}_t^2}\ \frac{1}{(p_{\bar{t}}+k)^2 - \bar{m}_t^2}\
%\right]
\eec
The result is proportional to the tree level amplitude --
 in the DPA, the virtual corrections 
due to interference factorize.
The proportionality factor includes the scalar four point function
(the integral on the second line of Eq. \ref{mbbtr}) 
${\cal{D}}^0_{b\bar{t}} = {\cal{D}}^0(-p_b,-p_t,p_{\bar{t}},0,m_b,\bar{m}_t,\bar{m}_t)$.
\footnote{
 For the scalar one-loop integrals appearing here
 we use the following notation:
$$ {\cal{D}}^0(p_1,p_2,p_3,m_0,m_1,m_2,m_3) = 
\int \frac{d^4 k}{i \pi^2}\ \frac{1}{N_0\ N_1\ N_2\ N_3}\ 
$$
$$ {\cal{E}}^0(p_1,p_2,p_3,p_4,m_0,m_1,m_2,m_3,m_4) = 
\int \frac{d^4 k}{i \pi^2}\ \frac{1}{N_0\ N_1\ N_2\ N_3\ N_4}\ 
$$
with the denominators :
$$ N_0 = k^2 - m_0^2 + i\epsilon, \hbox{~~~~~} 
N_i = (k+p_i)^2 - m_i^2 + i\epsilon, \hbox{~~~~~} i=1,...,4
$$
}

What can we tell about the singular behavior of the DPA amplitude
in (\ref{mbbtr})? 
Apparently, the result for ${\cal{M}}_{b\bar{t}}$ has a single pole at 
$p_{\bar{t}}^2 = m_t^2$ (the other pole  being canceled 
by the multiplicative term $ p_t^2-m_t^2 $).
 However, if the
top (or antitop) goes on-shell, the ${\cal{D}}^0$ function 
acquires an infrared singularity (in the zero top width limit;
this singularity is regularized by the top width). 
Since the infrared singular type terms have 
a logarithmic structure (this can also be reasoned from power 
counting arguments), this indicates that ${\cal{D}}^0$ has the 
following behavior close to the top resonances:
\bec{res_beh}
 {\cal{D}}^0_{b\bar{t}} \sim a_1 \hbox{log}(p_t^2 - \bar{m}_t^2) + 
a_2 \hbox{log}(p_{\bar{t}}^2 - \bar{m}_t^2) 
\eec
Here $a_1$ and $a_2$ are terms which are finite when {\it either}
the top or antitop quark go on-shell.

Formally, then, the overall resonant behavior of the 
interference amplitude $ {\cal{M}}_{b\bar{t}}$
in DPA is of type $pole \times logarithm$: 
$$ {\cal{M}}_{b\bar{t}} \sim \tilde{{\cal{M}}}_0\ \hbox{log}(p_t^2 - \bar{m}_t^2) \ 
\frac{1}{p_{\bar{t}}^2 - \bar{m}_t^2} $$
rather than $pole \times pole$,
as it is for the corrections to production or decay subprocesses.

%***
%5 point function\\

Using the same techniques, similar results are easily obtained for the
other two interference diagrams.
In the soft gluon approximation (and DPA):
\bec{mtbbr}
{\cal{M}}_{t\ab}(DPA+ESGA) = 
\frac{\alpha_s C_F}{4\pi}\
 {\cal{M}}_0 \ * \ (-4 p_t p_{\ab})( p_{\at}^2-\bar{m}_t^2 ) \
 {\cal{D}}^0_{t\ab}
\eec
\bec{mbbbr}
 {\cal{M}}_{b\bar{b}}(DPA+ESGA) = 
\frac{\alpha_s C_F}{4\pi}\ {\cal{M}}_0 \ * \ (-4 p_b p_{\bar{b}})( p_t^2-\bar{m}_t^2 )
( p_{\bar{t}}^2-\bar{m}_t^2 ) \ {\cal{E}}^0_{b\ab}
\eec
where ${\cal{D}}^0_{t\ab} = {\cal{D}}^0 (-p_{\ab},-p_{\at},p_t,0,m_b,\bar{m}_t,\bar{m}_t)$ and
\bec{e0}
{\cal{E}}^0_{b\ab} = {\cal{E}}^0
(-p_b,-p_t,p_{\bar{t}},p_{\bar{b}},\mu,m_b,\bar{m}_t,\bar{m}_t,m_b)
\eec
is the scalar five point function (here $\mu$ is the infinitesimally small
gluon mass  needed for the regularization of infrared divergent
behavior of ${\cal{E}}^0_{b\ab}$).

% It can also be shown that the resonant behavior of this
%function is 
%$$ {\cal{E}}^0_{b\ab} \sim 
%b_1\ \hbox{log}(p_t^2 - \bar{m}_t^2) \ \frac{1}{p_{\bar{t}}^2 - \bar{m}_t^2}
%+ b_2\ \hbox{log}(p_{\bar{t}}^2 - \bar{m}_t^2) \  \frac{1}{p_t^2 - \bar{m}_t^2} $$
%hence we get
%$$  {\cal{M}}_{b\bar{b}} \sim \tilde{{\cal{M}}_0} \left(
%b_1\ \hbox{log}(p_t^2 - \bar{m}_t^2) \ \frac{1}{p_{\bar{t}}^2 - \bar{m}_t^2}
%+ b_2\ \hbox{log}(p_{\bar{t}}^2 - \bar{m}_t^2) \  \frac{1}{p_t^2 - \bar{m}_t^2}
%\right)
%$$

We end this section with some comments on the numerical magnitude
of interference terms. Since the resonant behavior of these terms is
of $pole \times log$ type, it might be expected that they are less important
numerically that the double pole terms. However, 
analytic expressions for the ${\cal{D}}_0$ function (\cite{ddr},
\cite{mel_yak}, \cite{BBC})
show that, although the 
coefficients $ a_1, a_2$ in \ref{res_beh} are finite when one of 
the top or antitop quark goes on shell, they will diverge when both particles
go on-shell simultaneously:
$$ a_i\ \sim \ 
\frac{1}{c_{1i} (p_t^2 - \bar{m}_t^2) + c_{2i} (p_{\bar{t}}^2 - \bar{m}_t^2)}
$$
Therefore, the leading logarithms in the scalar 4 and 5-point functions
will be enhanced
by factors of order $m_t/\Gamma_t$ near the top, antitop quark mass resonances.

%--------------------------------------------------------------------------
%\newpage

\begin{figure}[ht!] % the general vertex
\centerline{\epsfig{file=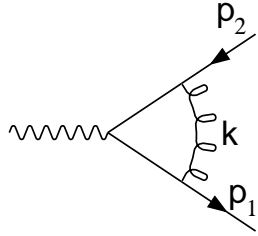,height=1.6in,width=1.6in}}
\caption{General vertex correction diagram.}
\label{ver_cor}
\end{figure}

\subsection{Vertex corrections}

 The results for the interference diagrams
%(Eq. (\ref{mbbtr}) and Eq. (\ref{mbbbr})) 
are completely analogous to
results obtained in the $W$ pair production computation.
However, for the off-shell vertex and self-energy
corrections diagrams, the results in the top case are different.
Consider for example, the amplitude for the general vertex correction
in Figure \ref{ver_cor}: 
$$ \delta \Gamma^{\mu} = \frac{\alpha_s}{4 \pi} \
\int \frac{d^4 k}{i \pi^2}\ \frac{1}{k^2}\ \gamma^{\nu}\
\frac{\not{p}_1 - \not{k} + m_1}{(p_1 - k)^2 - \bar{m}_1^2}\
\gamma^{\mu}(C_V + C_A \gamma^5 ) \
\frac{-\not{p}_2 - \not{k} + m_2}{(p_2 + k)^2 - \bar{m}_2^2}\
\gamma_{\nu}
$$
Upon evaluation (and keeping only the vector part)
% the tensorial structure of this expression is :
the result can be written in terms of eight form factors, each
of them multiplying a different tensor quantity:
$$
\delta \Gamma^{\mu}_V = \frac{\alpha_s}{4 \pi} C_V\
\left[ {\gamma^{\mu}} {F_2} + { (\not{p}_1 - m_1)\gamma^{\mu}} {F_4}
 + {\gamma^{\mu}(-\not{p}_2 - m_2)} {F_6} +
\right.
$$
\bec{vertex}
 \left. {(\not{p}_1 - m_1)\gamma^{\mu}(-\not{p}_2 - m_2)} {F_8} + 
{p_1^{\mu}} {F_1} +  
(\not{p}_1 - m_1) p_1^{\mu} F_3   +\ldots \right]
\eec
(expressions for the scalar form factors $F_1,\ldots, F_8$ can be found
in the appendix).
In the on-shell case, only the $F_2$ (electric dipole) and $F_1$ (magnetic dipole momentum) form factors contribute. It might be expected that in the 
double pole approximation we can drop the other terms, too, since
they  have a zero at 
$\not{p}_1 = m_1$ (or $\not{p}_2 = -m_2$) which will
cancel one pole (or both) in the amplitude. 
However, the 
form factors themselves
may have a resonant structure when the  particles go on shell.

Consider the top decay vertex correction. In this case,
$p_1 \rightarrow p_b , p_2 \rightarrow -p_t$, and only four terms survive
in Eq. \ref{vertex}; the corresponding form factors
contain terms which are proportional to the scalar three point function:
$$ F_i \ \sim \ C^0_{tb} \ =
\int \frac{d^4 k}{i \pi^2}\ \frac{1}{k^2}\ 
\frac{1}{(p_b - k)^2 - m_b^2}\
\frac{1}{(p_t - k)^2 - \bar{m}_t^2}\
\hbox{~~~~~~~} i= 1,2, 5,6 $$
which has 
a logarithmic resonant behavior:
$$C^0_{tb} \ \sim \ \hbox{log}(p_t^2 - \bar{m_t}^2).$$ 
Therefore, the contribution 
of $i=5,6$ terms to the top decay vertex correction is doubly
resonant, although of type $pole \times log$ rather than double pole.
Because these logarithms are not multiplied by large factors (as in the
case of the interference diagrams), we can expect these terms to be
numerically small; for consistency reasons it is still desirable to 
include them in the final result.

Similar results are obtained for the correction to the antitop decay 
vertex (we keep the $i=1,2,3,4$ terms in this case). In the case of the 
$t \ - \ \bar{t}$ vertex, though, both fermions are off-shell; as a consequence,
there are no resonant logarithms when either the top or antitop 
quark goes on-shell, and we keep only the $i=1,2$ terms.
 
 It follows that in the general expression (\ref{vertex})
we have to keep the terms which contain $F_1$ to $F_6$  (we can drop the 
 $F_7$ and $F_8$ terms), and we don't have factorization anymore. This is 
different from what happens in the $W$ pair production process,
where in DPA factorization holds even in the off-shell case. This
difference is due to the fact that in our process the intermediate particles
are fermions, and not bosons.

%--------------------------------------------------------------------------

\subsection{Renormalization and fermion self-energy}

\begin{figure}[t!] % the renormalized vertex
\centerline{\epsfig{file=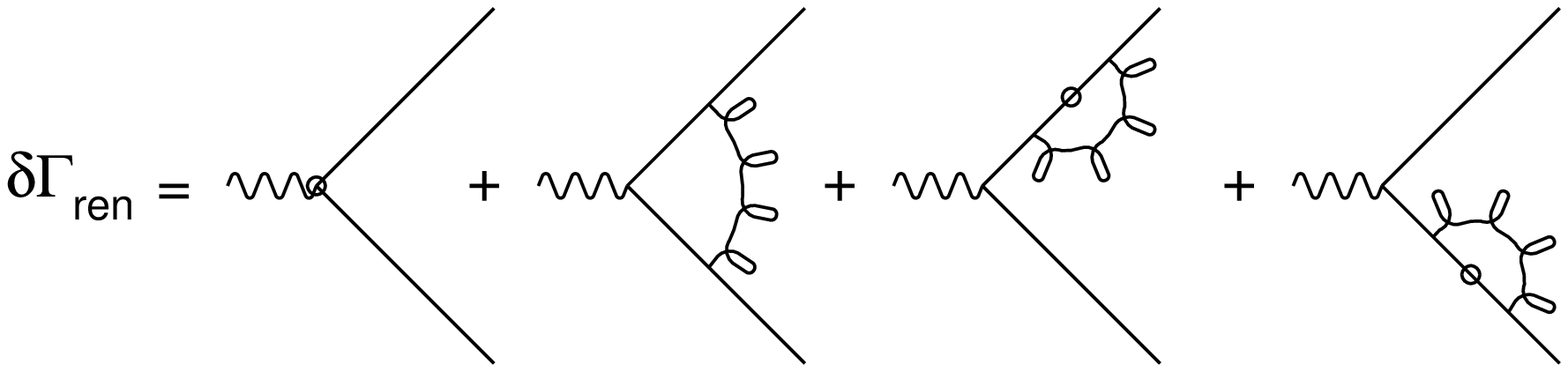,height=1.4in,width=5.in}}
\caption{Terms contributing to the renormalized vertex; the dots
 represent counterterm insertions.}
\label{ren_vert}
\end{figure}

Since we are performing a next-to-leading order computation, we have to 
deal with the issue of ultraviolet divergences and renormalization. 
We use the counterterm method (for a presentation of 
the essential features see for example \cite{peskin_schroeder}). What
renormalization amounts to in our case is replacing the bare vertex 
correction in Figure \ref{ver_cor} (which is UV divergent) with a 
finite renormalized vertex correction:
$$ \delta \Gamma^{\mu} \ \rightarrow \  \delta \Gamma^{\mu}_{ren}$$
\bec{ver_ren}
\delta \Gamma^{\mu}_{ren} \ = \
\delta \Gamma^{\mu} + \Gamma^{\mu} \delta Z_2 + 
\frac{1}{2}(-i  
\hat{\Sigma}_2(\not{p}_1) ) \frac{i}{\not{p}_1 - m_1} \Gamma^{\mu}
+ 
\frac{1}{2}\Gamma^{\mu}\frac{i}{-\not{p}_2 - m_2} (-i  
\hat{\Sigma}_2 ) (\not{p}_2)
\eec
in which we have included also the contributions of the fermion self-energy
diagram. The diagrams corresponding to
separate terms in Eq. \ref{ver_ren} are presented in Figure \ref{ren_vert}.

 The first two terms in Eq. \ref{ver_ren} are what is usually defined
as the renormalized vertex. The last two terms are one half of the 
renormalized fermion and antifermion self-energy:
\bec{ren_se}
  \hat{\Sigma}_2(\not{p}) = \Sigma_2(\not{p}) - 
(\pm \not{p} - m) \delta Z_2 -\Delta m
\eec
where $ \delta Z_2$ and $ \Delta m$ are coefficients of the counterterms
in the Lagrangian density 
(the $\pm$ sign corresponds to the particle/antiparticle case; one half
because a fermion propagator connects to two vertices). In the above 
equation, $\Sigma_2$ stands for the {\it bare} fermion self-energy:
\bec{bare_fse}
 \Sigma_2(\not{p}) = \frac{\alpha_s}{4 \pi} \
\int \frac{d^4 k}{i \pi^2}\ \frac{1}{k^2}\ \gamma^{\mu}\
\frac{\not{p} -\not{k} + m }{(p-k)^2 - \bar{m}^2}\ \gamma_{\mu}
\eec
Upon evaluation, we can write the result for Eq. \ref{bare_fse} in the
form:
$$ \Sigma(\not{p}) = (\pm \not{p} - m) \Sigma_a(p^2) \ + \ m \Sigma_b(p^2)$$
separating it into a scalar and a spinorial component. With these
notations, the renormalized vertex correction can be written as:
\bec{ren_ver2}
\delta \Gamma^{\mu}_{ren} \ = \
\delta \Gamma^{\mu} \ + \ \frac{1}{2}\ \Delta Z_2(p_1)\ \Gamma^{\mu}
\ + \ \frac{1}{2}\ \Gamma^{\mu}\ \Delta Z_2(p_2) 
\eec
with
\bec{del_Z2}
 \Delta Z_2(p)\ = \ \Sigma_a(p^2) + 
\frac{ m \Sigma_b(p^2) - \Delta m}{\pm \not{p} - m}
\eec
The counterterm coefficient $\Delta m$ is fixed by the 
on-shell renormalization condition: 
\bec{del_m}
 \hat{\Sigma}_2(\not{p}= m) \ = \ 0 \ \Rightarrow \ 
\Delta m =  m \Sigma_b(m^2)
\eec
Also, in the on-shell limit,
\bec{del_Z2_os}
 \Delta Z_2(p) \arrowvert_{\not{p}\rightarrow m}
= \left( \Sigma_a(p^2) + \left.
\frac{ \partial \Sigma_b(\not{p}\cdot\not{p})}{\partial\not{p}} 
\right\arrowvert_{\not{p}\rightarrow m}
 \right) =  \left. \frac{ \partial \Sigma_2(\not{p})}{\partial\not{p}} 
\right\arrowvert_{\not{p}\rightarrow m} = \delta Z_2
\eec
where for the last equality we have used the renormalization
condition
$$  \left.
\frac{ \partial \hat{\Sigma_2}(\not{p})}{\partial\not{p}} 
\right\arrowvert_{\not{p}\rightarrow m}
 = 0
$$

It is convenient to write the contribution of the fermion self-energy
diagrams in a form similar to that of Eq. \ref{vertex}. Using the resummed
top quark propagator:
\bec{top_prop}
\frac{i}{\pm \not{p} - m} \ \longrightarrow \
\frac{i(\pm \not{p} + m)}{ p^2 - \bar{m}^2}
\eec
we obtain the following result for $\Delta Z_2$ :
\bec{del_Z2_fin}
\Delta Z_2(p) =  \left( [\Sigma_a(p^2) + 2 \Sigma_{ir}(p^2)] \ + \
\frac{\Sigma_{ir}(p^2)}{m}( \pm \not{p} - m ) \right)
\eec
with 
$$  \Sigma_{ir}(p^2) = 
m \frac{ \Sigma_b(p^2) - \Sigma_b(m^2)}{p^2 - \bar{m}^2}. $$
The term in square brackets in Eq. \ref{del_Z2_fin} will multiply the 
Born cross section. The term proportional to $ ( \pm \not{p} - m )$ is 
identical to the like terms appearing in the expression for the 
vertex correction Eq. \ref{vertex}. Since $  \Sigma_{ir}$ is the part 
of the self-energy correction which would be infrared divergent on-shell
(which means that it has a logarithmic resonant behavior
$$ \Sigma_{ir}(p^2) \sim \hbox{log}(p^2 - \bar{m}^2)$$
in the off-shell case), we keep this term also.

%--------------------------------------------------------------------
\section{Gauge invariance and corrections to particular subprocesses}

The partial amplitudes appearing in Eq. (\ref{mvg_amp}) can be directly related 
to Feynman diagrams and are straightforward to evaluate. 
However, as mentioned before, they cannot be directly identified with 
corrections to particular subprocesses. For example, the top - antitop
production vertex diagram (Figure \ref{vir_diag}a)) contributes to the
correction to production vertex, as well as to interference between
production and decay and even to interference between top decay and 
antitop decay, depending on when the top quark propagators are closer
to the resonances. Moreover, the amplitudes in Eq. (\ref{mvg_amp})
are not gauge invariant one by one, although their sum is.

For  purposes related to gauge invariance, and in order to be able to perform 
comparisons with the on-shell computation,
 it is desirable to 
decompose the total amplitude into gauge invariant corrections to 
particular subprocesses, and interference between these. 
The aim is to rewrite  Eq. (\ref{mvg_amp}) as:
\bec{m1a_gi}
{\cal{M}}^{vg} = {\cal{M}}_{prod} + {\cal{M}}_{tdec} + {\cal{M}}_{\bar{t}dec} +
{\cal{M}}_{prod-tdec}^{intf} + {\cal{M}}_{prod-\bar{t}dec}^{intf} + 
{\cal{M}}_{tdec-\bar{t}dec}^{intf}
\eec
with each term being gauge invariant by itself.

 To this end, it is necessary to decompose the amplitudes ${\cal{M}}_{tt}, {\cal{M}}_{tb} ...$
into parts which contribute solely to corrections to production, decay, or
interference. 
This decomposition will be based on the top and antitop
propagator structure of the matrix element. 
Following the prescription introduced in section {\bf 3.1.1},
products of propagators
which go on-shell in different regions of the phase space can be decomposed 
as follows:
\bec{gi_dec}
\frac{1}{D(p_t)}\ \frac{1}{D(p_t-k)} \ = \
\frac{1}{D_0(p_t-k)}\ \frac{1}{D(p_t)}\ - 
\frac{1}{D_0(p_t-k)}\ \frac{1}{D(p_t-k)}
\eec
with 
\bec{prop_def}
 D(p) = p^2 - \bar{m}^2 \ , \ D_0(p-k) = (p-k)^2 - p^2
\eec
In Eq. \ref{gi_dec} the first term on the right hand side
is considered as a contribution to the production 
process and the second one a contribution to the decay process.
Furthermore, it is convenient to write the result in term of products
of gauge invariant currents (in a manner similar to \cite{ddr}).
For example, the ${\cal{M}}_{b\bar{t}}$ amplitude can be written
(using the extended soft gluon approximation):
\bec{gfgds}
{\cal{M}}_{b\bar{t}}(ESGA) = \frac{\alpha_s}{4\pi}\ {\cal{M}}_0 
\int \frac{d^4k}{i\pi^2}\ G_{\mu \nu}(k)\
\frac{ -2p_{\bar{t}}^{\mu} } { D(p_{\bar{t}}+k) }\
\frac{2p_b^{\nu}}{D_0(p_b-k)}\ \frac{D(p_t)}{D(p_t-k)}
\eec
where $G_{\mu \nu}(k)$ is the gluon propagator in an arbitrary gauge:
\bec{gkasd}
G_{\mu \nu}(k) \ = \ \frac{-i}{k^2 + i\epsilon}
\left( g _{\mu \nu} - (\xi-1) \frac{k_{\mu}k_{\nu}}{k^2} \right).
\eec
By using the currents :
\bec{gkfd}
j_{tdec}^{\ b ,\ \mu} = \frac{2p_b^{\mu}}{D_0(p_b-k)}\ \frac{D(p_t)}{D(p_t-k)}
\eec
$$
j_{prod}^{\ \bar{t},\ \mu} = \frac{-2p_{\bar{t}}^{\mu}}{D_0(p_{\bar{t}}+k)}\
\hbox{~~~,~~~}
j_{\bar{t}dec}^{\ \bar{t},\ \mu} = 
\frac{ -2p_{\bar{t}}^{\mu} }{ D_0(p_{\bar{t}}+k) }\
\frac{ D(p_{\bar{t}}) }{ D(p_{\bar{t}}+k) }
$$
we get :
\bec{mbbart_intf}
{\cal{M}}_{b\bar{t}}(ESGA) = \frac{\alpha_s}{4\pi}\ {\cal{M}}_0 
\int \frac{d^4k}{i\pi^2}\ G_{\mu \nu}(k)\
(\ j_{prod}^{\ \bar{t},\ \mu} - j_{\bar{t}dec}^{\ \bar{t},\ \mu} \ )
j_{tdec}^{\ b ,\ \nu} 
\eec
where the first term in parentheses contributes to production-decay
interference, and the second one contributes to decay-decay interference.

We can similarly define the currents :
\bec{glkvd}
j_{\bar{t}dec}^{\ \bar{b} ,\ \mu} = 
\frac{-2p_{\bar{b}}^{\mu}}{D_0(p_{\bar{b}}+k)}\ 
\frac{D(p_{\bar{t}})}{D(p_{\bar{t}}+k)}
\eec
$$
j_{prod}^{\ t,\ \mu} = \frac{2p_t^{\mu}}{D_0(p_t-k)}\
\hbox{~~~,~~~}
j_{tdec}^{\ t,\ \mu} = \frac{2p_t^{\mu}}{D_0(p_t-k)}\
\frac{ D(p_t) }{ D(p_t-k) }
$$
and the amplitudes for the other two interference diagrams can be written like:
\bec{mtbarb_intf}
{\cal{M}}_{t\bar{b}}(ESGA) = \frac{\alpha_s}{4\pi}\ {\cal{M}}_0 
\int \frac{d^4k}{i\pi^2}\ G_{\mu \nu}(k)\
(\ j_{prod}^{\ t,\ \mu} - j_{tdec}^{\ t,\ \mu} \ )
j_{\bar{t}dec}^{\ \bar{b} ,\ \nu}
\eec
\bec{mbbarb_intf}
{\cal{M}}_{b\bar{b}}(ESGA) = \frac{\alpha_s}{4\pi}\ {\cal{M}}_0 
\int \frac{d^4k}{i\pi^2}\ G_{\mu \nu}(k)\
j_{tdec}^{\ b ,\ \mu} \ j_{\bar{t}dec}^{\ \bar{b} ,\ \nu} \ \ .
\eec

Contributions to interference between subprocesses do not come solely
from the manifestly non-factorizable diagrams. The diagrams in which
the gluon contributes to vertex or self-energy corrections (Fig. 2a))
also contain interference terms. 
Since the decomposition into purely vertex (or self-energy) corrections
and interference corrections is not unique, we shall 
present our approach in some detail.

%These terms can be separated by using the following arguments :
The amplitude for the vertex correction diagram with off-shell particles
can be written as :
\bec{kaglkad}
 (\delta \Gamma)_{12} = \frac{\alpha_s}{4 \pi} \
\int \frac{d^4 k}{i \pi^2}\ G_{\mu \nu}(k)\ \gamma^{\nu}\
\frac{A(p_1,p_2) + k^{\alpha}B_{\alpha}(p_1,p_2) + 
k^{\alpha}k^{\beta} C_{\alpha \beta}(p_1,p_2)}
{( (p_1 - k)^2 - m_1^2 )\ (p_2 + k)^2 - m_2^2 )}\
\gamma_{\nu}
\eec
A corresponding on-shell approximation for this amplitude would be 
\bec{kfgg}
 (\delta \Gamma)_{os} = \frac{\alpha_s}{4 \pi} \
\int \frac{d^4 k}{i \pi^2}\ G_{\mu \nu}(k)\ \gamma^{\nu}\
\frac{A(p'_1,p'_2) + k^{\alpha}B_{\alpha}(p'_1,p'_2) + 
k^{\alpha}k^{\beta} C_{\alpha \beta}(p'_1,p'_2)}
{( -2p'_1 k + k^2  )\ ( 2 p'_2 k + k^2 )}\
\gamma_{\nu}
\eec
where $p'_1$ and $p'_2$ are some on-shell approximations for 
$p_1$ and $p_2$. Now, we can define the interference contribution through:
\bec{jdhgdso}
(\delta \Gamma)_{12} = (\delta \Gamma)_{12}^{os} + 
(\delta \Gamma)_{12}^{intf}
\eec
Note, however, that $(\delta \Gamma)_{12}^{os}$ is not unique, 
since $p'_1,p'_2$
are not unique; different choices for these momenta would yield different
results for $(\delta \Gamma)_{12}^{os}$.
 The uncertainty which arises is, of course, of order
$p^2-m^2$, so it can be neglected in the DPA. However, it allows us to 
choose the following definition for $(\delta \Gamma)_{12}^{os}$:
 $$
(\delta \Gamma)_{12}^{os} = \frac{\alpha_s}{4 \pi} 
\int \frac{d^4 k}{i \pi^2}\ G_{\mu \nu}(k) \left[
\frac{(2p_1^{\mu}) \ {\Gamma}\ (-2p_2^{\nu})}{ D_0(p_1-k) D_0(p_2+k)} +
\right.
$$
\bec{kgsvdkjl} \left.
\gamma^{\mu} \frac{ k^{\alpha}B_{\alpha}(p_1,p_2) + 
k^{\alpha}k^{\beta} C_{\alpha \beta}(p_1,p_2)}
{ D(p_1 - k) D(p_2 + k) } \gamma^{\nu} \right]
\eec
This choice means that the purely vertex correction (factorizable) part of the 
vertex diagram can be obtained by simply replacing the off-shell 
$C_0(p_1,p_2,0,m_1,m_2)$
function appearing in the expression for $(\delta \Gamma)_{12}$
with the on-shell, infrared divergent function 
$C_0(p_1,p_2,\mu,\sqrt{p_1^2},\sqrt{p_2^2})$

Conversely, the interference part of the off-shell vertex correction diagram
is:
\bec{kohdsi}
(\delta \Gamma)_{12}^{intf} = 
\frac{\alpha_s}{4 \pi} 
\int \frac{d^4 k}{i \pi^2}\ G_{\mu \nu}(k) \left[
\gamma^{\mu} \frac{ A(p_1,p_2) }{ D(p_1 - k) D(p_2 + k) }
\gamma^{\nu} - 
\frac{(2p_1^{\mu}) \ {\Gamma}\ (-2p_2^{\nu})}{ D_0(p_1 - k) D_0(p_2 + k)}\right]
\eec
with $ A(p_1,p_2) = (\not{p_1} + m_1)  {\Gamma} (-\not{p_2} + m_2)$.
For the $t \bar{t}$ production diagram, \nolinebreak[5] in \nolinebreak[5]
 DPA 
$$ \gamma^{\mu} (\not{p_t} + \bar{m}_t) \ {\Gamma_{t \at}}\ 
(-\not{p}_{\bar{t}} + \bar{m}_t) \gamma^{\nu} \rightarrow
(2p_t^{\mu}) \ {\Gamma_{t \at}}\ (-2p_{\bar{t}}^{\nu})
$$
leading to 
{\small \bec{lgdvkj}
{\cal{M}}_{t\bar{t}}^{intf} = \frac{\alpha_s}{4\pi}\ {\cal{M}}_0 
\int \frac{d^4k}{i\pi^2}\ G_{\mu \nu}(k)\ \left[
(- j_{tdec}^{\ t,\ \mu} )\ j_{prod}^{\ \bar{t},\ \nu}\ +
j_{prod}^{\ t,\ \mu} (- j_{\bar{t}dec}^{\ \bar{t},\ \mu} )\ +
(- j_{tdec}^{\ t,\ \mu} )\ (- j_{\bar{t}dec}^{\ \bar{t},\ \mu} )
\right]
\eec }

Things are  different for the decay vertices corrections, since
we have doubly resonant contributions which are not proportional
to the tree level matrix element. In the 
top decay case, the transformation:
$$ \gamma^{\mu} (\not{p_b} + \bar{m}_b) \ \not{\epsilon}_{W^+}\ 
 (\not{p_t} + \bar{m}_t) \gamma^{\nu} \rightarrow
(2p_b^{\mu}) \  \not{\epsilon}_{W^+}\  \left[ 2p_t^{\nu}
+ \gamma^{\nu} (-\not{p_t} + \bar{m}_t) \right]
$$
will lead to:
\bec{vhgdsj}
{\cal{M}}_{t b}^{intf} = \frac{\alpha_s}{4\pi}\
\int \frac{d^4k}{i\pi^2}\ G_{\mu \nu}(k)\ 
j_{tdec}^{\ b ,\ \mu}\ \left[ (-j_{prod}^{\ t,\ \nu}) {\cal{M}}_0 +
M_1^{t, \ \nu} \right]
\eec
where
\bec{gkfdsah}
M_1^{t, \ \mu}\ = \ 
\frac{-1}{D(p_t)\ D(p_{\bar{t}})}\ \left[
\bar{u}(b) \not{\epsilon}_{W^+}\
\gamma^{\mu}\ {\Gamma_{\gamma,Z_0}}\  (-\not{p}_{\bar{t}} + \bar{m}_t)
\not{\epsilon}_{W^-} v(\bar{b}) \right]
\eec
In a similar manner, the interference term ${\cal{M}}_{\bar{t} \bar{b}}^{intf}$ 
coming from the antitop vertex correction diagram can be written in terms 
of the currents $j_{\bar{t}dec}^{\ \bar{b} ,\ \mu},\ 
-j_{prod}^{\ \bar{t},\ \nu}$, and the matrix element $M_1^{\bar{t}, \ \nu}$.

Finally, the last diagrams to be split into on-shell and interference 
contribution are the top, antitop self-energy diagrams. Using the same 
approach as in the vertex case, we define:
$$ 
(\Delta Z)_{t}^{intf} = \frac{\alpha_s}{4\pi}\  \left\{
\int \frac{d^4k}{i\pi^2}\ G_{\mu \nu}(k)\ \left[
\gamma^{\mu} \frac{\not{p_t} + \bar{m}_t}{D(p_t-k)} \gamma^{\nu}\
-\ \frac{2p_t^{\mu} \gamma^{\nu}}{ D_0(p_t-k)} \right] \right\}\ 
\frac{\not{p_t} + \bar{m}_t}{D(p_t)}\ -
% - (\delta Z)_{t}
$$
\bec{jhvdsj}
%(\delta Z)_{t} = 
-\ \frac{\alpha_s}{4\pi}\ 
\int \frac{d^4k}{i\pi^2}\ G_{\mu \nu}(k)\
\frac{2p_t^{\mu}}{D_0(p_t-k)} \frac{2p_t^{\nu}}{D_0(p_t-k)}(-1)
\eec
where the quantity in the the curly brackets is the renormalized top
self-energy, and the quantity on the second line is the on-shell limit
of the quantity on the first line. 
%and $(\delta Z)_{t}$ is the on-shell limit of the first term :
This will lead to the following result for the interference 
contribution coming from the top self-energy diagram:
\bec{lkjgds}
{\cal{M}}_{tt}^{intf} = \frac{\alpha_s}{4\pi}\ 
\int \frac{d^4k}{i\pi^2}\ G_{\mu \nu}(k)\
\left[ j_{prod}^{\ t,\ \mu}\ {\cal{M}}_0\ -\ M_1^{t, \ \mu} \right]
j_{tdec}^{\ t,\ \nu}
\eec
and a similar one from the antitop self-energy diagram.

Now we have all the pieces needed to write down the interference terms.
The final result is:
$$
{\cal{M}}^{intf} = \frac{\alpha_s}{4\pi}\ 
\int \frac{d^4k}{i\pi^2}\ G_{\mu \nu}(k)\ \left[
(j_{prod}^{\ \mu} {\cal{M}}_0 + M_1^{t, \ \mu}) j_{tdec}^{\ \nu} - 
(j_{prod}^{\ \mu} {\cal{M}}_0 - M_1^{\bar{t}, \ \mu}) j_{\bar{t}dec}^{\ \nu}
\right.
$$
\bec{gi_intf_dec}
\left.
+ j_{tdec}^{\ \mu} j_{\bar{t}dec}^{\ \nu} {\cal{M}}_0 \right]
\eec
It is easy in this formula to identify the production-decay or decay-decay
interference terms. The currents:
\bec{dkscdsa} j_{prod} = j_{prod}^{\bar{t}} - j_{prod}^{t} \mbox{~~,~~}
j_{tdec} = j_{tdec}^b - j_{tdec}^t \mbox{~~,~~}
j_{\bar{t}dec} = j_{\bar{t}dec}^{\bar{b}} - j_{\bar{t}dec}^{\bar{t}}
\eec
are conserved, and gauge invariant in DPA
(that is, $k_{\mu} j^{\mu} = 0$, and the partial amplitudes in Eq. 
\ref{gi_intf_dec} are independent on the gauge paramenter $\xi$ in $G(k)$).
 Therefore, the total interference contribution
as well as the interference between subprocesses parts are gauge invariant
in the approximation used.

%--------------------------------------------------------------------------
\section{Computational Approach}

Once a consistent scheme for evaluating the virtual corrections 
to the top production and decay process (\ref{tree_p}) has been set up
(as described in the previous sections),
the next step is the implementation of this scheme
in a Monte Carlo program.  
In this section we give some details about the technical issues 
arising in the design of such a program, and how we choose to solve them.

There are two types of quantities involved in the evaluation of the
NLO amplitude: scalar quantities (form factors), which encode 
the contribution of loops, and spinorial quantities,
built from Dirac spinors and operators.
For example,
the contribution coming from the $t \bar{t}$ vertex correction can be written:
\bec{cdsa}
\tilde{\cal{M}}_{t \bar{t}} = 
\bar{u}(p_b) \not{\epsilon}_{W^+}  (\not{p}_t + m_t) \
\delta \Gamma^{t\at}_{ren} (-\not{p}_{\bar{t}} + m_t) 
\not{\epsilon}_{W^-}  v(p_{\bar{b}})
\eec
or, using the decomposition in Eq. \ref{vertex} :
\bec{prod_cont}
\tilde{\cal{M}}_{t \bar{t}}
 = \frac{\alpha_s}{4 \pi} \sum_{i = 1,2} \left[
C_V\ F_i^V\ T_i^V\ + C_A\ F_i^A\ T_i^A \right] 
\eec
(definitions for the quantities appearing in the above equation
can be found in Appendix {\bf B}).

Let's start by discussing the evaluation of the scalar form factors $F_i$.
Rather than compute analytic expressions for each form factor, we have
chosen to evaluate them in terms of Passarino-Veltman (PV)
functions \cite{passa}. This approach has the
advantage that we have to compute only a
 few quantities which contain logarithms 
and dilogarithms: the ${\cal{B}}_0$ two-point and ${\cal{C}}_0$ three-point 
scalar functions (all the rest can be
written as linear combination of these functions). In turn, for evaluating
the  PV 2 and 3-point scalar functions, we use the FF routines \cite{old}. 

To compute the amplitudes corresponding to the interference diagrams, we 
need to be able to evaluate the 4-point and 5-point scalar integrals in 
Eqs. \ref{mbbtr}, \ref{mtbbr}, \ref{mbbbr}.
 There are no published results (or routines) for the
general (complex masses) 4-point scalar integrals. We have build such
routines for the infrared finite ${\cal{D}}_0$
 function by using the general methods
described in \cite{thooft_si}. The results of these routines have been checked 
against analytical results in the soft gluon approximation published in 
\cite{BBC}. 

The 5-point scalar function ${\cal{E}}_0$
 has been computed by reduction to 4-point
functions, following the recipe in \cite{ddr}. The resulting infrared divergent
4-point functions have been evaluated using the analytic results published
in \cite{denner_ir4pt}.

Some comments on the treatment of the top width are needed here. One way of 
evaluating the scalar form factors in Eq. \ref{prod_cont} is to compute the
gluon integrals in the zero top width limit and introduce the finite width only 
in terms which are divergent on-shell (that is, replace $m_t^2$ with 
$\bar{m}_t^2 = m_t^2 - i m_t \Gamma_t$ in terms like log$(p_t^2 - m_t^2)$;
see for example \cite{ddr}).
The difference between this result and the one obtained by using
the complex top mass in all the terms is of order $\Gamma_t/m_t$, therefore
at about 1\% level. This would be acceptable if the radiative corrections 
would be small with respect with the tree level result (as is the case for the
$W$ production process), but in our case it turns out that the one-loop
QCD corrections are of the same order of magnitude as the tree level result
\footnote{the reason the {\it total } QCD corrections are of order 10 - 20\%
is because of large cancellations between the virtual corrections and
soft gluon real corrections.}. Therefore, order \% terms are important. Since
in the case of real gluon radiation the top width appears in all terms, for 
reasons of consistency we need to keep the width 
in all terms in the evaluation of the virtual corrections too. 

 The other elements needed in the evaluation of the amplitude (\ref{prod_cont})
are the spinor sandwiches $T_i$. We compute these quantities using
spinor techniques, as for the real gluon radiation case. Since 
this part of the computation is quite complex, and hence prone to errors, we 
have two different ways of performing it. In one approach, we express the
$T_i$'s in terms of basic spinor products
$ \bar{u}(p_i,s_i) u(p_j,s_j)$; this is the more involved
case (in terms of the work done by the programmer), geared  for
implementation in a Fortran routine, and which allows fast computation. 
The other approach uses C++ routines which allow the 
automated evaluation of general spinor sandwiches like 
$$ \bar{u}(p,s)(\not{p}_1 + m_1)(\not{p}_2 +m_2) \ldots u(p',s')$$
(To this purpose, we have 
constructed classes that describe $<bra|$ and $|ket>$ spinors,
and operators of type $ \not{p}_i \pm m_i $; in turn, 
these classes use the basic classes - 4-vector, complex number - 
defined in the 
Pandora event generator \cite{peskin}).
This method allows easy evaluation of $T_i$ expressions
(again from the programmer's viewpoint)
and is much more resistant to programming errors;
 but the computation is slower than in the previous method. Therefore, 
the main use of the results obtained from  the C++ routines is 
to check the Fortran results.

\section{On-shell DPA}

The issue of interference effects in the production and decay 
of heavy unstable particles has been the subject of extensive
studies in the past decade. Maybe the most important result
is a theorem, due to Fadin, Khoze and Martin \cite{FKM_0}, which states
that these interference effects are suppressed.  A stronger version
of this theorem \cite{FKM_1} claims that NLO interference effects
cancel in inclusive quantities up to terms of order $\alpha \Gamma/M$.
Following the methods used in \cite{FKM_1}, it is possible to 
define a framework
for the computation of interference corrections in which the total 
interference contribution to inclusive quantities is zero.

In this section we shall discuss this alternative approach 
(which we shall call on-shell DPA) to the computation
of NLO corrections to the production and decay of unstable particles.
%This approach has been initially used for the $W$ pair production case
Results obtained using this approach have been presented for 
the $W$ pair production case
at LEP II (\cite{ddr}, \cite{BBC}); recently, 
this approach has been also applied in the 
computation of interference (non-factorizable) 
corrections to the top production and decay process at $e^+ e^-$ as well
as at hadron colliders \cite{bbc_top}. 

The relevant features of this approach are two: first, the amplitudes
for corrections to subprocesses are computed in the on-shell approximation.
For example, the correction to the production process can be written in 
terms of the on-shell amplitude:
\bec{os_prod}
\tilde{\cal{M}}_{prod}^{os} \ = \ \sum_{\la,\la'}
{\cal{M}}_{\la,\la'}(e^+e^- \to t \at (g))\
{\cal{M}}_{\la}(t \to b W^+)\ {\cal{M}}_{\la'}(\at \to \ab W^-)
\eec
where $\la,\la'$ are the spins of the top quarks.
The difference between the above amplitudes and the ones used in our 
computation (Eqs. \ref{m1a_gi}, \ref{mrg_gi}) is due to non-doubly 
resonant terms, therefore acceptable in DPA.

The other characteristic feature of the on-shell DPA method
is that the interference due to real gluon radiation is computed
by using a semianalytic approach. This approach rests on
the observation that interference is due mainly to gluons of energies
of order $\Gamma_t$; therefore, we can use the (extended)
soft gluon approximation in the evaluation of interference terms.

There are two stages where this approximation comes into play. First,
we apply it at the matrix element evaluation level. For example,
consider interference between the diagrams where the gluon is radiated
from the bottom quark and from the antitop quark. Using the
formulas in the Appendix {\bf B}:
\bec{tbb_intf}
d\sigma^{rg}_{\at b}(p_b,p_W,\ldots,k) \sim 
2 \hbox{Re}\left[
\sum_{\epsilon_g}\ {\cal{M}}_{b}^{sg}\ ({\cal{M}}_{\at}^{sg})^* \right]
=
\eec
$$
\ = \ |{\cal{M}}_0(p_b,p_W,\ldots)|^2 \
2 \hbox{Re} \left[ \ \frac{4p_{\at} p_{b}}{2kp_{b}+i\epsilon}\
\frac{p_{t}^2 - \bar{m}_t^2}{(p_{t}+k)^2 - \bar{m}_t^2}\
\frac{1}{(p_{\at}+k)^2 - \bar{m}_t^{2*}} \ \right].
$$

 The second stage is the treatment of the final state phase space. In the 
soft gluon approximation, we can factorize it: 
$d\Omega_{b,W,\ldots,g} = d\Omega_{b,W,\ldots} \times d\Omega_g$ , and
perform the integration over the gluon momenta analytically:
$$
d\sigma^{rg}_{\at b}(p_b',p_W',\ldots)\ = \  
|{\cal{M}}_0(p_b',p_W',\ldots)|^2 \ \times
$$
\bec{tbb_sig}
\frac{\alpha_s}{\pi} \hbox{Re} \left[ \int \ \frac{d^3k}{2\pi \omega}
\ \frac{4p_{\at} p_{b}}{2kp_{b}+i\epsilon}\
\frac{p_{t}^2 - \bar{m}_t^2}{(p_{t}+k)^2 - \bar{m}_t^2}\
\frac{1}{(p_{\at}+k)^2 - \bar{m}_t^{2*}} \ \right]
\eec
where $p_b',p_W',\ldots$ are given by a suitable projection of the
off-shell momenta \\ $ p_b,p_W,\ldots $ onto the on-shell phase space (for
an example of how this projection might be accomplished see \cite{bbc_details}).
In the above equation we also have $p_t = p_b' + p_{W^+}', \ 
p_{\at} = p_{\ab}' + p_{W^-}'$, and the integral over gluon momenta
is allowed to go to infinity (since hard gluons contribute nonresonant
terms to the result).

The quantity on the second line of Eq. \ref{tbb_sig} can be evaluated 
analytically, through methods similar to those used to evaluate the
virtual 4-point functions. We will not give the results here 
(they can be found in \cite{ddr}, \cite{BBC}),
but there is an important comment to
make. If we use this procedure to compute the real gluon interference,
the total interference obtained by adding the virtual diagram contribution
(Eq. \ref{mbbtr}) to the above result and integrating over the top invariant
mass parameter is zero. The proof of this statement can be found in 
\cite{FKM_1}.
This cancellation works also for the other interference diagrams; therefore,
in this approach, the contribution of non-factorizable corrections is 
zero to the total cross section. However, this result depends on two 
things. First, it requires an inclusive treatment of real gluon radiation, with 
phase space integration extending to infinity. Second, both the virtual
and the real interference terms have to be treated in the soft gluon
approximation.

But, is the use of the soft gluon approximation justified in this case?
At the amplitude level (Eq. \ref{tbb_intf}), the answer is yes; the relevant 
gluon energy, being of order $\Gamma_t$, is much smaller than the other
momenta involved. However, this approximation does not seem to be acceptable
for the phase space factorization stage of the above approach. Here, problems
might arise when we try to perform the projection of 
the off-shell momenta onto the on-shell phase space. The reason for this is that
there is no single way to perform this projection; therefore, in the 
determination of the the on-shell momenta $p_b', p_W', \ldots$ there is an 
uncertainty of the order of the gluon energy, or $\Gamma_t$. Now, 
being close to the top resonances,
we are in a region of the phase space where the cross section varies
greatly over a range of energy of order $\Gamma_t$
(due to the top quark propagators); therefore such an 
uncertainty is not acceptable.

To illustrate the dependence of the result for real gluon interference
on the choice of the on-shell momenta $ p_b', p_W', \ldots$, let's presume
that instead of projecting $p_b$ into $p_b'$, we also take into account
the gluon momentum: $p_b + k \to p_b'$ (physically, this might be justified by 
the inclusion of the gluon jet in the bottom quark jet). 
Then, Eq. \ref{tbb_sig} becomes:
\bec{tbb_sig1}
d\sigma^{rg}_{\at b}(p_b',p_W',\ldots)\ = \  
|{\cal{M}}_0(p_b',p_W',\ldots)|^2 \ 
\frac{\alpha_s}{\pi} \hbox{Re} \left[ \int \frac{d^3k}{2\pi \omega}
\ \frac{4p_{\at} p_{b}}{2kp_{b}+i\epsilon}\
\frac{1}{(p_{\at}+k)^2 - \bar{m}_t^{2*}} \right]
\eec
The result for the above expression is different from the result 
for Eq. \ref{tbb_sig}, and the difference contains doubly resonant terms.
Therefore, in the on-shell DPA approach, the result for the interference
terms depends on how we perform the phase space factorization. A discussion
of this dependence for the $W$ pair production case can be found in \cite{ddr}.

%----------------------------------------------------------------------
\section{Results for virtual corrections and the total cross section}
 
In this section, we present some results on the total cross section
for the top production and decay process at linear colliders. We take into 
account the virtual corrections as well as contributions coming from 
real gluon radiation. Furthermore, we study the effect of 
interference (nonfactorizable)
terms on invariant top mass distributions and  we perform comparisons
with results previously published \nolinebreak[3] \cite{bbc_top}.

In obtaining the results presented in this section, 
the following set of parameters is used:
$$ m_t = 175\ \mbox{GeV,} \mbox{~~~~}
\alpha_s = 0.1, \mbox{~~~~}
\Gamma_{t}^0 = 1.55\ \mbox{GeV,} \mbox{~~~~}
\Gamma_{t} = 1.42\ \mbox{GeV,} \mbox{~~~~}
$$
where $\Gamma_{t}^0$ is the top width at the tree level, while
$\Gamma_{t}$ includes QCD radiative corrections.

\vspace{0.3cm}
We start by looking at the total cross section for our process. 
Table 4.1 presents results for the following quantities:
\begin{itemize}
\item 
$\sigma_0$ : cross section for the tree level process (\ref{proc1});
computed in the on-shell (narrow width) approximation,
using the zero-order top width.
\item 
$\sigma_1^{os}$ : cross section for the NLO process in the on-shell 
approximation (computed using NLO top width). 
%For this quantity, we use the same definition as \cite{schmidt};
%by construction, the total cross section in this approximation is
%equal to the NLO result for
%the production process $\sigma[e^+ e^- \rightarrow t \bar{t} (g)] $ 
\item 
$\sigma_1^{fact}$ : the main (factorizable) part of the DPA approximation
to the NLO process.
This quantity contains corrections to production and decay as defined
in section {\bf 4.2}. 
%The difference between $\sigma_1^{os}$ and 
%$\sigma_1^{fact}$ is due to the use of  
%that in the computation of the latter quantity 
%we use the off-shell phase space, and the corrected top width $\Gamma_{t}^1$.

\item 
$\sigma_1^{intf}$ : the interference (non-factorizable) part of the DPA
approximation to the NLO process, as defined in section {\bf 4.2}.
\end{itemize} 

We present results for three values of collision center-of-mass energies:
360 GeV, just above the $t \at$ production threshold, 500 GeV, the most 
common value used in linear collider studies, and 1 TeV, which can be relevant
for higher energy machines. Note that at 360 GeV  our results
are probably not good, being too close to the threshold; however, 
it is interesting to see the magnitude of the nonfactorizable corrections 
at fixed order in this energy range.

The NLO cross sections contain contributions from the virtual corrections as
well as from real gluon radiation.
We use a technical cut to separate the infrared from the real 
gluons $\epsilon = 0.1$ GeV; the results are independent
of the choice of this parameter.
No physical cuts have been imposed on the final phase space.

\begin{table}[!t]
\begin{center}
\begin{tabular}{|c|c|c|c|}
\hline
 & & & \\
 	& 360 GeV & 500 GeV & 1000 GeV \\
 & & & \\
\hline
 & & & \\
 $\sigma_0 $       &~ 0.386&~ 0.570 &~ 0.172\\
 & & & \\
 $\sigma_1^{os}$   &~ 0.700&~ 0.660 &~ 0.184 \\
 & & & \\
 $\sigma_1^{fact}$ &~ 0.676&~ 0.664  &~ 0.197 \\
 & & & \\
 $\sigma_1^{intf}$ & -0.032& -0.012 & -0.006 \\
 & & & \\
\hline
\end{tabular}
\end{center}
\caption{Total cross sections for top production at linear colliders 
(measured in picobarns), with no cuts on phase space.}
\label{table1}
\end{table}

 There are several comments to make concerning these results. First, let's
compare $\sigma_1^{os}$ with $\sigma_1^{fact}$. The difference between these
two quantities is due to non-doubly-resonant terms, therefore it could be expected 
to be small. This is indeed the case at 500 GeV; but at 1 TeV, this difference
is about 6\% of the cross section. The reason is that in obtaining these results, we have integrated over the complete kinematic range available for
the top quark invariant mass (that is, 
$ m_b + m_W < \sqrt{p_t^2} < W - (m_b + m_W)$), so we get contributions 
from regions of the phase space where the top quarks 
are far off-shell and non-resonant
terms are important. 

\begin{table}[!t]
\begin{center}
\begin{tabular}{|c|c|c|c|}
\hline
 & & & \\
 	& 360 GeV & 500 GeV & 1000 GeV \\
 & & & \\
\hline
 & & & \\
 $\sigma_1^{os}$   &~ 0.682&~ 0.627  &~ 0.175 \\
 & & & \\
 $\sigma_1^{fact}$ &~ 0.670&~ 0.629  &~ 0.178 \\
 & & & \\
 $\sigma_1^{intf}$ & -0.034& -0.007 & -0.002 \\
 & & & \\
\hline
\end{tabular}
\end{center}
\label{table2}
\caption{Total cross sections for top production at linear colliders
(measured in picobarns), 
with cuts on the top, antitop invariant mass.}
\end{table}

 In Table 4.2 we present the cross section results obtained 
with a cut on the $t, \bar{t}$ invariant mass 
$ | \sqrt{p_t^2},\sqrt{p_{\bar{t}}^2} - m_t | < 15$ GeV. The difference
between the two results for the main terms $\sigma_1^{os}$ and
$\sigma_1^{fact}$ is small in this case. Note 
that, since in the on-shell approach $p_t^2, p_{\bar{t}}^2 = m_t^2$,
 $\sigma_1^{os}$ in Table 4.1 and 4.2
contains a factor which simulates 
the effect of cuts (either from kinematic constraints or imposed ones)
on the top invariant mass. Note also that these cuts are not imposed
{\it ad hoc}, but they arise rather naturally in the process of defining 
a $t, \bar{t}$ production event; it makes sense to require that the reconstructed mass of the $b, W$ pairs is close to the top mass in the definition of such an event. In this context, it is also worth noting that
the contribution coming from the phase space region where
either the $t$ or $\bar{t}$ is far off-shell (more that ten times 
the width) is quite sizable (around 5\% of the total cross section for CM energies greater than 500 GeV).   

\vspace{0.3cm}
Another quantity of interest is the differential interference cross section
as a function of the top invariant mass. Even if the total interference 
contribution to the cross section is small (at about 1\% level),
 it can have larger effects in differential distributions since
it can be positive in certain regions of the phase space and negative 
in others. In 
particular, it can be important in the reconstruction of the top invariant
mass; since $d \sigma_1^{intf}$ tends to decrease as $\sqrt{p_t^2}$ increases,
%$d \sigma_1^{intf} < 0$ for $\sqrt{p_t^2}>m_t^2$, and
%$d \sigma_1^{intf} > 0$ for $\sqrt{p_t^2}<m_t^2$, 
the net effect
would be to shift the position of the Breit-Wigner peak to smaller invariant
mass values. This effect can be quantified by the following equation:
the shift in the mass is
\bec{mass_shift}
\Delta M_t \ =\ \left. \left( \frac{d \delta_{nf}}{dM_t} \right)
\right|_{M_t = m_t}
\frac{\Gamma_t^2}{8}
\eec  
where $M_t = \sqrt{p_t^2}$, and $\delta_{nf}$ is the ratio of the 
non-factorizable (interference) part of the cross section to the
Born cross section:
$$ \delta_{nf} \ =\ \frac{d\sigma_1^{intf}}{d\sigma_0}$$

\begin{figure}[!t] % interference
\centerline{\epsfig{file=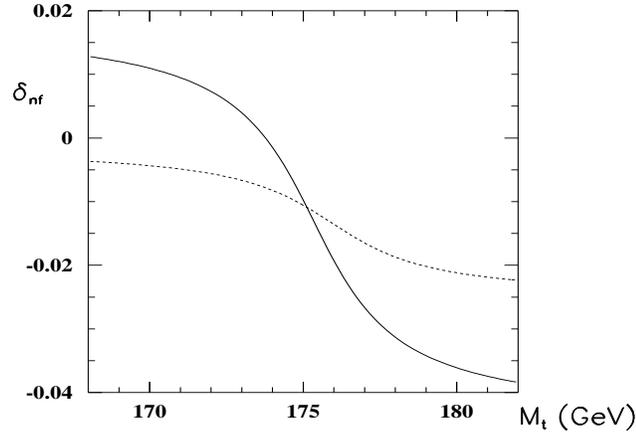,height=2.5in,width=3.in}}
\caption{ The relative nonfactorizable correction to the invariant
mass distribution; the solid line is the contribution of terms proportional
to the tree level amplitude, while the dashed line contains also the 
$M_1$ terms in Eq. \ref{gi_intf_dec}. }
\label{interf_rat}
\end{figure}

In Figure \ref{interf_rat} we present the differential distribution
for the relative non-factorizable correction $\delta_{nf}(M_t)$ at center
of mass energy 500 GeV. The dashed line is the result which takes into 
account the full interference corrections in Eq. \ref{gi_intf_dec}; the 
solid line is obtained by taking into account only the terms
proportional to the Born amplitude. Note that,
 although the contribution of the 
$M_1$ terms in Eq. \ref{gi_intf_dec} to the total cross section
is very close to zero, they have 
a sizable effect on the differential distribution in Fig. \ref{interf_rat}. 
Using Eq. \ref{mass_shift}, we conclude that the shift 
in the position of the peak in the top invariant mass distribution 
due to interference effects is very small (of order of a few MeV).

\begin{figure}[!t] % real interference
\centerline{\epsfig{file=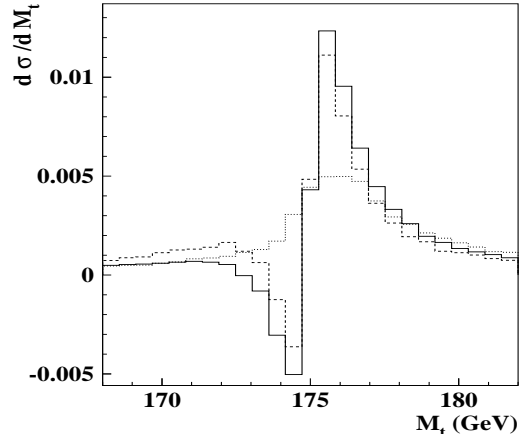,height=2.5in,width=3.in}}
\caption{ Real gluon interference: the $\sigma_{b\at}$ term. The
solid line corresponds to the semianalytic approach; the dashed line
is obtained through numerical evaluation with $M_t = \sqrt{p_{bW}^2}$;
the dotted line is obtained through numerical evaluation with $M_t$
given by Eq. \ref{recomb}.}
\label{real_interf}
\end{figure}

The results in Table 4.1 and 4.2 indicate that
the contribution of interference terms to the total cross section is
of order 1\%, in agreement with the $\Gamma_t/m_t$ order
of magnitude  expected from naive arguments. However, it is not zero, as 
implied by results presented in \cite{bbc_top}, which use the on-shell
DPA method.
We have argued in section {\bf 4.4} that this difference is due to the way
in which the radiation of real gluon with energies of order $\Gamma_t$
is treated. In Figure \ref{real_interf},
we present the results for the real interference 
between the diagram where the gluon is radiated from the bottom quark
and the diagram where the gluon is radiated from the antitop quark.
The solid line is the result of the semianalytical approach described
in section {\bf 4.4}. The other two lines are the result of the exact off-shell
computation (where the integration over the gluon momenta is performed
numerically). The two lines correspond to two different ways in which
the gluon momentum is treated in the reconstruction of the 
invariant top mass. For the dashed line, the gluon momentum is ignored
in the top mass reconstruction: $M_t = \sqrt{p_{bW}^2}$. Note that in this 
case, the result is quite close to that of the semianalytical computation, 
which is natural, since the gluon momentum is treated in both cases the same
way.

To obtain the dotted line, we have followed a more realistic approach, in 
which the gluon is included in the top mass reconstruction if
it happens to be radiated close enough to the top quark:
\bec{recomb} M_t \ = \ \left\{
\begin{array}{ll}
\sqrt{p_{bWg}^2} & \hbox{if }\ \hbox{cos}\theta_{tg}<\pi/3 \\
\sqrt{p_{bW}^2} & \hbox{otherwise}
\end{array} \right.
\eec
Although the total cross section is the same as for the other exact 
evaluation case, the differential cross section differs by quite a bit.

The total cross section corresponding to the 
 interference term presented in
Figure \ref{real_interf} has the value $\sigma_{b\at} = 0.121\ pb$ for the
semianalytical result, and $\sigma_{b\at} = 0.124\ pb$ for the numerical one.
Note that, since the diagram set contributing to this interference term is not
gauge invariant, this result has no physical meaning by itself. However, 
from these numbers we can get some insight  concerning the 
evaluation of interference corrections. First, note that the contribution
of this single diagram is much bigger (about two orders of magnitude) than
the total result for the interference terms. This means that there
are large cancellations taking place between the real and virtual interference 
contributions. This is quite natural, in accordance 
with the discussion in section {\bf 4.4}; however, this also means that small
uncertainty (order percent) in evaluating one of this contributions
(the one coming from the real gluon interference, for example) can lead 
to large uncertainties in the evaluation of the total interference
contribution.  

\begin{figure}[!t] % interference comparison
\centerline{\epsfig{file=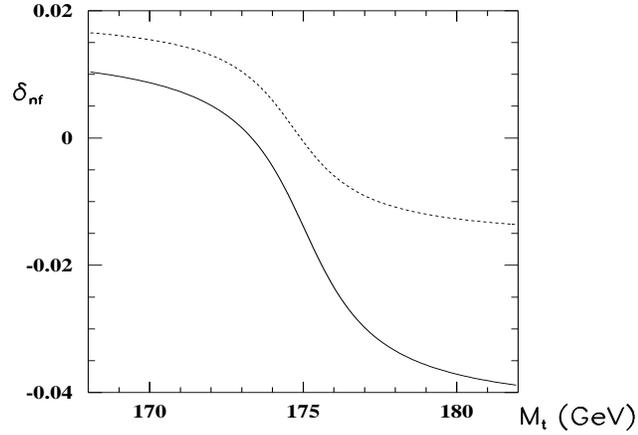,height=2.5in,width=3.in}}
\caption{ The relative nonfactorizable correction to the invariant
mass distribution; comparison between the semianalytical (dashed line) 
and the numerical (solid line) approach. }
\label{ir_comp}
\end{figure}

In Figure \ref{ir_comp} we present the comparison between the 
relative non-factorizable corrections computed in the semianalytical
approximation (dashed line) and the complete off-shell approach (solid line).
For the purpose of this comparison, we consider only the terms 
proportional to the Born amplitude in the exact computation, since
only these terms are taken into account in the semianalytical approach.
Note that the total interference cross section integrates to zero in the
latter case, and, as discussed above, the complete off-shell distribution
contains contributions that do not cancel in the total cross section.

\chapter{Conclusions}

{

In this thesis, we have discussed in some detail the evaluation of 
next to leading order QCD corrections to the top production and decay
process at a linear collider.
Since a full computation of the NLO amplitudes contributing to the 
process $e^+ e^- \rightarrow\ b\ W^+\ \bar{b}\ W^-$ is not feasible,
we have employed the double pole approximation. In our case this 
means taking into account only the diagrams which contain two intermediate top
quarks. Unlike most of the previous treatments, we allow for the two top
quarks to be off-shell, and include also the corrections due to 
interference between the top production and decay processes.
}

 This thesis can be roughly split into two parts: the first one deals with the
radiation of a real gluon in top production and decay, and the second one
deals with the computation of virtual corrections and the radiation
of soft gluons in this process.  For the real gluon radiation case, 
we give a method to split the contributing amplitudes into parts 
which can be thought of as associated with the top production or decay 
subprocesses. This allows for the separation of the cross section into
%a part describing gluon radiation in top production, 
%a part describing gluon radiation in top decay, 
parts describing gluon radiation in top production, top decay,
and interference.
We discuss the properties of the gluon radiation, and analyze the impact
the gluon has on top mass reconstruction. We pay special attention  to  
the analysis of interference terms; since the magnitude of these terms 
depends on the top width, they might provide a way to measure this 
quantity at energies above the threshold. Even if this is not possible,
observing experimental evidence
for interference between the radiation at various stage is an interesting 
goal by itself. Although further studies are needed, 
our analysis indicates that, while difficult, this goal is not {\it a priori} 
unreachable.  
 
For the virtual gluon corrections case, we first discuss the evaluation 
of contributing amplitudes in the double pole approximation.
A parallel is drawn between our computation and the DPA evaluation 
of QED corrections to the $W$ pair production and decay process at LEP.
While there are many similarities between the two computations, 
there are also some differences; maybe the most important one is that
there are nonfactorizable (interference) corrections no longer proportional
to the Born amplitude in the top quark case.
 Previous analyses state that the interference
terms cancel out completely in inclusive quantities 
(like the total cross section). We discuss the evaluation of the 
real gluon interference terms using analytic methods and we point out
the shortcomings of this approach.
The total magnitude of nonfactorizable corrections in our computation 
is found to be of order 1\% of the cross section.
We also present results for the total cross section
and the relative nonfactorizable correction to the top invariant 
mass distribution. The effect of nonfactorizable corrections on the 
top mass reconstruction is found to be very small.  

The computations in this thesis are valid at collision energies
above the top -antitop production threshold.
Since we are interested in differential distributions of final state kinematic
variables, the approach used to obtain
the results presented throughout the paper is that of numerical
simulations. The amplitudes are evaluated using spinor techniques,
and the integration over the final state variables is performed using
Monte Carlo techniques. This approach has the added advantage
that it allows for the inclusion of experimentally relevant 
selection criteria on the final state phase space.

The calculation presented here is entirely at the parton level. For more
realistic simulations, it is necessary to take into account 
initial state related issues, like initial state radiation (ISR), beam
energy spread and beamstrahlung. The hadronization of the final 
state partons also has to be modeled. In order to address these issues,
we plan to provide an interface of our code with Pandora (a general
physics event generator for linear collider studies
which includes ISR and beamstrahlung effects) and Pythia (a Monte Carlo
which simulates final state parton shower and hadronization). 

For the future, we plan to extend our computations by including
all the lowest order diagrams contributing to our process. Also,
the evaluation of NLO corrections to the singly resonant diagrams
in the on-shell approximation should prove feasible. Beyond QCD,
we plan to take into account electroweak radiative corrections to the 
top production and decay process, and maybe SUSY corrections too. 
The framework we use for our computations is flexible enough 
to build on the results presented
in this thesis with the final goal of constructing a comprehensive Monte 
Carlo for top related issues at future linear colliders.

%----------------------------------------------------------------------

% ----------------------------------------------------------------------

\appendix
%\chapter{}

\chapter{Spinor techniques}
The standard technique for computing a cross section used to
rely on the evaluation of the square of the amplitude with the
help of trace formulas. More recently, since the complexity of the 
processes being analyzed has increased, this approach has proven
to be too cumbersome. To understand why this is so, note that if
the total amplitude gets $N$ contributions from $N$ different Feynman
diagrams, the evaluation of the square amplitude through trace techniques 
requires the computation of $N^2$ terms.
Thus, in recent years the emphasis has shifted toward the evaluation 
at the amplitude level, using spinor techniques.

These techniques have been developed by a number of people over a number
of years (for some examples, see \cite{CALKUL}). We will use the particular
scheme proposed by Kleiss and Stirling in \cite{kleiss}; for the sake 
of completeness, we review here the general features. 

\section{Spinors describing massive and massless \\ fermions}

A spinor describing a massless particle of momentum $p$ and helicity
$\la$ can be constructed with the help of two auxiliary 4-vectors $k_0$ and
$k_1$ and the basic spinor $u_-(k_0)$:
\bec{fer_0}
 u_{\la}(p)= \frac{\not p}{\sqrt{2pk_0}} u_{-\la}(k_0) 
\eec
The $k_0$ and $k_1$ vectors have to satisfy the following conditions :
\bec{c1}
k_0 k_0 =0, \ \ k_1 k_1 = -1, \ \ k_0 k_1 = 0
\eec
$k_1$ is necessary in order to define the relative complex phase of the 
positive helicity $k_0$ spinor in relation to the negative one : 
$u_+(k_0) = \not{k_1} u_-(k_0)$.

 The basic elements from which the total amplitude can be 
ultimately constructed 
are the products of two massless spinors:
$$ \bar{u}_{\la _1}(p_1) u_{\la _2}(p_2) $$
These basic spinor products can be expressed in terms of a single complex
function $s(p_1,p_2)$:
\bec{a2}
 \bar{u}_+(p_1) u_-(p_2) = s(p_1,p_2)
 \eec
 $$ \bar{u}_-(p_1) u_+(p_2) = t(p_1,p_2)=[s(p_2,p_1)]^*$$
$$\bar{u}_{\la} (p_1) u_{\la} (p_2)=0$$
The exact form of function $s(p_1,p_2)$ depends on the choice for
the auxiliary vectors $k_0$ and $k_1$.
Using the values suggested in \cite{kleiss}:
\bec{k_vals}
k_0=(1,1,0,0), \ \  k_1=(0,0,1,0)
\eec
we have :
\bec{s_expr}
s(p_1,p_2)= (p_1^y-i p_1^z)\frac{\eta(p_2)}{\eta(p_1)}
-   (p_2^y-i p_2^z)\frac{\eta(p_1)}{\eta(p_2)}, \hbox{~~~~ with~~}
\eta(p)=\sqrt{2pk_0} 
\eec

The spinor which describes a massive fermion of momentum $q$ can be similarly
defined, with an added degree of complexity: since the number of inner degrees
of freedom is four for a massive fermion as opposed to two for a massless one,
we have an extra degree of freedom in our definition:
\bec{fer_1}
u(q,s) = \frac{1}{m}(\not{p_1} + \not{p_2} \pm m)u_-(p_2)
\eec
(the minus sign in the parentheses above corresponds to the antiparticle case)
where $p_1$ and $p_2$ are any two momenta which satisfy the constraints:
\bec{ctrs}
p_1^2 = p_2^2 = 0, \ \ \ p_1 + p_2 = q \ .
\eec
What this extra freedom amounts to is the liberty to choose along 
which axis the spin of the fermion points. It can be shown that the spin 
vector is given by 
\bec{spin_v}
s = (p_1 - p_2)/m \ .
\eec
An appropriate choice for the vectors $p_1, p_2$ will allow decomposition 
in the helicity basis, for example. However, since in our 
case we are not interested in specific spin states for the massive fermions
(the spin of the $b$ quarks is not observable, and the top quark is off-shell),
we have chosen a definition which, while not particularly relevant physically,
it is computationally convenient:
\bec{p_def}
p_1 = q - \frac{m^2}{2qk_0} k_0, \ \ \ p_2 = \frac{m^2}{2qk_0} k_0
\eec
Now, since $\not{k_0} u(k_0) = 0$ and $u(\alpha k_0) = \sqrt{\alpha}\ u(k_0)$
(from normalization constraints), we get:
\bec{my_def}
u(q,\pm s) = \frac{\not{q} + m}{\sqrt{2qk_0}}u_{\mp}(k_0), \ \ \
v(q,\pm s) = \frac{\not{q} - m}{\sqrt{2qk_0}}u_{\pm}(k_0)
\eec
$$ \hbox{ with ~~~~} s = \frac{q}{m} - \frac{m}{qk_0}k_0$$
(note that for the antiparticle spinor, the sign of the spin is reversed, as
it should be, since in the massless limit, the antiparticle is identical 
to the particle of opposite helicity).

With these definitions, we get the following expressions for the 
massive spinor products:
\bec{mass_spin}
 \bar{u}_+(q_1) u_-(q_2) = s(q_1,q_2), \ \ \ 
 \bar{u}_-(q_1) u_+(q_2) = t(q_1,q_2)
 \eec
$$\bar{u}_+ (q_1) u_+ (q_2) = \bar{u}_- (q_1) u_- (q_2)
= m_1 \frac{\eta(q_2)}{\eta(q_1)} + m_2 \frac{\eta(q_1)}{\eta(q_2)} $$
with the $s$ and $t$ functions the same as in Eqs. \ref{a2}, \ref{s_expr}.
If one spinor represents an antiparticle, the corresponding mass changes
sign in the second line of the equation above.

\vspace{0.5cm}
The total amplitude can be written in term of the basic spinor products
 discussed above. Terms like
$$ \bar{u}(p_1) {\not p} _i {\not p} _j \ldots u(p_2)$$ 
are evaluated using the completeness relations
\bec{gfkj}
\sum_{\la} u(p) \bar{u}(p) = \not p + m, \ \ \
  \sum_{\la} v(p) \bar{v}(p) = \not p - m \eec
while terms of the form:
$$  [\bar{u}_{\la _1}(p_1) \gamma ^{\mu} u_{\la _2}(p_2)]
   [\bar{u}_{\la _3}(p_3) \gamma _{\mu} u_{\la _4}(p_4)]$$
can be evaluated using the Chisholm identity; for massless spinors:
\bec{a5}
    \bar{u}_{\la _1}(p_1) \gamma ^{\mu} u_{\la _2}(p_2)  \gamma _{\mu} = \left\{
\begin{array}{cl}
2\ [  u_{\la }(p_2) \bar{u}_{\la }(p_1) +  u_{-\la}(p_1) \bar{u}_{-\la}(p_2)]
& \  \hbox{if}\ \la _1 = \la _2 = \la \\
0 &  \ \hbox{if}\ \la _1 \neq \la _2 
\end{array} \right.
\eec

\section{ Spinors describing massless bosons}

 The only massless external boson contributing to our process is the gluon. The
polarization vector for the gluon can be built with the help of an auxiliary
vector $k_a$:
\bec{glu_pol}
\epsilon_{\la }^{\mu} = \frac{1}{\sqrt{4 k k_a} }
\bar{u}_{\la }(k) \gamma^{\mu} u_{\la }(k_a)\ .
\eec 
Here, $\la $ stands for the helicity of the gluon ( + or - ); we take into account
only states of physical, transverse polarization. The freedom given by
the possibility to choose $k_a$ corresponds to freedom in the choice
of the gluon gauge.

 In relation to the gluon gauge choice, it is worth mentioning here an important point concerning numerical instabilities. When the 
gluon momentum and the auxiliary vector for the gluon gauge are
almost parallel ($k k_a \approx 0$), the normalization factor in Eq.
\ref{glu_pol} will become very large. If the total amplitude is gauge invariant,
this is of no concern (besides the fact that there will be large cancellations
between different terms in the amplitude). However, if the total amplitude
contains terms which are not gauge invariant, even if they are usually small,
these terms will be enhanced. In other words, the cancellations
between large terms won't be exact anymore. Obviously this is a problem;
the solution is either to use a completely gauge invariant 
amplitude in our computation, or make sure that the auxiliary vector
$k_a$ points in different direction from the gluon momentum.

\section{ Spinors describing massive bosons}

The treatment of massive boson ($W$) polarization states
actually amounts to letting the $W$ decay into two massless 
particles (an electron and an antineutrino, for example) and integrating over
the momenta of these particles.
The sum over polarizations can be evaluated as follows:
\bec{mass_pol}
\sum_{\epsilon} \epsilon ^{\mu} \epsilon ^{*\nu} \rightarrow 
\frac{3}{8\pi m^2} \int d\Omega\ a^{\mu}a^{*\nu} ~~~ ;
\ a^{\mu}= \bar{u}_{-}(r_1) \gamma ^{\mu} u_{-}(r_2) 
\eec
$r_1$ and $r_2$ being two lightlike vectors
(the momenta of the two massless particles) which add up to the momentum of
the  massive boson, and the integral is over the direction of $r_1$.

In our computation, the polarization vector for the $W$ boson stands for:
\bec{W_pol}
 \not{\epsilon}_W = 
\left( \frac{-i g_W}{\sqrt{2}} \gamma^{\mu} P_L \right) 
\ \frac{-i}{p_W^2 - M_W^2 + i M_W \Gamma_W} \
\left[ \bar{u}(\nu)\ \frac{-i g_W}{\sqrt{2}} \gamma_{\mu} P_L \ u(e^+) \right]
\eec
where $P_L = (1-\gamma^5)/2 $ is the projector on left state helicity. 
Treating the $W$ boson in the narrow width approximation means that, at the
amplitude square level, the term coming from the $W$ propagator is:
$$ \frac{1}{(p_W^2 - M_W^2)^2 + M_W^2 \Gamma_W^2} \ \longrightarrow \
\frac{\pi}{M_W \Gamma_W}\ \delta(p_W^2 - M_W^2)
$$
Therefore, we can replace the quantity in Eq. \ref{W_pol} with an effective
polarization:
\bec{W_pol_eff}
 \not{\epsilon}_W \ = \ 
\left( \frac{g_W^2}{2} \sqrt{\frac{\pi}{M_W \Gamma_W}} \right)
 \gamma^{\mu} P_L 
\left[ \bar{u}_-(\nu)\ \gamma_{\mu}  \ u_-(e^+) \right]
\eec
at the amplitude level, while in the differential cross section:
\bec{W_crsec}
d\Omega_W = \left. \int\  \frac{d^3 p_{\nu}}{(2\pi)^3 2 E_{\nu}}\
		\frac{d^3 p_{e^+}}{(2\pi)^3 2 E_{e^+}}\ 
		\delta(p_W - p_{\nu} - p_{e^+})
\right|_{p_W^2 = M_W^2}
\eec
%-------------------------------------------------------------------
\chapter{Amplitudes and cross sections formulas}

In this section, we will present some formulas for amplitudes and
cross sections.

\section{Tree level amplitudes}

For the tree level diagrams corresponding to the lowest order process
in Figure \ref{tree_level}, the amplitude
can be written as :
\bec{tree_lev_amp}
{\cal{M}}_0 = \bar{u}(p_b) \not{\epsilon}_{W^+} 
\frac{\not{p}_t + m_t }{p_t^2 - \bar{m}_t^2}
\ \Gamma_{\gamma,Z_0} \ 
\frac{-\not{p}_{\at} + m_t }{p_{\at}^2 - \bar{m}_t^2}
\not{\epsilon}_{W^-} v(p_{\ab})
\eec
with the following notations:
{ %\small
\bec{gam_v}
\Gamma_{\gamma,Z_0} = \left. [ \bar{u}(p_2) (i e \gamma^{\mu}) u(p_1) ]\
\frac{-i\ g_{\mu \nu}}{q^2} \ (-i e Q_t \gamma^{\nu}) \ + \right.
\eec
$$ \left[ \bar{u}(p_2) \left(-i \frac{g_W}{\hbox{cos}\ \theta_W } \gamma^{\mu}
\frac{1}{2}(V^e - A^e \gamma^5) \right) u(p_1) \right]\
\frac{-i\ g_{\mu \nu}}{q^2 - M_Z^2 + i M_Z \Gamma_Z }\ \times
$$
$$
\left(-i \frac{g_W}{\hbox{cos}\ \theta_W } \gamma^{\nu}
\frac{1}{2}(V^t - A^t \gamma^5) \right)
$$
}
(the first line represents the contribution of the photon exchange
diagram, while the second line the contribution of the $Z_0$ boson 
exchange diagram). Here $g_w$ is the weak coupling constant, $\theta_W$
is the Weinberg angle, and $Q_t = 2/3$ is the electric charge of the
top quark in units of positive electron charge $e$. $p_1$ and $p_2$ are 
the momenta of the initial state electron and positron; $q^2 = (p_1 + p_2)^2$
is the square of the total energy available for the process. The couplings
of the electron and top quark to the gauge boson $Z_0$ are given by the 
vector and axial coupling parameters:
$$ V^e = -\frac{1}{2} + 2\ {\hbox{sin}^2 \theta_W } , \ \ \
A^e = -\frac{1}{2} $$
$$ V^t = \ \ \frac{1}{2} - \frac{4}{3}\ {\hbox{sin}^2 \theta_W } , \ \ \
A^t = \ \ \frac{1}{2} $$
We shall usually denote by $\tilde{\cal{M}} $ the amplitudes multiplied
by the denominators of the fermion propagators; for example:
$$ {\cal{M}}_0 = \tilde{{\cal{M}}_0} \ 
\frac{1 }{p_t^2 - \bar{m}_t^2} \ \frac{1 }{p_{\at}^2 - \bar{m}_t^2}
$$
Throughout the paper, we also use the definitions:
$$ p_t = p_{W^+} + p_b \ , \ 
p_{\at} = p_{W^-} + p_{\ab} \ , \ 
\bar{m}_t^2 = m_t^2 -i m_t \Gamma_t \ , $$
and $k$ is the momentum of the gluon.

For the process with a real gluon radiated, the amplitudes are obtained
by inserting the corresponding gluon polarization vector in the appropriate
place, taking into account the extra fermionic propagator, and modifying 
the momenta in the appropriate way.
 For example, for the diagram with the gluon radiated from the top
quark:
{\small 
\bec{mtopg}
{\cal{M}}_t = \bar{u}(p_b) \not{\epsilon}_{W^+} 
\frac{\not{p}_t + m_t }{p_t^2 - \bar{m}_t^2}
(-i g_s \not{\epsilon}_g) 
\frac{\not{p}_t + \not{k} + m_t }{(p_t+k)^2 - \bar{m}_t^2}
\ \Gamma_{\gamma,Z_0} \ 
\frac{-\not{p}_{\at} + m_t }{p_{\at}^2 - \bar{m}_t^2}
\not{\epsilon}_{W^-} v(p_{\ab})
\eec}
while for the diagram with the gluon radiated from the antibottom
quark:
{\small
\bec{mabg}
{\cal{M}}_{\ab} = \bar{u}(p_b) \not{\epsilon}_{W^+} 
\frac{\not{p}_t + m_t }{p_t^2 - \bar{m}_t^2}
\ \Gamma_{\gamma,Z_0} \ 
\frac{-\not{p}_{\at} - \not{k} + m_t }{(p_{\at}+k)^2 - \bar{m}_t^2}
\not{\epsilon}_{W^-} 
\frac{-\not{p}_{\ab} - \not{k} + m_b }{(p_{\ab}+k)^2 - m_b^2}
(-i g_s \not{\epsilon}_g) v(p_{\ab})
\eec}
Using the decomposition of the top propagator products described in section
{\bf 3.1.1}, we obtain the following expressions for the gauge invariant 
amplitudes in Eq \ref{mrg_gi} :
\bea
{\cal{M}}_{prod} & = & \left(\ \  \frac{ \tilde{\cal{M}}_t}{2 p_t k}\ + \ 
\frac{ \tilde{\cal{M}}_{\at} }{2 p_{\at} k} \right)\
\frac{1}{D(p_t)}\ \frac{1}{D(p_{\at})}
\\
{\cal{M}}_{tdecay} & = & \left( -\frac{ \tilde{\cal{M}}_t}{2 p_t k}\ + \ 
\frac{ \tilde{\cal{M}}_b }{2 p_b k} \right)\
\frac{1}{D(p_t+k)}\ \frac{1}{D(p_{\at})} \nonumber
\\
{\cal{M}}_{{\at}decay} & = & 
\left( \ \ \frac{ \tilde{\cal{M}}_{\at} }{2 p_{\at} k}\ - \ 
 \frac{ \tilde{\cal{M}}_{\ab}}{2 p_{\ab} k} \right)\
\frac{1}{D(p_t)}\ \frac{1}{D(p_{\at}+k)} \nonumber
\eea
where $D(p) = p^2 - \bar{m}_t^2$.

We also give here the amplitudes in the (extended) soft gluon approximation.
The evaluation in this case proceeds as in Eqs. \ref{gi_tprod1}, 
\ref{gi_tdec1}, and we drop the $\not{\epsilon}_g \not{k}$ terms in the 
denominators of Eqs. \ref{gi_tprod2}, \ref{gi_tdec2}. Then
\bec{fja} \tilde{\cal{M}}_{prod,tdecay}^{(t)} (ESGA) \ = \ (-i g_s) \
(2 \epsilon_g p_t)\ \tilde{\cal{M}}_0 \eec
and, evaluating the other amplitudes in a similar manner, we get:
\bea
{\cal{M}}_{prod} (ESGA) & = & (-ig_s \epsilon_{g\mu})\ \tilde{\cal{M}}_0 
\left(\ \ \frac{ p_t^\mu }{ p_t k}\ - \ 
\frac{ p_{\at}^\mu}{ p_{\at} k} \right)\
\frac{1}{D(p_t)}\ \frac{1}{D(p_{\at})}
\\
{\cal{M}}_{tdecay} (ESGA) & = & (-ig_s \epsilon_{g\mu})\ \tilde{\cal{M}}_0
\left( -\frac{ p_t^\mu}{ p_t k}\ + \ 
\frac{ p_b^\mu }{ p_b k} \right)\
\frac{1}{D(p_t+k)}\ \frac{1}{D(p_{\at})} \nonumber
\\
{\cal{M}}_{{\at}decay} (ESGA) & = & (-ig_s \epsilon_{g\mu})\ \tilde{\cal{M}}_0
\left( - \frac{ p_{\at}^\mu }{ p_{\at} k}\ + \ 
 \frac{ p_{\ab}^\mu }{ p_{\ab} k} \right)\
\frac{1}{D(p_t)}\ \frac{1}{D(p_{\at}+k)} \nonumber
\eea
 Neglecting the gluon momentum in the denominator of the propagators in the
above equations (a valid approximation if the gluon energy is much smaller than
the top width) we obtain the expression (Eq. \ref{mrg_ir})
used in the evaluation of the 
infrared singular part of the cross section:
\bec{fdsa}
{\cal{M}}_{tot}^{sg} = (-ig_s \epsilon_{g\mu})\
{\cal{M}}_0 \ \left( \frac{p_b^{\mu}}{k p_b} - 
\frac{p_{\ab}^{\mu}}{k p_{\ab}} \right)\ 
\eec
%----------------------------------------------------------------------------
\section{Virtual corrections amplitudes}

The amplitude for the general vertex correction in Figure \ref{ver_cor}
 can be written as: 
\bec{gen_gam}
 \delta \Gamma^{\mu} = \frac{\alpha_s}{4 \pi} \
\int \frac{d^4 k}{i \pi^2}\ \frac{1}{k^2}\ \gamma^{\nu}\
\frac{\not{p}_1 - \not{k} + m_1}{(p_1 - k)^2 - \bar{m}_1^2}\
\gamma^{\mu}(C_V + C_A \gamma^5 ) \
\frac{-\not{p}_2 - \not{k} + m_2}{(p_2 + k)^2 - \bar{m}_2^2}\
\gamma_{\nu}
\eec
Upon evaluation of the integral, the result can be written in terms of two
sets of form-factors; one for the vectorial part of the vertex correction, one
for the axial part. The number of form factors needed depends on the specific 
constraints on the process; in our case, when the momenta $p_1, p_2$ are 
off-shell, we need eight$\times 2$ form factors:
\bec{gen_gam_res}
\delta \Gamma^{\mu}_V = \frac{\alpha_s}{4 \pi}\
\sum _{i=1,8} \left( C_V F_i^V T_i^{V \mu} \ + \ C_A F_i^A T_i^{A \mu} \right)
\eec

The definition of these form factors depends on the choice of the spinorial
elements in terms of which the result is written. In our case, we shall 
define:
\bec{del_gam_res}
\delta \Gamma^{\mu}_V = \frac{\alpha_s}{4 \pi} C_V\ \times
\eec
$$
\begin{array}{rclcccccl}
          [        & p_1^\mu & & F_1^V &  + 
 		&                   &  \gamma^{\mu} & & F_2^V \ + \\
(\not{p}_1 - m_1) & p_1^\mu & & F_3^V &  + 
		& (\not{p}_1 - m_1) &  \gamma^{\mu} & & F_4^V \ + \\
                  & p_1^\mu & (-\not{p}_2 - m_2) & F_5^V &  + 
 		& &  \gamma^{\mu} & (-\not{p}_2 - m_2) & F_6^V \ + \\
(\not{p}_1 - m_1) & p_1^\mu & (-\not{p}_2 - m_2) & F_7^V &  + 
& (\not{p}_1 - m_1) &  \gamma^{\mu} & (-\not{p}_2 - m_2)& F_8^V \ ]
\end{array}
$$
for the vectorial part of the vertex correction; for the axial one, replace
$p_1^\mu, \ \gamma^\mu$ with 
$p_1^\mu \gamma^5, \ \gamma^\mu \gamma^5$ in the expression above. This
definition has the advantage that when the particle $i$ is on-shell, the
terms which contain the $ \pm \not{p}_i + m_i$ drop out. Also, we have 
made use of the fact that, if $\delta \Gamma^{\mu}$ multiplies $A_{\mu}$ in 
the full matrix element ($A_{\mu}$ can be thought of as the polarization
vector of the weak gauge boson in diagram \ref{ver_cor}), then 
$(p_1+p_2)^{\mu} A_{\mu} = 0$.

 We shall evaluate and write the results for the form factors 
in terms of Passarino - Veltman functions
(for the definition of these see section Appendix {\bf C}):

\bec{f_vect}
F_1^V = 4[\ m_1(C_{12}-C_{11}+C_{23}-C_{21}) - m_2(C_{12}+C_{23}) \ ]
\eec
$$ F_2^V = -2[\ 2 p_1 p_2(C_0 + C_{11}) + 2(C_{24}-1/4) + 2B_0^{12} -1  
$$
$$
- m_1 m_2 C_{11}
+ p_1^2(-C_{11}+C_{12}) - p_2^2 C_{12} \ ] $$
$$ F_3^V = -4 ( C_0 + 2C_{11} - C_{12} + C_{21} - C_{23} ) \ \ , \ \ 
 F_4^V =  2 m_2 (C_0 + C_{11}) $$
$$ F_5^V = -4 ( C_0 + C_{11} + C_{23} + C_{12} ) \ \ , \ \ 
 F_6^V = 2 m_1 (C_0 + C_{11}) $$
 $$
F_7^V = 0 \ \ , \ \ F_8^V = 2 ( C_0 + C_{11} )
 $$

For the computation of the axial form factors, we can shift $\gamma^5$ in  
Eq. \ref{gen_gam} to the right, and then perform the same evaluations as in 
the vectorial case.
In the result for the form factors, this amounts to changing the sign of $m_2$,
$ m_2 \rightarrow -m_2$, and multiply the $F_5, \ldots, F_8$
 with $(-1)$.
 Then :
\bec{f_axial}
 F_1^A = F_1^V + 8 m_2 (C_{12}+C_{23}) \ \ , \ \ 
 F_2^A = F_2^V - 4m_1 m_2 C_{11}
\eec
$$ F_3^A = \ \ F_3^V \ \ , \ \ 
 F_4^A = -F_4^V $$
$$ F_5^A = -F_5^V \ \ , \ \ 
 F_6^A = -F_6^V $$
$$ F_7^A = -F_7^V \ \ , \ \ F_8^A = -F_8^V$$

The result \ref{del_Z2_fin} for the fermion self-energy corrections:
\bec{del_Z2_fin2}
\Delta Z_2(p) =  \left( [\Sigma_a(p^2) + 2 \Sigma_{ir}(p^2)] \ + \
\frac{\Sigma_{ir}(p^2)}{m}( \pm \not{p} - m ) \right)
\eec
 can
be similarly written in terms of Passarino-Veltman functions:
\bec{f_fse}
\Sigma_a(p^2) = \frac{\alpha_s}{4 \pi} (1 + 2B_1(p^2,\bar{m}^2))
\eec
$$
\Sigma_{ir}(p^2) = \frac{\alpha_s}{4 \pi} \frac{m^2}{p^2 - \bar{m}^2}
\left[ 4 \Delta B_0(p^2,\bar{m}^2) + 4 \Delta B_1(p^2,\bar{m}^2) \right]
$$
with 
$$ \Delta B_n(p^2,\bar{m}^2) = B_n(p^2,\bar{m}^2) - B_n(\bar{m}^2,\bar{m}^2)
\ , \ n = 0,1$$
If we further define the $X_0, X_1$ form factors through:
\bec{del_Z2_fin3}
\Delta Z_2(p) = 2\ \frac{\alpha_s}{4 \pi} \left[ X_0(p^2) \ + \ X_1(p^2)
( \pm \not{p} - m ) \right]
\eec
the renormalized vertex correction in Eq. \ref{ren_ver2}:
\bec{ren_ver3}
\delta \Gamma^{\mu}_{ren} \ = \
\delta \Gamma^{\mu} \ + \ \frac{1}{2}\ \Delta Z_2(p_1)\ \Gamma^{\mu}
\ + \ \frac{1}{2}\ \Gamma^{\mu}\ \Delta Z_2(p_2) 
\eec
can be obtained by making the following redefinitions of form-factors in 
Eq. \ref{f_vect}, \ref{f_axial}:
\bec{ren_f_fact}
F_2^{V,A} \ \to \ F_2^{V,A} + X_0(p_1^2) + X_0(p_2^2)
\eec
$$ F_4^{V,A} \ \to \ F_4^{V,A} + X_1(p_1^2)$$
$$ F_6^{V,A} \ \to \ F_6^{V,A} + X_1(p_2^2)$$

These general results are easily translated for the specific cases which
appear in our computation.  Take 
$$ F_{i,t \at}^{V,A} \ = \ F_i^{V,A}(
 p_1 = p_t, \ p_2 = p_{\bar{t}}, \ m_1 = m_2 = m_t ) $$
for the correction to the top, antitop production vertex,
$$ F_{i,t b}^{V,A} \ = \ F_i^{V,A}(
p_1 = p_b, \ p_2 = -p_t ,\ m_1 = m_b, \ m_2 = m_t ) $$
for the correction to the top decay vertex, and
$$ F_{i,\at \ab}^{V,A} \ = \ F_i^{V,A}(
p_1 = -p_{\bar{t}}, \ p_2 = p_{\bar{b}} ,\ m_1 = m_t, \ m_2 = m_b )$$
for the correction to the antitop decay vertex. 
%With the definitions: 
Decomposing the top production and top decay interaction vertices in a
vectorial and an axial part:
\bec{gam_dec}
 \Gamma_{\gamma,Z_0} \ = \ V^t\ \Gamma_{\gamma,Z_0}^{V,\mu} \gamma_{\mu}\
  - \ A^t \ \Gamma_{\gamma,Z_0}^{A,\mu} \gamma_{\mu}\gamma^5
\eec \nopagebreak[3]
$$ \not{\epsilon}_{W^+,W^-} \ = \ 
\frac{1}{2} \not{\epsilon}_{W^+,W^-}^{V,\mu} \gamma_{\mu}\
 -\ \frac{1}{2} \not{\epsilon}_{W^+,W^-}^{A,\mu} \gamma_{\mu}\gamma^5
$$
(see also Eqs. \ref{gam_v}, \ref{W_pol_eff}) we can write:
\bec{gen_m_vir}
\tilde{\cal{M}}_{t \at} = \frac{\alpha_s}{4 \pi} \sum_{i} \left[
C_V^{t \at}\ F_{i,t \at}^V\ T_{i,t \at}^V\ + 
C_A^{t \at}\ F_{i,t \at}^A\ T_{i,t \at}^A \right] 
\eec
$$
\tilde{\cal{M}}_{t b} = \frac{\alpha_s}{4 \pi} \sum_{i} \left[
C_V^{t b}\ F_{i,t b}^V\ T_{i,t b}^V\ + 
C_A^{t b}\ F_{i,t b}^A\ T_{i,t b}^A \right] 
$$
$$
\tilde{\cal{M}}_{\at \ab} = \frac{\alpha_s}{4 \pi} \sum_{i} \left[
C_V^{\at \ab}\ F_{i,\at \ab}^V\ T_{i,\at \ab}^V\ + 
C_A^{\at \ab}\ F_{i,\at \ab}^A\ T_{i,\at \ab}^A \right]
$$
where
\bec{gen_T_vir}
T_{i,t \at}^{(V,A)} = \bar{u}(p_b) \not{\epsilon}_{W^+}  (\not{p}_t + m_t) \
\Gamma_{\gamma,Z_0}^{(V,A),\mu} T_{i,t \at, \mu}^{(V,A)} \ 
(-\not{p}_{\bar{t}} + m_t) 
\not{\epsilon}_{W^-}  v(p_{\bar{b}})
\eec
$$
T_{i,t b}^{(V,A)} = \bar{u}(p_b)\ \not{\epsilon}_{W^+}^{(V,A),\mu}
T_{i,t b, \mu}^{(V,A)}\ (\not{p}_t + m_t) \ \Gamma_{\gamma,Z_0} \
(-\not{p}_{\bar{t}} + m_t) 
\not{\epsilon}_{W^-}  v(p_{\bar{b}})
$$
$$
T_{i,\at \ab}^{(V,A)} = \bar{u}(p_b)\ \not{\epsilon}_{W^+}\ 
(\not{p}_t + m_t) \ \Gamma_{\gamma,Z_0} \
(-\not{p}_{\bar{t}} + m_t) \
\not{\epsilon}_{W^-}^{(V,A),\mu} T_{i,\at \ab, \mu}^{(V,A)}\
v(p_{\bar{b}})
$$
and
$$ 
C_V^{t \at} = V^t \ ,\ C_A^{t \at} = -A^t \ ,\
C_V^{t b} = C_V^{\at \ab} = \frac{1}{2} \ ,\ 
C_A^{t b} = C_A^{\at \ab} = -\frac{1}{2}\ .
$$
%----------------------------------------------------------------------------
\section{Cross sections and color factors}

The cross section for the generic process $(p_1 p_2) \rightarrow (p_i p_j ...)$
can be written as:
\bec{crsgen}
d \sigma_{i\rightarrow f} = \frac{(2\pi)^4}{2 E_1 2 E_2 | v_1 - v_2 |}\
\sum_{spins} |M_{i\rightarrow f}|^2 \ d \Omega_f
\eec
where $d \Omega_f$ is the differential volume element in the final 
state phase space:
\bec{omega}
d \Omega_f = \prod_{final states} \frac{d^3 p_f}{(2\pi)^3 2 E_f} 
\ \delta^4(P)
\eec
The sum over spins in Eq. \ref{crsgen} is performed over the spins of the 
final state particles. If we consider unpolarized electrons in the 
initial state, we sum over the helicities of these particles as well and
multiply by a $1/4$ average spin factor.
 
 In the process under consideration here, the initial collision takes
place in the center-of-mass frame. Therefore $E_1 = E_2 = W/2$, where
$W$ is the total available energy for the collision. Also, since the 
energies involved are much bigger than the electron mass (which is 
considered zero for all purposes in our computation), the speed of particles
in the initial state is $c$.

 Actually, in the formulas for the amplitudes presented in the previous
sections there is an element missing. Since our final state particles
include quarks and gluons, the amplitude for the process will contain
a color index as well. However, because the experimental observables are
colorless quantities, in the computation of the final cross section we
have to sum over these color indices.

Since the initial state is colorless, the bottom and antibottom quarks
in the final state of the lowest order process have the same color: 
$$ {\cal{M}}^0 \ \rightarrow \ {\cal{M}}^0_{ij} = {\cal{M}}^0\ \delta_{ij} $$
(here $i$ would be the color index of the bottom quark, and $j$ the
color index of the antibottom quark). Summing over the colors in the
cross section will give us a factor of 3:
$$ \sigma^0 \sim \sum_{i,j} |{\cal{M}}^0_{ij}|^2 \ = \ 3\ |{\cal{M}}^0|^2$$
In the case of a real gluon in the final state, taking the gluon
coupling to a pair of quarks with color indices $i,j$ to be 
$-i g_s T^a_{ij}$ , we have:
$$ {\cal{M}}^{rg} \ \rightarrow \ ({\cal{M}}^{rg})^{a}_{ij}
\ = \  {\cal{M}}^{rg}\ T^a_{ij}$$
where $a$ is the color index of the gluon. The matrices $T^a_{ij}$ form
a representation of the gauge group of QCD -- SU(3). In order to perform
the sum over the color indices, we use the relation:
$$ Tr [ T^a\ T^b ] = \frac{1}{2}\delta_{a,b}$$  
then:
$$ \sigma^{rg} \sim \sum_{a,i,j} \ |{\cal{M}}^{rg}|^2\ T^a_{ij}\ (T^a_{ij})^*
= \sum_{a} |{\cal{M}}^{rg}|^2\ Tr[ T^a\ T^a ] = 4 \ |{\cal{M}}^{rg}|^2
$$
The cross section in this case acquires a factor of 4 due to color
summation.

In the case of virtual corrections, all the contributing amplitudes
get a factor:
$$  {\cal{M}}^{vg} \ \rightarrow \ ({\cal{M}}^{vg})_{ij} \ = \ 
({\cal{M}}^{vg})\ \sum_{a,k} T^a_{k,i} T^a_{k,j}
\ = \ C_F\ {\cal{M}}^{vg}\ \delta_{ij}
$$
Here $k$ is the color index of the quarks inside the gluon loop. For the
sum over color indices in the equation above, we have used:
$$ \sum_{a}\ [T^a\ T^a]_{ij} = C_F\ \delta_{ij}, \ \ \ 
C_F = \frac{N^2-1}{2N} = \frac{4}{3} \ \ \hbox{for~~} N = 3
$$
In the cross section, the virtual gluon contribution gets a factor of 3
from the sum over the color indices of the quarks in the final state.
Note that the $C_F$ factor is not included in the expressions for
${\cal{M}}^{vg}$ throughout the paper.

%-----------------------------------------------------------------------
\chapter{Passarino-Veltman functions}

The method of Passarino-Veltman (PV) functions \cite{passa} provides a 
systematic way of evaluating the tensor integrals appearing in the
computation of loop corrections. For example, the result for the 
vertex correction integral \ref{gen_gam} can be written in 
terms of the following tensor functions:
\bec{c_tensor}
{\cal{C}}^{ \{0,\mu,\mu \nu \} } =  
\int \frac{d^4 k}{i \pi^2}\ \frac{ \{1,k^\mu,k^\mu k^\nu \} }
{(k^2 - m_1^2)\ ((k+p_1)^2 - m_2^2)\ ((k+p_1+p_2)^2-m_3^2)}\
\eec
Following the Passarino-Veltman method, these functions can be 
written in term of scalar functions:
\bea
\label{c_decom}
{\cal{C}}^{\mu} & = & p_1^{\mu} {\cal{C}}_{11} + p_2^{\mu} {\cal{C}}_{12} \\
{\cal{C}}^{\mu \nu}& = & p_1^{\mu} p_1^{\nu} {\cal{C}}_{21} + 
p_2^{\mu} p_2^{\nu} {\cal{C}}_{22} + 
\left(p_1^{\mu} p_2^{\nu} + p_2^{\mu} p_1^{\nu}\right){\cal{C}}_{23}
+ g^{\mu \nu} {\cal{C}}_{24} \nonumber
\eea
Furthermore, the ${\cal{C}}_{ij}$ functions 
in the above equations can be written
in terms of the scalar one-, two- and three-point
integrals ${\cal{A}}^0, {\cal{B}}^0 $ and
$ {\cal{C}}^0$. For completeness, we give here the formulas used to compute
these functions 
(these formulas can be easily derived
by multiplying the Eqs. \ref{c_decom} by $p_{1\mu}, p_{2\mu},
g_{\mu \nu}$ and solving the resulting systems of equations;
they can also be found in \cite{passa}):
\bec{c_res1}
\left( \begin{array}{c} {\cal{C}}_{11} \\ {\cal{C}}_{12} \end{array}
\right) \ = \ X^{-1}\ \left( 
\begin{array}{c} {\cal{B}}_0^{(13)} - {\cal{B}}_0^{(23)} + f_1 {\cal{C}}^0 \\
	{\cal{B}}_0^{(12)} - {\cal{B}}_0^{(13)} + f_2 {\cal{C}}^0
	\end{array} \right)
\eec
\bec{c_res2}
{\cal{C}}_{24}\ = \ \frac{1}{4} + \frac{1}{4} {\cal{B}}_0^{(23)} +
\frac{1}{2} m_1^2 {\cal{C}}^0 - 
\frac{1}{4} (f_1 {\cal{C}}_{11}+ f_2 {\cal{C}}_{12})
\eec
\bea
\label{c_res3}
\left( \begin{array}{c} {\cal{C}}_{21} \\ {\cal{C}}_{23} \end{array}
\right) & = & X^{-1}\ \left( 
\begin{array}{c} {\cal{B}}_1^{(13)} + {\cal{B}}_0^{(23)} + 
		f_1 {\cal{C}}_{11} - 2 {\cal{C}}_{24} \\
	{\cal{B}}_1^{(12)} - {\cal{B}}_1^{(13)} + f_2 {\cal{C}}_{11}
	\end{array} \right) 
\\
\left( \begin{array}{c} {\cal{C}}_{23} \\ {\cal{C}}_{22} \end{array}
\right) & = & X^{-1}\ \left( 
\begin{array}{c} {\cal{B}}_1^{(13)} - {\cal{B}}_1^{(23)} + 
		f_1 {\cal{C}}_{12} \\
	 - {\cal{B}}_1^{(13)} + f_2 {\cal{C}}_{12} - 2 {\cal{C}}_{24}
	\end{array} \right) \ . \nonumber
\eea

Here we have used the following notations:
\bec{hic}
X \ = \ 2 \left( 
\begin{array}{cc} p_1^2 & p_1 p_2 \\ p_1 p_2 & p_2^2 \end{array} 
\right)
\eec
\bea
f_1 & = & m_2^2 - m_1^2 -p_1^2 \\
f_2 & = & m_3^2 - m_2^2 -(p_1+p_2)^2 + p_1^2
\nonumber
\eea
\bea
{\cal{B}}^{ \{0,\mu \} } (p_1,m_1,m_2) & = &  
\int \frac{d^4 k}{i \pi^2}\ \frac{ \{1,k^\mu \} }
{(k^2 - m_1^2)\ ((k+p_1)^2 - m_2^2) }
\\
 {\cal{B}}^{\mu} (p_1,m_1,m_2) & = & p_1^{\mu} {\cal{B}}_1 (p_1^2,m_1^2,m_2^2) 
\nonumber
\\
2p_1^2 {\cal{B}}_1 (p_1^2,m_1^2,m_2^2) & = & 
{\cal{A}}^0(m_1^2) - {\cal{A}}^0(m_2^2)
+ f_1 {\cal{B}}_0 (p_1^2,m_1^2,m_2^2) \nonumber
\eea
\bea
{\cal{B}}_n^{(12)} & = & {\cal{B}}_n(p_1^2,m_1^2,m_2^2) \\
{\cal{B}}_n^{(13)} & = & {\cal{B}}_n((p_1+p_2)^2,m_1^2,m_3^2) \nonumber \\
{\cal{B}}_n^{(23)} & = & {\cal{B}}_n(p_2^2,m_2^2,m_3^2) \ , \ \ \ \ \ \ n = 0,1
\nonumber
\eea
and, finally,
\bec{gfd}
{\cal{A}}^0(m^2) \ = \  \int \frac{d^4 k}{i \pi^2}\ \frac{1} {(k^2 - m^2)} \ .
\eec
The ultraviolet divergent integrals in the above expressions are evaluated
with the help of the dimensional regularization method.
 
The above equations follow the standard definition of the 
PV functions; we shall denote the ${\cal{C}}$ 
functions in Eq. \ref{c_tensor}, \ref{c_decom}
by ~ ${\cal{C}}^{\{\mu,\mu \nu \}}(p_1,p_2,m_1,m_2,m_3)$
and the ${\cal{C}}$ 
functions in Eqs. \ref{c_res1}, \ref{c_res3}, \ref{c_res3} by ~
${\cal{C}}_{ij}(p_1^2,p_2^2,m_1^2,m_2^2,m_3^2)$. 
The $C$ function used in Appendix {\bf B} are defined by:
\bec{cc_decom}
C^{ \{0,\mu,\mu \nu \} } =  
\int \frac{d^4 k}{i \pi^2}\ \frac{ \{1,k^\mu,k^\mu k^\nu \} }
{(k^2 + i\epsilon)\ ((k - p_1)^2 - m_1^2)\ ((k + p_2)^2-m_2^2)}\
\eec
\bea
C^{\mu} & = & -p_1^{\mu} C_{11} + (p_2-p_1)^{\mu} C_{12}  \\ 
C^{\mu \nu}& = & p_1^{\mu} p_1^{\nu} C_{21} + 
(p_2-p_1)^{\mu} (p_2-p_1)^{\nu} C_{22} - 
\left[p_1^{\mu} (p_2-p_1)^{\nu} + (p_2-p_1)^{\mu} p_1^{\nu}\right]C_{23}
\nonumber\\ 
 & & +\ g^{\mu \nu} C_{24} \nonumber
\eea
therefore:
\bea
C^{ \{\mu,\mu \nu \} }  & = & 
	{\cal{C}}^{\{\mu,\mu \nu \}}(-p_1,p_2-p_1,0,m_1,m_2) \\
C_0,C_{ij} & = & {\cal{C}}^0, {\cal{C}}_{ij}(p_1^2,(p_2 - p_1)^2,0,m_1^2,m_2^2)
 .
\nonumber
\eea
Moreover, the function $B_0^{12}$ in Eq. \ref{f_vect} is given by:
\bec{jfdg}
B_0^{12} \ = \ {\cal{B}}^0((p_1+p_2)^2,m_1^2,m_2^2)
\eec 
and in Eqs. \ref{f_fse}:
\bec{fhgkh}
B_n(p^2,m^2) = {\cal{B}}_n(p^2,0,m^2) \ , \ \hbox{for  } n = 0 , 1.
\eec

%\section{Phase-space generators}

\end{document}